\documentclass[11pt]{article}

\let\phi\varphi

\usepackage{lipsum}
\usepackage{amsmath}
\usepackage{amsfonts}
\usepackage{amsthm}
\usepackage{amssymb}
\usepackage{relsize}
\usepackage[utf8]{inputenc}
\usepackage{graphicx}
\usepackage[T1]{fontenc}
\usepackage{setspace}
\usepackage{fancyhdr}
\usepackage{afterpage}
\usepackage{placeins}
\usepackage{hyperref}
\usepackage[capitalize,nameinlink]{cleveref}
\usepackage[english]{babel}

\usepackage[font=small,labelfont=bf]{caption}
\usepackage{url}
\usepackage[left=2cm,right=2cm,top=2.5cm,bottom=2.5cm]{geometry}

\usepackage{tikz}
\usetikzlibrary{arrows.meta}
\usetikzlibrary{decorations.pathmorphing}
\usetikzlibrary{shapes}
\tikzstyle{causalarrow} = [draw=black,thick,decorate,decoration={snake,amplitude=.7mm,pre length=1mm,post length=2mm}]
\tikzstyle{affectsarrow} = [draw=black,thick]
\usepackage{tikz-3dplot}
\tdplotsetmaincoords{80}{45}
\tdplotsetrotatedcoords{-90}{180}{-90}

\usepackage{verbatim}

\usepackage{bbold}
\usepackage{fdsymbol}

\let\subsetneq\bettersubsetneq

\usepackage{braket}
\usepackage{csquotes}

\usepackage[backend=bibtex,sorting=none,giveninits=true]{biblatex}
\usepackage{stmaryrd}
\SetSymbolFont{stmry}{bold}{U}{stmry}{m}{n}
\usepackage[header]{appendix}
\usepackage{tabularx}
\usepackage{mathtools}
\usepackage{mathrsfs}
\usepackage{booktabs}
\usepackage{mdframed}
\usepackage{xcolor}
\DefineBibliographyExtras{english}{}

\numberwithin{equation}{section}

\usepackage{xstring}
\usepackage{subcaption}
\usepackage{multirow}

\graphicspath{
	{./tikz/}
	{./pics/}
}

\newcommand{\CC}{\ensuremath{\mathcal{C}}}

\let\AA\relax
\newcommand{\AA}{\ensuremath{\mathcal{A}}}
\newcommand{\BB}{\ensuremath{\mathcal{B}}}
\newcommand{\DD}{\ensuremath{\mathcal{D}}}
\newcommand{\TT}{\ensuremath{\mathcal{T}}}
\newcommand{\WW}{\ensuremath{\mathcal{W}}}
\newcommand{\XX}{\ensuremath{\mathcal{X}}}
\newcommand{\YY}{\ensuremath{\mathcal{Y}}}
\newcommand{\ZZ}{\ensuremath{\mathcal{Z}}}
\newcommand{\SC}{\ensuremath{\mathcal{S}}}
\newcommand{\RC}{\ensuremath{\mathcal{R}}}
\newcommand{\Fut}{\bar{\mathcal{F}}}

\newcommand{\abs}[1]{\left| #1 \right|}

\newcommand{\indep}{\ensuremath{\upvDash}}
\newcommand{\unord}{\ensuremath{\not\preceq \not\succeq}}

\newcommand{\editorial}[1]{}

\DeclareMathOperator{\spann}{span}

\DeclareMathOperator{\parent}{par}
\DeclareMathOperator{\child}{child}
\DeclareMathOperator{\anc}{anc}
\DeclareMathOperator{\desc}{desc}
\DeclareMathOperator{\doo}{do}

\DeclareMathOperator{\affects}{\models}
\DeclareMathOperator{\naffects}{\not\models}

\DeclareMathOperator{\cause}{\twoheadrightarrow}

\DeclareMathOperator{\dircause}{\longrightsquigarrow}
\DeclareMathOperator{\diresuac}{\longleftsquigarrow}
\DeclareMathOperator{\potcause}{\vdash}

\DeclareMathOperator{\actpotcause}{\mathrel{\enspace\mkern-6mu{\cause}\mkern-20mu{\color{white}\bullet}\mkern-16mu\dashrightarrow\enspace}}

\DeclareMathOperator{\given}{\,|\,}
\newcommand{\hin}{\enquote{$\Rightarrow$}:\ }
\newcommand{\rueck}{\enquote{$\Leftarrow$}:\ }

\newtheoremstyle{note}
{1.2em}
{1.2em}
{}
{}
{\bfseries}
{.}
{.5em}
{}
\theoremstyle{note}

\newtheorem{theorem}{Theorem}[section]
\newtheorem{definition}[theorem]{Definition}
\newtheorem{notation}[theorem]{Notation}
\newtheorem{lemma}[theorem]{Lemma}
\newtheorem{corollary}[theorem]{Corollary}

\newtheorem{proposition}[theorem]{Proposition}

\newtheorem{example}[theorem]{Example}

\newtheorem{principle}[theorem]{Principle}
\newtheorem{remark}[theorem]{Remark}

\newtheorem{conjecture}[theorem]{Conjecture}

\numberwithin{theorem}{section}
\numberwithin{figure}{section}

\makeatletter
\renewcommand\listoffigures{
        \@starttoc{lof}
}
\makeatother

\begin{document}

\author{Maarten Grothus}
\title{
    Compatibility of Cyclic Causal Structures with Spacetime
    in General Theories
    with Free Interventions}
\begin{titlepage}
    \vspace*{-2cm}
    \raggedleft
    \includegraphics[width=0.2\textwidth]{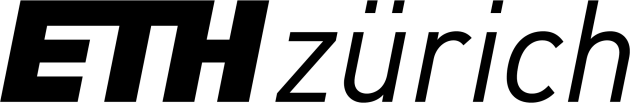}
    \hfill
    \vspace*{0.8cm}
    \hrule

    \makeatletter
    \begin{center}
        \vspace*{1cm}

        \Huge
        \textbf{\@title}


        \vspace{2cm}

        \LARGE
        \@author

        \vfill

        A thesis presented for the degree of\\
        Master of Science

        \vspace{1cm}
        September 2022

        \vspace{2cm}

        \Large
        Supervision by: \\
        Dr.~V.~Vilasini \\
        Prof.~Dr.~Renato Renner

        \vspace{1.5cm}

        \Large
        Quantum Information Theory Group\\
        Institute for Theoretical Physics\\
        ETH Zürich

        \vspace{0.5cm}

    \end{center}
    \makeatother
\end{titlepage}
\setcounter{page}{2}

\newpage

\begin{center}
\begin{minipage}[c]{36em}
    \begin{center}
        \textbf{Abstract}
    \end{center}
    By relating and ordering events, causality constitutes a pivotal feature of our world.
    However, different notions of causality exist, whose relation is not completely understood so far.
    In particular, we may consider both information-theoretic causality, covering the operational idea of information processing, and relativistic causality, linked to a light cone structure limiting signalling to the future.
    In this work, we improve on various results on the connection between both notions, as studied by V.~Vilasini and R.~Colbeck in \cite{VVC}\cite{VVC_Letter}, in particular for questions of cyclicity.

    In the first part, we take an information-theoretic point of view, reviewing general, potentially cyclic or fine-tuned causal models. 
    Here, the most general way of signalling is given by a concept of generalized affects relations, which use interventions on the model  to uncover relations between nodes in these graphs.
    Building from their results, we study the properties of these affects relations and establish new ways to use them to characterize causal structures. focusing on higher-order (HO) affects relations in particular, we can use knowledge of the absence of affects relations for causal inference.
    Further, we demonstrate a complete and constructive way to detect causal loops from a set of affects relations.

    In the second part, we embed these causal structures into a generic spacetime whose causal structure forms a partial order. Here, it was shown in \cite{VVC_Letter} that limiting signalling to the relativistic future does not suffice to generally rule out operationally detectable causal loops.
    In light of this, we propose additional stability conditions on the spacetime embedding and find that this can rule out a class of operationally detectable loops that cannot be ruled out by the principle of no-signalling (outside the relativistic future) alone. We then propose a number of order-theoretic properties that we conjecture to hold in Minkowski spacetime with $d \geq 2$ spatial dimensions.
    This would imply that in contrast to our result for generic spacetimes, in that Minkowski case, the no-signalling principle is indeed sufficient for ruling out this class of loops.
    Finally, we deduce novel restrictions for compatibility for certain HO affects relations.
\end{minipage}
\end{center}

\newpage

\begin{center}
\begin{minipage}[c]{36em}
    \setlength\parindent{17pt}
    \section*{Acknowledgements}

    First and foremost, I would like to express my deepest gratitude to Vilasini for all her time and effort which has allowed me to successfully work on this project.
    Her excitement and commitment to the project have always encouraged me to dive further into its various facets, ever increasing my enthusiasm over its framework.
    Even in the final weeks and days of this project, she has provided me invaluable feedback an uncountable number of times and was always available for last-minute questions and discussions.
    Beyond that, I thoroughly enjoyed discussing a variety of fascinating foundational questions not directly related to this thesis in our meetings.

    Furthermore, I am grateful to Prof.~Renato Renner and the Quantum Information Theory Group as a whole, allowing me to learn about the foundations of this world, culminating (for now) in this thesis, in such a stellar environment with lectures, seminars, and of course the Discussion Group on the Foundations and Philosophy of Physics.

    I would especially like to thank Leon Geiger, who has accompanied me on this journey through theoretical physics since the first day of our Bachelor studies, for countless chats about physics, philosophy, leisure and life,
    and Anna Lena Stetter, for her personal and academic support and her close companionship over the last few months.
    By their interest in my research, their thorough proof reading and high standards for mathematical precision, logical coherence and comprehensibility, they have had considerable impact on the ultimate presentation of my results in this thesis.

    Further, I am thankful to my friends and my fellow students, showing interest in my research and providing some helpful comments regarding this work. In particular, these include Lars Esser, Tanja Küfner, Vincent Grande and Lukas Schmitt.

    Finally, I would like to express my deep gratitude to my family for their loving and unwaivering support, which is a constant source of stability in my life and has allowed me to study and succeed at ETH in the first place.
\end{minipage}
\end{center}

\newpage

\tableofcontents
\newpage

\section{Introduction} \label{sec:intro}
Causality is a concept of utmost importance in science \cite{VVC}. In contrast to just observing correlations in arbitrary scenarios, it allows to ask \textit{why} these occur, differentiating between cause and effect.
Applications of this approach are manifold, including economic models \cite{Pearl2009}, ecological relationships \cite{Schoolmaster2020}\cite{Laubach2021}, medical trials \cite{Kleinberg2011}\cite{Raita2021}, machine learning \cite{Arti2020} and data science in general.
The respective mathematical framework has been developed by J.~Pearl et al. since the 1980s \cite{Pearl1990}\cite{Pearl2009}, and is receiving an increasing amount of attention in various scientific communities.

In particular, causality -- just like locality -- is one of the most pivotal notions of foundational physics. It plays an important role not only in relativistic physics \cite{Kronheimer1967}\cite{Penrose1972}\cite{Malament1977}, but also for quantum theory and related fields.
However, an implication of Bell’s theorem \cite{Bell2004} is that classical causal models cannot satisfactorily explain quantum correlations \cite{Henson2014}\cite{Wood2015} i.e., in a faithful manner that does not allude to certain fine-tuned causal parameters and superluminal communication.
After some early results in the 1990s and 2000s \cite{Tucci1995}\cite{PhysRevLett.74.2619}\cite{Laskey2007}, research on causal modelling of quantum systems really started to take off over the 2010s \cite{PhysRevLett.104.170401}\cite{Bancal2012}\cite{Fritz2012}.
With this, the use of directed graphs representing causality information also became popular in the foundational community, as established earlier in information theory.
Over the following years, the respective approaches developed and diversified, bringing forth remarkable insights, some of which we will review over the course of this work.
Nonetheless, to the best of the author's knowledge, \cite{VVC} suggests the first framework which clearly separates the notions of relativistic and information-theoretic causality to explore their connections.

This work builds particularly on generalized notions of signalling for models with free interventions, which have been thoroughly studied in \cite{VVC}.
This approach is almost entirely theory independent, focusing mainly on the operational interface of the theory which is in-principle accessible via experiments. More precisely, it assumes that experimenters can perform free interventions through their free choice of classical input parameters (such as measurement settings) and collect statistics over certain classical outputs (such as measurement outcomes).
Our statements will therefore have broad generality and apply to a wide class of potential physical theories, including possible post-quantum theories.
Also, there is a wide range of applicability: For example, we can identify processes in different theories as implementing the same type of signalling between agents, thereby representing physical analogues of one another. This may also enhance an intuitive, yet precise understanding of signalling in different theories. Additionally, we can potentially apply general transformation rules of this framework to derive both general features of communication and new results for inferring causation through free interventions in these theories.

By further taking the causal structure of spacetime into account, we may study compatibility conditions of these affects relations with spacetime, in particular with regard to cyclicity of the information-theoretic causal structure. Here, \cite{VVC_Letter} reaches the remarkable result that generally the acyclicity of spacetime does not suffice to rule out causal loops which are operationally detectable.
Contrary to common expectations, the cyclicity of information flow is therefore in no general contradiction to well-defined notions of future and past.
Even for the physically relevant example of Minkowski spacetime with 1 spatial dimension, an example where such a causal loop is embeddable is known.
Evaluating to which extent and under which conditions compatibility of cyclic causal structures with partially ordered spacetimes is possible will therefore be a central topic of this work.
We derive a number of new results in this regard, both on the purely information-theoretic characterization of causal loops as well as on the embedding of such loops in partially ordered spacetimes.
These results are summarized in the remainder of this section.

\subsection{Content of this Thesis}
After a brief introduction, this thesis will be split into two main parts, thereby separating results which refer to the information-theoretic causal structure alone from those which cover compatibility with spacetime.
Due to the tight interconnectedness of various new results with known ones, we will not strictly split the review of existing results and new results into different sections. In \cref{sec:contrib}, we will give a Summary of Contributions to explicitly list all new results, while focusing on giving a rough outline for a novice reader in this section.

In the first part, we evaluate information-theoretic notions of causality and signalling.
In \cref{sec:causality-info}, we start with introducing the causal modelling framework in \cref{sec:causal-structures}, linking probability distributions with a directed graph which codifies the causal dependences.
Within this framework, we cover the concept of 0th- and higher-order (HO) affects relations, which give a general model for signalling through interventions on observable nodes in a causal structure, in \cref{sec:affects} and \cref{sec:affects-new}.
Here, we focus on their transformation properties and their implications for the causal structure, proposing a new property for higher-order affects relations, which we call indecreasability, analyzing its applications for causal inference.
After establishing these results, we use them to detect causal loops from affects relations in \cref{sec:acl}.

Afterwards, in \cref{sec:heavy} we turn to the heavier new results which can be derived using this framework. As most of these will not be necessary for understanding the later parts of this work, it may be advisable to skip this part when reading the thesis for the first time, or when primarily interested into questions of compatibility with a spacetime structure.
In \cref{sec:completeness}, we will derive some results regarding the maximal amount of information that can be extracted from a given causal structure using affects relations, under different conditions.
In particular, we will show the completeness of the causal inference rules derived in the previous section, and use this to find a complete way to identify the causal information represented by affects relations in \cref{sec:graph}. This will culminate in a method to identify all causal loops which are operationally detectable if only a set of affects relations is known.

In the second part, we will add the causal structure of spacetime, and relate it to the information-theoretic notions introduced before.
In \cref{sec:poset}, we will introduce our general model of spacetime, which is represented by a general partially ordered set to again be applicable to a wide class of physical theories.
We will then introduce the notion of an embedding in \cref{sec:orvs}, relating both notions for a causal structure with each other, and review a compatibility condition in \cref{sec:compat}.
This condition will ensure that the affects relations of the embedded causal model can not be used to signal outside the spacetime's future.
Building on this, we will introduce both stronger compatibility conditions in \cref{sec:stable-aff} and additional properties of specific posets like higher-dimensional Minkowski spacetime in \cref{sec:minkowski-prop}, both restricting which sets of affects relations may be embedded.

Using these compatibility properties, we will return to the concept of operationally detectable causal loops in \cref{sec:loops} and present some results how their presence impacts compatible embeddability both generally (\cref{sec:general-loops}) and for Minkowski spacetime in particular (\cref{sec:minkowski-loops}).
Finally, in \cref{sec:compat-indec}, we study how compatibility (which relates information-theoretic and spacetime causal structures) interacts with indecreasability (which is a purely information-theoretic property).
Thereby, we will uncover remarkable parallels between the implications of signalling for the information-theoretic causal structure and regions in spacetime.

We will conclude by discussing our results and stating open questions in \cref{sec:discussion}. In our Appendix, we will give not only our references and a list of figures, but also a list of abbreviations introduced over the cause of this thesis, to aid the reader.

\subsection{Summary of Contributions} \label{sec:contrib}

In total, we contribute the following new results in this work:

\begin{itemize}
    \item In \cref{sec:affects-new}, we introduce the notion of \textit{indecreasable} affects relations. In contrast to previous works, which focus on detecting causation using the presence of affects relations, indecreasability provides a way to leverage knowledge regarding the absence of affects relations to infer additional causal information. Thereby, it allows to use higher-order affects relations to derive further information regarding the causal structure.
    \item \cref{sec:acl} mostly reviews known results from \cite[Sec.~IV.]{VVC}. However, we introduce a new, generalized class of causal loops that are operationally detectable through affects relations. We call this class ACL6a and prove that it indeed certifies the cyclicity of the underlying causal structure.
    \item In \cref{sec:completeness}, we study how much information can be extracted from a causal structure using higher-order affects relations and show that for causal structures without unobserved nodes, they allow for full causal discovery even in the presence of fine-tuning. Further, we consider how this changes in the presence of unobserved nodes, show the completeness of the causal inference rules for affects relations, and give a characterization for affects causal loops.
    \item In \cref{sec:graph}, we devise a simple graphical method to capture the complete causal inference possible from a set of affects relations in a \textit{potential cause graph}. As a derivative concept, we introduce a \textit{loop graph}, which gives a graphical way to infer whether or not a given set of affects relations certifies the cyclicity of any underlying causal structure that could give rise to them.
    \item In \cref{sec:stable-aff}, we introduce and briefly compare different notions for stability of compatible embeddings, focusing on the notion of \textit{support stability}. These will be used to restrict which sets of affects relations may be embedded.
    \item In \cref{sec:minkowski-prop}, we suggest various properties which may be fulfilled by a partially ordered set representing spacetime, and give proof sketches that these are indeed fulfilled by Minkowski spacetime with more than 1 spatial dimension.
    \item In \cref{sec:loops}, we show some results regarding how compatibility and stability can be used to restrict which sets of affects relations may be embedded into spacetime. In particular, we find that so-called complete affects loops and similar structures never admit a stable embedding. Further, for a wide class of spacetimes, we get even stronger restrictions. For higher-dim. Minkowski spacetime in particular, we present hints how these and other restrictions apply based on arguments presented before.
    \item In \cref{sec:compat-indec}, we show that the novel notion of indecreasable affects relations implies to a tighter compatibility condition for signalling. For higher-dim. Minkowski and similar spacetimes, we show this condition simplfies such that it matches the structure of the causal structure implied by the affects relations. This is not the case for the original compatibility condition.
\end{itemize}

\newpage
\part{Information-theoretic Causality and Interventions}
\section{Causal Structures and Affects Relations} \label{sec:causality-info}
In this section, we will introduce the causal modelling framework defined by V.~Vilasini and R.~Colbeck in \cite{VVC}. Their model builds on the causal modelling framework developed by J.~Pearl and others for the classical, acyclic case \cite{Pearl1990}, and generalizes their approach to the cyclic case in a theory-independent matter.

In particular, we will focus on the notion of affects relations, which aim to capture the concept of signalling between random variables (RVs),
and study which information on the causal structure may be derived using only a set of known affects relations.
After introducing the respective concepts, we will revisit the relevant theorems provided in \cite[IV.C]{VVC} and occasionally provide stronger statements, clearly indicating the enhancements.
Since most proofs do not give too much additional insight, they will usually not be provided explicitly if already given in \cite{VVC}.

\subsection{Causal Structures} \label{sec:causal-structures}

In this section we will introduce the concept of causal structures, which use directed graphs (DGs) to endow random variables, encoding operationally accessible physical information, with a notion of information-theoretic causality. As required by the aspirations of this work, we will thereby deviate from the standard treatment presupposing acyclicity for the directed graphs.
Therein, we will follow ideas presented in Section~II and~III of \cite{VVC}, although significantly condensed and enriched with points from additional sources.
Eventually, we will link causal structures with probability distributions following \cite[Sec.~IV.A]{VVC}. To do so, we will specify a compatibility condition using the notion of d-separation, which will be defined within this section.

To begin, we give a formal definition of a causal structure:

\begin{definition}[Causal Structure with Observed and Unobserved Nodes]
    A \emph{causal structure} is a directed graph $\mathcal{G}$, where each node in the graph is classified as either being an \emph{observed node} (or \emph{observable node}) or an \emph{unobserved} node.
    Observed nodes correspond to classical RVs $X_1, \ldots, X_n$ from a set $S$, while unobserved nodes $\Lambda_1, \ldots, \Lambda_n$ may correspond to objects from arbitrary sets.
    For unobserved nodes, relevant choices for objects include classical RVs ($\Lambda_1 \ldots \Lambda_n \in S_\text{unobs}$), quantum Hilbert spaces ($\rho_\Lambda \in \mathscr{H}$ etc.) or systems of generalized probabilistic theories (GPT). A causal structure is called classical, quantum or GPT if all unobserved nodes are associated with objects from the respective theories.
\end{definition}

While this definition may appear a bit daunting at first, we will go through its elements and its technical terms one at a time over the remainder of this section. Here, we start with the case of classical acyclic causal structures without unobserved nodes, and then generalize by first allowing for non-classical and unobserved nodes, and finally by allowing for cyclicity.

\subsubsection{Classical Acyclic Causal Structures} \label{sec:dag}
As is known both mathematically and from day-to-day examples, a collection of random variables (RVs) without further structure is only good to make claims about correlation, not causation.
Therefore, we require additional mathematical structure to represent which RVs are a cause of one another. A way to do so has been realized by H.~Reichenbach \cite{Reichenbach1956} with his principle of common cause.

\begin{principle}[Reichenbach's Principle of Common Cause]
    \label{def:reichenbach}
    If two random variables $X$ and $Y$ are correlated, either $X$ is a cause of $Y$, $Y$ is a cause of $X$ or $X$ and $Y$ share a common cause $Z$ which is influencing both $X$ and $Y$. Further, having both a common cause and a direct influence, as given by each of the first two options, is possible \cite{PhysRevX.7.031021}.
\end{principle}

Extending this to the interplay of larger sets of RVs, as has been done by J.~Pearl and others, leads to the theory of arbitrary causal structures, which are usually modeled as directed acyclic graphs (DAGs) \cite{Pearl2009}.

\begin{notation}
    For a directed graph $\mathcal{G}$, we denote a directed edge between two nodes $N_i$ and $N_j$ as $N_i \dircause N_j$.
\end{notation}

\begin{definition}[Cause {\cite[Def.~II.1]{VVC}}]
    \label{def:cause}
    Given a causal structure represented by a finite directed (acyclic) graph $\mathcal{G}$, we call $N_i$ a \emph{cause} of $N_j$ if there is a directed path
    $N_i \dircause \ldots \dircause N_j$ in $\mathcal{G}$. If the path is of the form $N_i \dircause N_j$, it is called a \emph{direct cause}.
    Generally, for sets of nodes $S_1$ and $S_2$, if
    $\exists N_i \in S_1, N_j \in S_2 \ \colon \ N_i \ \text{cause of} \ N_j$, we say $S_1$ is a cause of $S_2$.
\end{definition}

\begin{definition}[Parents, Children, Ancestors and Descendants]
    \label{def:parents}
    Given a directed graph $\mathcal{G}$, we call the set of nodes $N_i$ which satisfy $N_i \dircause N_j$ the \emph{parents} of $N_j$, denoted by $\parent(N_j)$.
    Analogously, we call the set of $N_j$ which satisfy this relation the \emph{children} of $N_i$, denoted by $\child(N_i)$.
    Similarly, we call the set of $N_i$ for which there is a directed path to $N_j$ the \emph{ancestors} of $N_j$, denoted by $\anc(N_j)$,
    and the set of $N_j$ to which there is a directed path from $N_i$ the \emph{descendants} of $N_i$, denoted by $\desc(N_i)$.
\end{definition}

\begin{definition}[Exogenous nodes]
    \label{def:exogenous}
    Given a directed graph $\mathcal{G}$, we call a node $N_j$ \emph{exogenous} if there exists no $N_i$ which is a direct cause of $N_j$, or equivalently $\parent(N_j) = \emptyset$.
    A node like this is sometimes called \textit{freely chosen}.
\end{definition}

Of course, we still need a condition relating this causal structure with a probability distribution $P(N_i, N_j, \ldots)$ to ultimately arrive at a full \textit{causal model}, for which we will provide a complete definition at the end of this section.
In the literature on classical acyclic causal models, the \emph{causal Markov condition} is used for this purpose:
\begin{equation}
    \label{eq:causal-markov}
    P(X_1, \ldots, X_n) = \prod_{i=1}^n P(X_i|\parent(X_i)) \,
\end{equation}
Hence, the probability distribution of the RV associated with each node is dependent only of the distribution of its direct causes. Therefore, the arrows correspond to memoryless classical channels.

This approach is sufficient for describing causation in acyclic causal structures where all nodes are observed, since they correspond to RVs which are part of the observed probability distribution.
Here, any information is principally observable, even without perturbing the causal structure.\footnote{
    This still holds even if we consider affects relations later, which modify the causal structure using interventions. We will elaborate on this point in \cref{sec:affects-intro}.
}
Naturally, this poses the question whether the Markov condition is still applicable in the presence of unobserved nodes in the causal structure.
Continuing, we will consider this and other peculiarities that causal structure allow for, some of which will require the use of a modified, more general condition instead of \cref{eq:causal-markov}.

Before returning to the case of unobserved nodes, we start off with a rather mundane point. While we assume that correlation implies some form of causation via Reichenbach's principle, the reverse of this statement does not necessarily hold: Generally, it is possible for two random variables to be uncorrelated, but still be causally related.
Commonly, this scenario is excluded, considering only so-called \textit{faithful} causal models.
Specifically, unfaithful models correspond to the phenomenon of \textit{fine-tuning}, as the causal parameters must be perfectly fitted to not surface in the correlations.
While this is often considered undesirable for modelling physical situations, relevant use cases exist, for example in cryptography.
Therefore, we will consider this scenario to retain generality.

\begin{figure}[t]
	\centering
    \begin{tikzpicture}
        \begin{scope}[every node/.style={circle,thick,draw,inner sep=0pt,minimum size=0.9cm}]
            \node (M) at (0,0) {$M$};
            \node (K) at (4,0) {$K$};
            \node (M') at (2,2) {$M'$};
        \end{scope}

        \begin{scope}[>={Stealth[black]},
                      every node/.style={fill=white,circle},
                      every edge/.style=causalarrow]
            \path [->] (M) edge (M');
            \path [->] (K) edge (M');
        \end{scope}
    \end{tikzpicture}
	\caption[The causal structure associated with the One-Time Pad.]{The causal structure associated with the One-Time Pad, as given in \cref{ex:otp}, where $M' \cong M \oplus K$.}
	\label{fig:otp}
\end{figure}
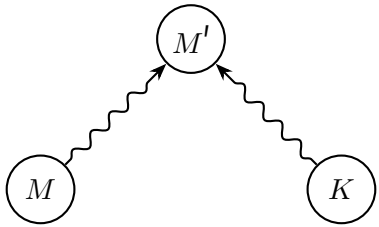

Working with causal models, we will use the respective $X_i$ to refer both to the nodes and to the RVs associated with them. Therefore, we introduce special notation to refer to the equality of the values of the respective RVs, which makes no statement regarding the identity of nodes.

\begin{notation}
    For a causal model over a set $S$ of RVs, where $X, Y \in S$, we will write $X \cong Y$ if both RVs do always have the same value, being perfectly correlated with each other.
    Only if additionally, $X$ and $Y$ refer to the same node in the respective causal structure $\mathcal{G}$, we write $X = Y$.
    Therefore, $\cong$ is a equivalence relation between nodes.
\end{notation}

\begin{example}[One-Time Pad]
    \label{ex:otp}
    As a simple example for a causal structure with fine-tuning from classical cryptography, we can consider the \textit{One-Time Pad} protocol, which realizes information-theoretically secure symmetric encryption by classical means.
    Here, we consider a binary RV $M$, representing an $1$-bit message to be sent, and a binary $K$, representing the key used for encryption. Assuming both to be statistically uniform, we also get that the resulting encrypted message $M' \cong M \oplus K$ is distributed uniformly.
    Therefore, $M$, $K$ and $M'$ are statistically independent with regard to each other, even though they are connected to a common causal structure, as depicted in \cref{fig:otp}.
\end{example}

\subsubsection{Unobserved Nodes and Non-Classical Acyclic Causal Structures} \label{sec:gpt}
More significantly, we can return to the possibility of unobserved nodes in the causal structure, which are frequently denoted by $\Lambda$, with values $\lambda$, which may correspond either to classical random variables or to different objects.

Such nodes are operationally inaccessible and hence no arguments of the observed probability distribution $P(X_1, \ldots, X_n)$.
Therefore, the causal Markov condition can no longer be applied \textit{directly} on the observed distribution to arrive at the observed distribution. However, when we assume that the unobserved nodes also correspond to (\enquote{hidden}) random variables, we can still impose the causal Markov condition given in \cref{eq:causal-markov} for the joint distribution over all nodes (observed and unobserved). Then, we recover the observed distribution $P$ by marginalizing over the unobserved nodes.

Otherwise, we obtain the possibility to associate unobserved nodes and the linked arrows with mathematical objects other than random variables and classical channels.
As an important physical model, we can consider quantum causal models, where nodes and arrows can be associated with density matrices $\rho_\Lambda$ and quantum channels \cite{PhysRevX.7.031021}\cite{arxiv.1906.10726}.
Even more universally, we can model generalized probabilistic theories (GPTs), the research of which has made considerable progress over the last decade \cite{Henson2014}.
In a GPT, only operational notions like input and output (or in terms more familiar from quantum theory, preparation and measurement) for an experiment are required to be modelled as random variables, while their dynamics can be described by processes and states from any theory admitting a well-defined composition \cite{PhysRevA.75.032304}.

While of course, classical and quantum theory are special cases of GPTs, a wide variety of alternative theories exists, which are seemingly not realized in nature \cite{PhysRevA.75.032304}\cite{Dakic2014}.
In finite dimensions, we observe shared central properties with quantum theory, as given by the no-cloning-theorem \cite{PhysRevLett.99.240501}, the existence of incompatible measurements \cite{PhysRevA.94.042108}, and, most recently, the presence of superposition and entanglement \cite{PhysRevLett.128.160402}, in any non-classical GPT.
By contrast, other properties single out quantum mechanics from other GPTs \cite{Masanes2011}\cite{PhysRevLett.109.090403}.
While the research of general properties of GPTs is an active field of research, ultimately, in this work we will only be interested in the operationally detectable relations between observed nodes. Therefore, we will not go into further detail here.

\begin{figure}[t]
    \hspace{0.06\textwidth}
    \begin{subfigure}[t]{0.28\textwidth}
        \centering
        \begin{tikzpicture}
            \begin{scope}[every node/.style={circle,thick,draw}]
                \node (A) at (-1.8,0) {$A$};
                \node (B) at (2,0) {$B$};
                \node (X) at (-1.8,2) {$X$};
                \node (Y) at (2,2) {$Y$};
            \end{scope}
            \node (L) at (0,-1.5) {$\Lambda$};
            \node (SD) at (0, -2) {$\vphantom{\Lambda_{SD}}$};

            \begin{scope}[>={Stealth[black]},
                          every node/.style={fill=white,circle},
                          every edge/.style=causalarrow]
                \path [->] (A) edge (X);
                \path [->] (L) edge (X);
                \path [->] (B) edge (Y);
                \path [->] (L) edge (Y);
            \end{scope}
        \end{tikzpicture}
        \caption{}
        \label{fig:bell-naive}
    \end{subfigure}%
    \hspace{0.02\textwidth}
    \begin{subfigure}[t]{0.28\textwidth}
        \centering
        \begin{tikzpicture}
            \begin{scope}[every node/.style={circle,thick,draw}]
                \node (A) at (-1.8,0) {$A$};
                \node (B) at (1.8,0) {$B$};
                \node (X) at (-1.8,2) {$X$};
                \node (Y) at (1.8,2) {$Y$};
            \end{scope}
            \node (L) at (0,-1.5) {$\Lambda$};
            \node (SD) at (0, -2) {$\vphantom{\Lambda_{SD}}$};

            \begin{scope}[>={Stealth[black]},
                          every node/.style={fill=white,circle},
                          every edge/.style=causalarrow]
                \path [->] (A) edge (X);
                \path [->] (L) edge (X);
                \path [->] (B) edge (Y);
                \path [->] (L) edge (Y);
                \path [->] (A) edge (Y);
                \path [->] (B) edge (X);
            \end{scope}
        \end{tikzpicture}
        \caption{}
        \label{fig:bell-fine}
    \end{subfigure}%
    \hspace{0.02\textwidth}
    \begin{subfigure}[t]{0.28\textwidth}
        \centering
        \begin{tikzpicture}
            \begin{scope}[every node/.style={circle,thick,draw}]
                \node (A) at (-1.8,0) {$A$};
                \node (B) at (1.8,0) {$B$};
                \node (X) at (-1.8,2) {$X$};
                \node (Y) at (1.8,2) {$Y$};
            \end{scope}
            \node (L) at (0,-0.5) {$\Lambda$};
            \node (SD) at (0, -2) {$\Lambda_{SD}$};

            \begin{scope}[>={Stealth[black]},
                          every node/.style={fill=white,circle},
                          every edge/.style=causalarrow]
                \path [->] (A) edge (X);
                \path [->] (L) edge (X);
                \path [->] (B) edge (Y);
                \path [->] (L) edge (Y);
                \path [->] (SD) edge (L);
                \path [->] (SD) edge (A);
                \path [->] (SD) edge (B);
            \end{scope}
        \end{tikzpicture}
        \caption{}
        \label{fig:bell-sd}
    \end{subfigure}%
    \hspace{0.06\textwidth}
	\caption[Different causal structures to explain Bell correlations.]{
        Different causal structures to explain bipartite Bell correlations.
        Here, $\Lambda$ is the hidden common cause associated with the quantum state, $A$ and $B$ are the choices of measurement, and $X$ and $Y$ are the respective measurement outcomes.
        As usual, $A$ and $X$ are presumed to be spacelike separated from $B$ and $Y$.
        a) The naive approach for a causal structure, which is not compatible with the probability distribution if we presume $\Lambda$ to be classical. In a quantum causal model however, this is a compatible causal structure, modelling $\Lambda$ bya quantum state $\rho_\Lambda$ instead of a RV.
        b) Superluminal fine-tuned causation is used to explain the correlations between the measurement choice $A$ and the outcomes $Y$, and respectively for $B$ and $C$.
        c) Superdeterminism: Measurement results are predetermined by an ancient common cause $\Lambda_{SD}$, which renders $A$, $B$ and $\Lambda$ non-exogenous.
    }
	\label{fig:bell}
\end{figure}
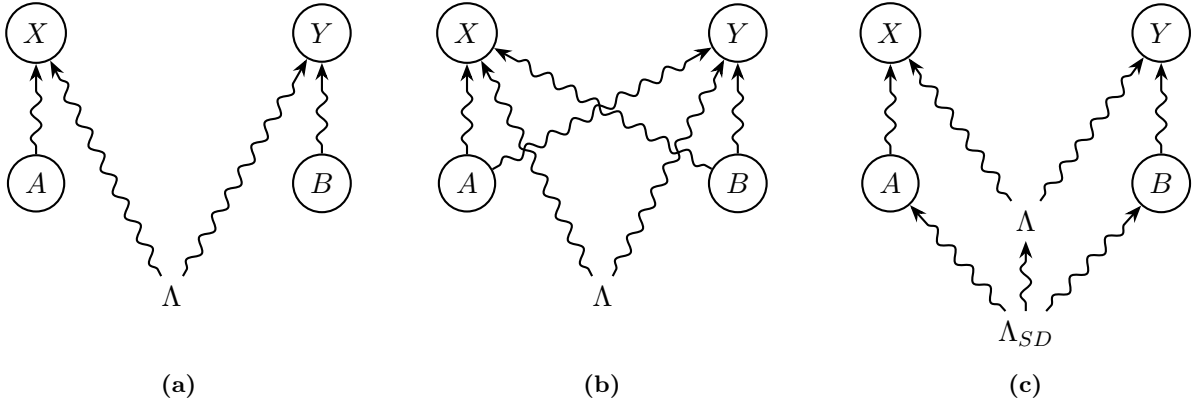

Using the language of causal structures, we can understand Bell's theorem \cite{Bell2004} and generalizations precisely as a restriction on the correlations which are realizable within a given theory using a given causal structure, as we have already seen for the example of quantum theory.
More precisely, \cite{Henson2014} finds that only for certain causal structures the space of correlations which can be realized depends on the nature of unobserved nodes.
Specifically, this is the case for the causal structure naturally associated with the Bell experiment, as depicted in \cref{fig:bell-naive}. Here, there exists no probability distribution $P(XY)$ over the experimental outcomes which satisfies
\begin{equation}
    P (XY) = P(X|\Lambda) P(Y|\Lambda) P(\Lambda) \, .
\end{equation}
Therefore, if this causal structure is to describe the Bell experiment, $\Lambda$ can not be associated with a classical random variable.

By suitably generalizing Reichenbach's principle to quantum theory, we can fundamentally extend causality into the realm of quantum theory. This allows to explain the respective correlations, carrying over the principle of common cause from random variables to density matrices \cite{PhysRevX.7.031021}\cite{arxiv.1906.10726}.
For Bell's theorem in particular, one can see that the causal structure depicted in \cref{fig:bell-naive} can indeed explain the respective correlations if $\Lambda$ is associated with a quantum state $\rho_\Lambda$.
For classical $\Lambda$, we are forced to amend the causal structure using fine-tuning to explain the respective correlations instead.

This leads to options like using fine-tuned superluminal transmission of hidden variables \cite{PhysRev.85.166}, an ancient common cause predetermining our measurement results in a fine-tuned way (superdeterminism) \cite{Wood2015}, or retrocausal influences where our measurement results influence the choice of measurement via causation from the future to the past \cite{RevModPhys.92.021002}. Examples for this are depicted in \cref{fig:bell}b and c.
Independent from our assessment of these models, we are still able to represent them in a concise and mathematically rigorous way using general causal structures.

To study causal models in the light of these properties, we can choose one of two complementary approaches.
Bottom-up approaches for causality require a notion of composition, usually from some circuit structure. This compositional structure then fully specifies the causal structure.
By contrast, in the top-down approach we will use here, we are not assuming any underlying theory, and especially do not require a notion of sub-systems or composition. Here, we only require a notion of free interventions, which we will discuss in \cref{sec:affects}, and some probabilities of observations encoded by observable RVs.
For example, this concept could be applied to an algebraic quantum field theory (aQFT) \cite{Ruep}.

\subsubsection{Cyclic Causal Structures} \label{sec:cyclic-structures}

Finally, we generalize by removing the acyclicity condition from the causal structure.
Using causal cycles, we can model feedback processes \cite{arxiv.1302.3595}, as exemplarily present for predator-prey relationships in ecology, the dynamic share prices on the stock market, or a PID controller, which regulates the value of a variable based on its \textit{p}roportional, \textit{i}ntegral and \textit{d}erivative terms. We will review two simple examples for cyclic causal structures.

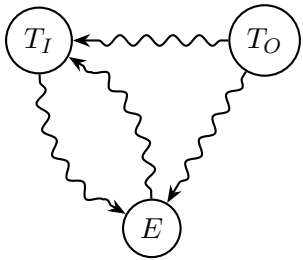
\begin{figure}[t]
	\centering
    \begin{tikzpicture}
        \begin{scope}[every node/.style={circle,thick,draw}]
            \node (T_I) at (0,0) {$T_I$};
            \node (T_O) at (3,0) {$T_O$};
            \node (E) at (1.5,-2.5) {$E$};
        \end{scope}

        \begin{scope}[>={Stealth[black]},
                      every node/.style={fill=white,circle},
                      every edge/.style=causalarrow]
            \path [->] (T_O) edge (T_I);
            \path [->, bend right] (E) edge (T_I);
            \path [->, bend right] (T_I) edge (E);
            \path [->] (T_O) edge (E);
        \end{scope}
    \end{tikzpicture}
	\caption[Causal structure of Friedman's thermostat.]{Causal structure of Friedman's thermostat, as given in \cref{ex:friedman}.}
	\label{fig:friedman}
\end{figure}

\begin{example}[Friedman's thermostat]
    \label{ex:friedman}
    Consider a (not perfectly isolated) house with an ideal thermostat, effecting a constant inside temperature $T_I$ at all times using energy $E$.
    Then even though $T_I$ is constant, it is causally influenced by both the outside temperature $T_O$ and $E$ via fine-tuning.
    Further, consider that the thermostat is realized as a heat pump, which energy consumption $E$ depends on $T_I$ and $T_O$.
    In total, we arrive at the cyclic causal structure depicted in \cref{fig:friedman}.
\end{example}

\begin{example}[Cyclic Dependences between Demand and Price]
    \label{ex:economics}
    Consider the economics of demand and price of blackboard chalk.
    A decrease in price at time $t_1$ may lead to an increase in demand at time $t_2 > t_1$, which again may lead to an increase of price at time $t_3 > t_2$.
    In this case, the cyclicity of the causal structure originates from considering the value of variables over a period of time.
\end{example}

While these every-day examples can be distilled into an acyclic causal structure by considering values at different points in time, this is not the case a priori. In a certain sense, we will return to this point in \cref{sec:orvs}, where we will associate RVs with locations in spacetime. This will remove the cyclicity from cases like \cref{ex:economics}.

With this generalization, we face certain difficulties already on a level of definitions, as becomes apparent when considering the simple-most example of two observed RVs, which are direct causes of each other: $X \leftsquigarrow\mkern-14.25mu\dircause Y$.
In contrast to the acyclic case, it is possible here to choose classical channels for the respective causal model which lead to a paradox, as is for instance the case for two binary variables $X$ and $Y$ with $X \cong Y \ \land \ Y \cong X \oplus 1$.
To avoid inconsistencies like this in our framework, we exclude such cases where no compatible observed probability distribution exists.
Further, the causal Markov condition given by \cref{eq:causal-markov} is unable to cover this case in full generality, as it would resolve to $P(XY) = P(X|Y)P(Y|X)$.
This, however, would imply a deterministic distribution of RVs in a causal loop, which seems unnecessarily restrictive.

Therefore, we choose a more general compatibility condition based on the concept of \textit{d-separation}, as developed by D.~Geiger, T.~Verma and J.~Pearl \cite{Pearl1990}, which is a weaker version of the causal Markov condition and sometimes called the \textit{directed global Markov property}.
Here, the \textit{d} stands for \textit{directed}.
For directed acyclic graphs with observed nodes only, it is equivalent to the causal Markov condition given in \cref{eq:causal-markov} \cite{arxiv.1710.08775}.
This construct has been shown to retain its applicability for acyclic causal structures with unobserved nodes, even if these nodes are associated with objects from quantum theory \cite{arxiv.1906.10726} or even elements from GPTs \cite{Henson2014}. Further, d-separation has been shown to be applicable to a wide class of classical cyclic causal models \cite{arxiv.1302.3595}\cite{arxiv.1710.08775}.
Therefore, it constitutes a natural candidate to consider for a compatibility condition applicable to the combination of both properties.

\begin{definition}[Blocked Paths {\cite[Def.~II.2]{VVC}}]
    \label{def:blocked}
    Let $\mathcal{G}$ be a directed graph, where $X$ and $Y$ are distinct nodes and $Z$ is a set of nodes not containing $X$ and $Y$.
    An (undirected) path from $X$ to $Y$ is \emph{blocked} by $Z$ if the path contains $A, B$ such that either
    $A \dircause W \dircause B$,
    $A \diresuac W \dircause B$ with $W \in Z$, or
    $A \dircause V \diresuac B$ with neither $V$ nor any child of $V$ in $Z$.
\end{definition}

\begin{definition}[d-separation {\cite[Def.~II.3]{VVC}}]
    \label{def:d-sep}
    Let $\mathcal{G}$ be a directed graph, with $X, Y, Z$ being disjoint subsets of nodes.
    $X$ and $Y$ are \emph{d-separated} by $Z$ in $\mathcal{G}$, denoted as $(X \perp^d Y \,|\, Z)_\mathcal{G}$, if every path from an element of $X$ to an element of $Y$ is blocked by $Z$.
    Otherwise, $X$ is \emph{d-connected} to $Y$ given $Z$. If obvious from context, the index $\mathcal{G}$ may be suppressed.
\end{definition}

After formally introducing d-separation, we can finally introduce the compatibility condition for causal structures and probability distributions which we will assume for the remainder of this work.

\begin{definition}[Compatibility of Probability Distributions with a Causal Structure {\cite[Def.~IV.1]{VVC}}]
    \label{def:compat-stoch}
    Let $S := \{ X_1, \ldots X_n \}$ be a set of RVs associated with the observed nodes of a directed graph $\mathcal{G}$, which may also have unobserved nodes, and $P(X_1, \ldots, X_n)$ be the probability distribution over them.
    Then $P$ and $\mathcal{G}$ are \emph{compatible} with each other if for all $X, Y, Z \subset S$ disjoint
    \begin{equation}
        X \perp^d Y \,|\, Z \quad \implies \quad
        X \indep Y \,|\, Z \,  \quad :\iff \quad P(XY|Z) = P(X|Z) P(Y|Z) \, .
    \end{equation}
\end{definition}

However, d-separation is by no means the only applicable criterion for compatibility. In fact, \cite{arxiv.1710.08775} derives a more general notion called $\sigma$-separation for potentially cyclic causal structures built from (potentially unobserved) classical nodes, that is significantly more complex.
This would allow to capture an even wider class of cyclic causal models, examples of which have already been found before in \cite{Neal2000}.
Going forward,  we will restrict ourselves to the d-separation case if necessary.
It remains an open question how the properties of d- and $\sigma$-separation generalize to cyclic \textit{quantum} causal models, as have been studied in \cite{Barrett2021} and \cite{VVR}. Uncovering these relations is subject of ongoing research \cite{Carla}, and to the author's best knowledge, no research about cyclic GPT models has been done up until now.

We conclude this section by synthesizing a compatible pair of causal structure and probability distribution into a single concept.

\begin{definition}[Causal Model {\cite[Def.~IV.2]{VVC}}]
    \label{def:causal-model}
    A \emph{causal model} over a finite set of observed RVs $S := \{ X_1, \ldots X_n \}$ consists of a causal structure, given by a directed graph $\mathcal{G}$ over them (and potentially, further GPT systems as unobserved nodes) and a joint probability distribution $P_\mathcal{G} (X_1, \ldots, X_n)$ compatible with each other, in the sense of \cref{def:compat-stoch}.
\end{definition}

With this, we can also give a formal definition for \textit{faithfulness} of general causal models, as discussed in \cref{sec:dag} for the acyclic case.

\begin{definition}[Faithful and Fine-tuned Causal Models]
    Consider a causal model over a set of RVs $S$ with a causal structure $\mathcal{G}$. Then this causal model is \emph{faithful} if
    \begin{equation}
        X \perp^d Y \,|\, Z \quad \iff \quad
        X \indep Y \,|\, Z \, .
    \end{equation}
    Otherwise, we have
    \begin{equation}
        X \perp^d Y \,|\, Z \quad \not\Longleftarrow \quad
        X \indep Y \,|\, Z \, .
    \end{equation}
    and we call the causal model \emph{unfaithful} or \emph{fine-tuned}.
\end{definition}

\subsection{Affects Relations: Introduction and Review} \label{sec:affects}

Having introduced all of this additional complexity to causal structures, it is natural to wonder how we can extract information about the causal structure present using operational means.
In this section, we will focus on capturing the information-theoretic concept of signalling between random variables, by defining the notion of \textit{affects relations} and studying their properties and transformations on the one hand and their potential for causal inference on the other hand.
To do so, we will summarize the most relevant results of Section IV.B and IV.C of \cite{VVC}.

\subsubsection{Introduction} \label{sec:affects-intro}
Imagine two agents, Alice and Bob, who are able to perform probabilistic experiments in their respective labs.
Hence, as modelled by the framework of GPTs briefly discussed in \cref{sec:gpt}, they choose some experimental setup, where they can probabilistically perform some preparations, usually understood as initial states, as well as a procedure to observe the probability distribution of outcomes.
Now how can we determine if there is communication, or signalling, from Alice to Bob? For this to be the case, the outcomes of Bob must somehow depend on the preparation performed by Alice.

Imagining the experiment as a causal structure $\mathcal{G}$ allows to capture this directedness in contrast to considering the observable correlations $P$ only.
On the one hand, independent preparation choices correspond with a set of exogenous nodes $X$ (cf. \cref{def:exogenous}). On the other hand, the probability distribution $P(Y)$, associated with a set of observed nodes $Y$, corresponds to some outcomes of the joint experiment. We say $X$ \textit{affects} $Y$ if such a dependence $P(Y|X) \neq P(Y)$ is present. In this case, we also expect $X$ to be a cause of $Y$.

Further, we generalize this concept to non-exogenous nodes $X$ by wondering how the experimental outcomes would differ when $X$ would have a different probability distribution. This can be imagined as modifying the experimental setup, and therefore, the causal structure $\mathcal{G}$, by including additional independent steps of preparation.

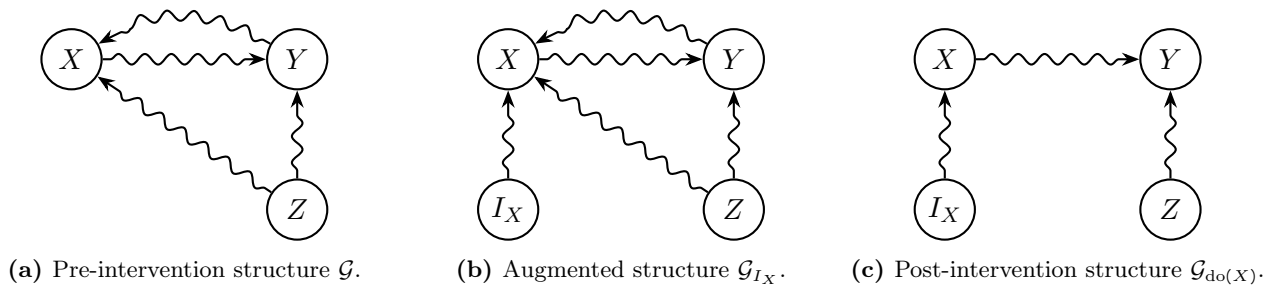
\begin{figure}[t]
	\centering
    \begin{subfigure}[b]{0.33\textwidth}
        \centering
        \begin{tikzpicture}
            \begin{scope}[every node/.style={circle,thick,draw,inner sep=0pt,minimum size=0.8cm}]
                \node (X) at (-3,0) {$X$};
                \node (Y) at (0,0) {$Y$};
                \node (Z) at (0,-2) {$Z$};
            \end{scope}

            \begin{scope}[>={Stealth[black]},
                          every node/.style={fill=white,circle},
                          every edge/.style=causalarrow]
                \path [->] (X) edge (Y);
                \path [->] (Y) edge[bend right] (X);
                \path [->] (Z) edge (X);
                \path [->] (Z) edge (Y);
            \end{scope}
        \end{tikzpicture}
        \caption{Pre-intervention structure $\mathcal{G}$.}
        \label{fig:pre-intervention}
    \end{subfigure}%
    \begin{subfigure}[b]{0.33\textwidth}
        \centering
        \begin{tikzpicture}
            \begin{scope}[every node/.style={circle,thick,draw,inner sep=0pt,minimum size=0.8cm}]
                \node (X) at (-3,0) {$X$};
                \node (Y) at (0,0) {$Y$};
                \node (Z) at (0,-2) {$Z$};
                \node (I_X) at (-3, -2) {$I_X$};
            \end{scope}

            \begin{scope}[>={Stealth[black]},
                          every node/.style={fill=white,circle},
                          every edge/.style=causalarrow]
                \path [->] (X) edge (Y);
                \path [->] (Y) edge[bend right] (X);
                \path [->] (Z) edge (X);
                \path [->] (Z) edge (Y);
                \path [->] (I_X) edge (X);
            \end{scope}
        \end{tikzpicture}
        \caption{Augmented structure $\mathcal{G}_{I_X}$.}
        \label{fig:augmented}
    \end{subfigure}%
    \begin{subfigure}[b]{0.33\textwidth}
        \centering
        \begin{tikzpicture}
            \begin{scope}[every node/.style={circle,thick,draw,inner sep=0pt,minimum size=0.8cm}]
                \node (X) at (-3,0) {$X$};
                \node (Y) at (0,0) {$Y$};
                \node (Z) at (0,-2) {$Z$};
                \node (I_X) at (-3, -2) {$I_X$};
            \end{scope}

            \begin{scope}[>={Stealth[black]},
                          every node/.style={fill=white,circle},
                          every edge/.style=causalarrow]
                \path [->] (X) edge (Y);
                \path [->] (Z) edge (Y);
                \path [->] (I_X) edge (X);
            \end{scope}
        \end{tikzpicture}
        \caption{Post-intervention structure $\mathcal{G}_{\doo(X)}$.}
        \label{fig:post-intervention}
    \end{subfigure}%
	\caption[Pre-intervention, augmented and post-intervention causal structures.]{
        An example for pre-intervention, augmented and post-intervention causal structures.
        For $I_X = \text{idle}$, (a) and (b) give the same effective model, for $I_X = \doo(x)$, the same is the case for (b) and (c).
        Notice that after the intervention, the causal structure is no longer cyclic.
    }
	\label{fig:interventions}
\end{figure}
These \textit{interventions} are represented by removing all incoming arrows of a node $X$ in $\mathcal{G}$ and replacing them with an arrow from an intervention node $I_X = \doo (x)$, which is defined to force $X = x$.
The modified causal structure is denoted as $\mathcal{G}_{\doo(X)}$. This definition naturally carries over to sets of nodes, where a joint intervention consists of a set of individual interventions on each node.\footnote{
    Additionally, one may consider correlated interventions between nodes, which would thereby not correspond to exogenous nodes.
    Since that does only restrict the information to be gained through the intervention however, this case will not be discussed further.
}
With this definition, we capture the notion of free interventions: The values of the respective intervention nodes are freely chosen, as they are exogenous in the causal structure.

The relation between pre- and post-intervention causal model is made mathematically precise by introducing an augmented causal structure $\mathcal{G}_{I_X}$, used to relate the respective probability distributions with each other.
Here, the intervention can be switched on and off. The respective procedure is depicted in \cref{fig:interventions} and is defined by the following set of equations:
\begin{align}
    P_{\mathcal{G}_{I_X}} (Y | I_X = \text{idle}) &= P_\mathcal{G} (Y) \\
    P_{\mathcal{G}_{I_X}} (Y | I_X = \doo(x)) &= P_{\mathcal{G}_{\doo(X)}} (Y | I_X = \doo(x)) = P_{\mathcal{G}_{\doo(X)}} (Y | X = x) \quad \forall x \\
    P_{\mathcal{G}_{I_X}} (Y | I_X = \doo(x), X = x) &= P_{\mathcal{G}_{I_X}} (Y | I_X = \doo(x)) \quad \forall x \\
    P_{\mathcal{G}_{I_X}} (I_X = \doo(x), X = x') &= 0 \quad \forall x, x' \ \text{with} \ x \neq x' \, .
\end{align}
For further details regarding the definition of interventions, we refer to \cite[p.~14]{VVC} and Pearl's general rules of do-calculus \cite{Pearl2009} for further reading.

In other words, interventions allow us to model how the probability distribution would change by enforcing a specific value $x$ for a RV $X$, replacing the RV which has been present in the causal structure before with an exogenous one (i.e. without any parents).\footnote{
    Looking for an analogue in cryptography, we find some parallels to a completely naive man-in-the-middle attack at our causal structure.
}
Perhaps surprisingly, in causal structures with unobserved nodes, the post-intervention probability distribution can generally not be determined from the pre-intervention distribution alone \cite{Pearl2009}. This fact is sometimes refered to as the impossibility of \textit{counterfactual} inference, which is exhibited by some causal models \cite[p.~49]{VVphd}.
Therefore, the intervention needs to actually be performed to characterize the post-intervention distribution.
However, in the absence of unobserved nodes, the post-intervention distribution can counterfactually be determined at least for distributions satisfying the causal Markov condition given in \cref{eq:causal-markov} \cite[p.~42ff.]{VVphd}.

This formalism allows to codify the concept of signalling between random variables: $X$ \textit{affects} $Y$ indicates that intervening (enforcing a specific value) on $X$ is detectable on $Y$, as it changes the probability distribution of $Y$ in any way. More generally, we can state:

\begin{definition}[Affects Relations {\cite[Def.~IV.5]{VVC}}]
    \label{def:affects}
    Consider a causal model over a set of $S$ observed nodes, associated with a causal structure $\mathcal{G}$.
    For pairwise disjoint subsets $X, Y, Z, W \subset S$, with $X, Y$ non-empty, we say
    \begin{equation}
        X \, \text{affects} \, Y \, \text{given} \, \{\doo(Z), W \} \, ,
    \end{equation}
    which we alternatively denote as
    \begin{equation}
        X \models Y \,|\, \{\doo(Z), W \} \, ,
    \end{equation}
    if there exist values $x$ of $X$, $z$ of $Z$ and $w$ of $W$ such that
    \begin{equation}
        P_{\mathcal{G}_{\doo(XZ)}} (Y | X=x, Z=z, W=w) \neq
        P_{\mathcal{G}_{\doo(Z)}} (Y | Z=z, W=w)
    \end{equation}
    For $W \neq \emptyset$, we speak of a \emph{conditional affects relation}, denoted by $X \affects Y \given W$.
    If $Z \neq \emptyset$, we have a \emph{higher-order (HO) affects relation}, denoted by $X \affects Y \given \doo(Z)$.
    More specifically, it is also called a \emph{$\abs{Z}$th-order affects relation}.
    The trivial case of $W = Z = \emptyset$ is called an \emph{unconditional 0th-order affects relation}, denoted by $X \affects Y$,
    or for short, a \emph{simple affects relation}.
\end{definition}

This generalizes the concept of signalling by allowing conditioning on both other intervention data $\doo (Z)$ and the values of other nodes $W$.
Within this framework, this is the most general conceivable notion of an affects relation, and therefore the most general way for two agents to signal to one another.
Conceptually, the case of conditional affects relations may be associated with post-selecting on the respective random variables.
Higher-order affects relations, on the other hand, are particularly interesting since they allow to transform a causal model which is cyclic before the respective intervention to be acyclic afterwards.

Since we allow for unfaithful (i.e. fine-tuned) and cyclic causal models, many stronger relations between affects relations, correlations and the causal structure, which seem natural at first, can generally not be assumed in this framework.

\begin{figure}[t]
	\centering
    \begin{subfigure}[b]{0.3\textwidth}
        \centering
        \begin{tikzpicture}
            \begin{scope}[every node/.style={circle,thick,draw}]
                \node (A) at (-2,0) {$A$};
                \node (B) at (0,-1.3) {$B$};
                \node (C) at (2,0) {$C$};
                \node (D) at (0,1.3) {$D$};
            \end{scope}
            \node (L) at (0,-3) {$\Lambda$};

            \begin{scope}[>={Stealth[black]},
                          every node/.style={fill=white,circle},
                          every edge/.style=causalarrow]
                \path [->] (B) edge (A);
                \path [->] (B) edge (C);
                \path [->] (A) edge (D);
                \path [->] (C) edge (D);
                \path [->] (L) edge (A);
                \path [->] (L) edge (C);
            \end{scope}
        \end{tikzpicture}
        \caption{}
        \label{fig:causal-structure-examples-7}
    \end{subfigure}%
    \hspace{0.1\textwidth}
    \begin{subfigure}[b]{0.3\textwidth}
        \centering
        \begin{tikzpicture}
            \begin{scope}[every node/.style={circle,thick,draw}]
                \node (A) at (-2,0) {$A$};
                \node (B) at (0,-1.3) {$B$};
                \node (C) at (2,0) {$C$};
            \end{scope}
            \node (D) at (0,1.3) {};
            \node (L) at (0,-3) {$\Lambda$};

            \begin{scope}[>={Stealth[black]},
                          every node/.style={fill=white,circle},
                          every edge/.style=causalarrow]
                \path [->] (B) edge (A);
                \path [->] (B) edge (C);
                \path [->] (L) edge (A);
                \path [->] (L) edge (C);
            \end{scope}
        \end{tikzpicture}
        \caption{}
        \label{fig:causal-structure-examples-6}
    \end{subfigure}%
	\caption[Two examples for causal structures.]{Two examples for causal structures.
        a) Causal structure of \cref{ex:IV.7}. Here, $B \affects D$ even though there are no simple affects relations along the respective arrows.
        b) Causal structure of \cref{ex:IV.6} (Jamming). Here, $B \affects AC$, but there are no simple affects relations between any individual nodes.
    }
	\label{fig:causal-structure-examples}
\end{figure}
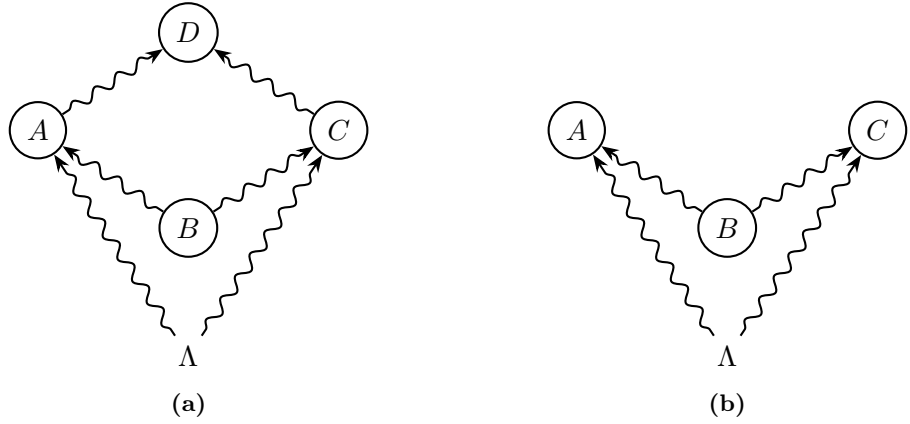

\begin{example}[{\cite[Ex~IV.7]{VVC}}]
    \label{ex:IV.7}
    Consider a causal model over the variables $S = \{ A, B, C, D \}$, whose causal structure is depicted in \cref{fig:causal-structure-examples-7}.
    For the compatible binary probability distribution $B \cong A \oplus C \cong D$, none of the individual variables affect each other, but still $B \affects D$.
    This distribution is compatible with the d-separation property by the introduction of an additional unobserved node $\Lambda$.
    Further, $AC \affects D$ and a variety of conditional and higher-order affects relations holds.
\end{example}

Hence, affects relations may arise from indirect causation:
\begin{equation}
    X \affects Y \quad \not\implies \quad X \dircause Y \quad \ \text{for} \ X, Y \in S \, .
\end{equation}

Further, simple affects relation give no information about the correlations between two RVs in the pre-intervention graph.
While we consider it probable, to the author's best knowledge it is an open question whether the presence of a correlation in the pre-intervention graph does imply the presence of a corresponding general affects relation.
However, the relation between correlations and affects relations will not be particularly relevant for this work, which focuses on implications for the causal structure.
For further information, we therefore refer to Section IV of \cite{VVC}.

\begin{example}[One-Time Pad {\cite[Sec.~II.A, App.~A.1]{VVC}}]
    We return to the example of the one-time pad, also known as fine-tuned collider, depicted in \cref{fig:otp}, and look for general affects relations.
    We easily see that $M \naffects M'$ and $K \naffects M'$, as for both interventions, $M'$ will still be uniform.
    Further, we see that intervening on the message or the key modifies the joint probability distribution of the other two RVs, respectively, and therefore $K \affects MM'$ and $M \affects KM'$.
    Additionally, we have $MK \affects M'$, corresponding to the fact that we can freely choose the probability distribution of the encrypted message by controlling all of its causes.

    Further, we can observe multiple conditional and higher-order affects relations. As the causal structure is symmetric under permuting $M$ and $K$, we will restrict ourselves to conditioning or intervening on $K$ here. Then, we get $M \affects M' \given \doo(K)$ and $M \affects M' \given K$, which are equivalent due to $K$ being exogenous (i.e. not having any parents) and assuming all values, but also $M \affects K \given M'$ and $K \affects M \given M'$. This backwards-directed affects relation, which is typical for collider causal structures, reflects the decryption procedure, understood as signalling process: Since $M'$ is available, one can use it to generate the original message $M$ as a signal from the key.
\end{example}

\begin{example}[Jamming {\cite[Ex.~IV.6]{VVC}}]
    \label{ex:IV.6}
    For comparison, we consider the jamming scenario, depicted in \cref{fig:causal-structure-examples}b. Here, we have an unobserved node given by $\Lambda$ in addition to the observed nodes $A, B, C \in S$ uniform and (unfaithfully) related by $B \cong A \oplus C$. Hence, the correlation between $A$, $B$ and $C$ can only be found by an agent having access to all three RVs.
    Via the hidden common cause $\Lambda$, this satisfies d-separation compatibility (\cref{def:compat-stoch}).

    Therefore, the complete set of affects relations in this scenario is given by
    $B \affects AC$, $B \affects C \given A$ and $B \affects A \given C$. In particular, we observe neither simple affects relations between individual RVs in $S$ nor higher-order affects relations.
\end{example}

These examples suggest that conditional and higher-order affects relations might generally be related to simple affects relations.
We will show general results in this direction in the upcoming section. However, generally the information given by these can not be expressed by a set of unconditional 0th-order affects relations.
When restricting to acyclic faithful causal models without unobserved nodes, all non-implications pointed out here turn into implications.
Further research is needed to figure out which implications are regained by reducing generality, for example by restricting to faithful cyclic models.

\subsubsection{Properties of Affects Relations and Causal Inference} \label{sec:affects-rules}

Here, we review the transformative properties of affects relations and their implications for the causal structure, which have been originally shown in \cite{VVC}.
As this is not clear immediately due to the proofs not being given explicitly here, it is valuable to point out that all statements here retain their validity no matter which kind of compatibility for probability distributions (cf. \cref{def:compat-stoch}) is chosen, as both the notion of cause (cf. \cref{def:cause}) and the definition of affects relations itself is invariant under that choice.

To gain insight regarding the causal structure from the affects relations, we want to differentiate nodes which are actually relevant for the affects relation in question from those which are superfluous in an affects relation.
For example, if $X \affects Y$, usually for most $Z$ in the same causal model $XZ \affects Y$ holds.\footnote{
    In \cref{ex:IV.4}, we will consider an instance where this is not the case.}
In particular, this is the case for all $Z$ with no outgoing arrows, which could mediate it being a cause of $Y$.

Treating this on the same footing would confuse not only causal discovery (i.e. inference of the causal structure), but also potential external conditions we could consider for signalling (as will be done in \cref{sec:compat}).
Hence, we will usually exclude these cases by recognizing such affects relations as \textit{reducible} using the following definition.

\begin{definition}[Reducible and Irreducible Affects Relations {\cite[Def.~IV.6]{VVC}}]
    \label{def:reducible}
    Let $S$ be a set of RVs in a causal model and $X, Y, Z, W \subset S$ disjoint.
    An affects relation defined as in \cref{def:affects} is called \emph{reducible} if $\exists s_X \subsetneq X$ such that
    \begin{equation}
        s_X \naffects Y \given \{ \doo(Z \tilde{s}_X), W \} \, ,
    \end{equation}
    where $\tilde{s}_X := X \setminus s_X$. Otherwise, it is called \emph{irreducible}.
\end{definition}

\begin{lemma}[{\cite[Lem.~IV.6]{VVC}}]
    \label{thm:reducible-first}
    Let $S$ be a set of RVs in a causal model and $X, Y, Z, W \subset S$ disjoint.
    For every reducible affects relation $X \affects Y \given \{ \doo(Z), W \}$, there exists $\tilde{s}_X \subsetneq X$ with
    $\tilde{s}_X \affects Y \given \{ \doo(Z), W \}$.
    Further, with $s_X := X \setminus \tilde{s}_X$, it satisfies $s_X \naffects Y \given \{ \doo(Z \tilde{s}_X), W \}$.
    \begin{proof}
        See \cite[p.~49]{VVC}.
    \end{proof}
\end{lemma}

This concept effectively allows implicitly using information about higher-order affects relations without referring to them explicitly.
Using this will allow us to tighten the following statements, which can be proven by using d-separation, to perform improved causal inference:

\begin{notation}
    \label{not:affects-cause}
    If the presence of a causal relation is implied by an individual affects relation, we will denote the respective causal relation by $\cause$.
\end{notation}

\begin{lemma}[{\cite[Lem.~IV.3]{VVC}}]
    \label{thm:affects-to-cause-first}
    Let $S$ be a set of RVs in a causal model and $X, Y, Z, W \subset S$ disjoint. Then
    \begin{enumerate}
        \item $X \affects Y \given \doo(Z) \implies X \cause Y \, .$
        \item $X \affects Y \given \{ \doo(Z), W  \} \implies X \cause Y \ \lor \ X \cause W \, .$
    \end{enumerate}
    \begin{proof}
        See \cite[p.~48f.]{VVC}.
    \end{proof}
\end{lemma}

Both to enhance this notion and as an end in itself, we will now turn to transformative properties of affects relations.
Usually, these are understood best by disregarding the conditional or higher-order part if unchanged under the transformation.

\begin{lemma}[{\cite[Lem.~IV.4]{VVC}}]
    \label{thm:ZO-first}
    Let $S$ be a set of RVs in a causal model and $X, Y, Z, W \subset S$ disjoint. Then
    \begin{equation}
        X \affects Y \given \{ \doo(Z), W  \} \implies Z \affects Y \given W \ \lor \ XZ \affects Y \given W \, .
    \end{equation}
    \begin{proof}
        See \cite[p.~49]{VVC}.
    \end{proof}
\end{lemma}

We will give an example for the case that $X \affects Y \given \doo(Z)$, but neither $X \affects Y$ nor $XZ \affects Y$.

\begin{figure}[t]
	\centering
    \begin{tikzpicture}
        \begin{scope}[every node/.style={circle,thick,draw}]
            \node (A) at (-2,0) {$A$};
            \node (B) at (0,-1.3) {$B$};
            \node (C) at (2,0) {$C$};
            \node (D) at (0,1.3) {$D$};
        \end{scope}

        \begin{scope}[>={Stealth[black]},
                      every node/.style={fill=white,circle},
                      every edge/.style=causalarrow]
            \path [->] (A) edge (B);
            \path [->] (A) edge (D);
            \path [->] (B) edge (C);
            \path [->] (B) edge (D);
            \path [->] (C) edge (D);
        \end{scope}
    \end{tikzpicture}
	\caption[Example for a causal structure with a higher-order affects relation without simple companion.]{
        Causal structure of \cref{ex:IV.4}, where we find a higher-order affects relation $B \affects D \given \doo(C)$, but both $B \naffects D$ and $BC \naffects D$ hold.
    }
	\label{fig:causal-structure-examples-4}
\end{figure}

\begin{example}[{\cite[Ex.~IV.4]{VVC}}]
    \label{ex:IV.4}
    Consider a causal model over the variables $S = \{ A, B, C, D \}$, whose causal structure is depicted in \cref{fig:causal-structure-examples-4}.
    We specify a compatible causal model by assuming the variables to be binary and related by
    $D \cong A \oplus B \oplus C$, $C \cong B$, $B \cong A$. By additionally assuming $A$ to be uniformly distributed, we get the same value for all four RVs.
    In this graph, we observe $C \affects D$, since under this intervention, $C \cong D$.
    However, $BC \naffects D$, as due to $D \cong A \oplus B \oplus C$, $P_{\mathcal{G}_{\doo(BC}} (D|BC)$ still remains uniform for each value of the intervention.
    Nonetheless, we observe $B \affects D \given \doo(C)$ as a higher-order affects relation.
\end{example}

\begin{lemma}[{\cite[Lem.~IV.8]{VVC}}]
    \label{thm:conditional-first}
    Let $S$ be a set of RVs in a causal model and $X, Y, Z, W \subset S$ disjoint. Then
    \begin{enumerate}
        \item $X \affects Y \given \{ \doo(Z), W  \} \implies X \affects YW \given \doo(Z) \, .$
        \item $X \affects Y \given \{ \doo(Z), W  \} \ \text{irreducible} \implies X \affects YW \given \doo(Z) \ \text{irreducible.}$
        \item $X \affects YW \given \doo(Z) \iff X \affects Y \given \{ \doo(Z), W  \} \ \lor \ X \affects W \given \doo(Z) \, .$
    \end{enumerate}
    \begin{proof}
        See \cite[p.~50]{VVC}.
    \end{proof}
\end{lemma}

With this lemma, we can give the following perspective on the strength of conditional affects relations:
\begin{equation}
    X \affects YW \ \land \ X \naffects W \implies
    X \affects Y \given W \implies
    X \affects YW
\end{equation}
with $Z = \emptyset$ for simplicity. The opposite directions of these implications do not hold. For the second implication, this follows directly by setting $Y$ to a set superficial to the causal model, i.e. without any incoming or outgoing arrows. For the first one, we can consider an example.

\begin{example}
    Consider a causal structure with $W \cong X \oplus Y$ with $X$ uniform and $Y$ non-uniform. Then we get $X \affects Y \given W$, $X \affects YW$ and $X \affects W$ while $X \naffects Y$ due to the latter being uniform. Therefore, $X \affects Y \given W \not\implies X \affects YW \ \land \ X \ \naffects W$.
\end{example}

Nonetheless, the causal inference rules we derive with the following corollary will give the same inferred causal relations for all three statements. Therefore, we can disregard the case of conditional affects relations for the process of causal inference.

\begin{corollary}[{\cite[Cor.~IV.3]{VVC}}]
    \label{thm:affects-to-cause}
    Consider a causal model over a set of RVs $S$ where $X, Y, Z, W \subset S$ disjoint. Then
    \begin{enumerate}
        \item $X \affects Y \given \doo(Z)$ irreducible $\implies$ each $e_X \in X$ is a cause of at least one element $e_Y \in Y$.
        \item $X \affects Y \given \{ \doo(Z), W \}$ irreducible $\implies$ each $e_X \in X$ is a cause of at least one element $e_{YW} \in YW$.
    \end{enumerate}
    \begin{proof}
        See \cite[p.~51]{VVC}.
    \end{proof}
\end{corollary}

Finally, this equips us with the ability to deduce information regarding the presence of individual \enquote{causal paths}, consisting of causal arrows, from affects relations alone.
However, the higher-order interventional data given by $Z$ does not enter here.

\subsection{Affects Relations: New Results} \label{sec:affects-new}

In this section, we show some original results, complementing the ones before. These will allow us to expand our toolset for causal discovery by also taking the information given by higher-order interventional data into account. Before turning to the respective results, we supply an additional technicality to \cref{thm:reducible-first}, which is required here.

\begin{lemma}
    \label{thm:reducible}
    For every reducible affects relation $X \affects Y \given \{ \doo(Z), W \}$, there exists some (non-empty) $\tilde{s}_X \subsetneq X$ with
    $\tilde{s}_X \affects Y \given \{ \doo(Z), W \}$, which is irreducible.
    \begin{proof}
        The existence of such a $\tilde{s}_X$ is guaranteed by \cref{thm:reducible-first}. We show irreducibility by contradiction:

        If there would be no irreducible affects relation implied, the respective affects relation would still be given, but be reducible.
        Therefore, applying the first part again, we again have $\tilde{t}_X \subsetneq \tilde{s}_X$ with
        $\tilde{t}_X \affects Y \given \{ \doo(Z), W \}$.
        If this affects relation is irreducible, we are done, otherwise, we repeat.
        Since the cardinality of the affecting set shrinks with each iteration, we will ultimately reach an affects relation $\tilde{u}_X \affects Y \given \{ \doo(Z), W \}$ which is irreducible.
        This is due to the fact that if for some $e_X \in X$ the relation $e_X \affects Y \given \{ \doo(Z), W \}$ holds, it is trivially irreducible, and we always reach it if no irreducible affects relation is encountered before.
        Since $\tilde{u}_X \subsetneq X$, this poses a contradiction. $\lightning$
    \end{proof}
\end{lemma}

While the results from the previous section have focused on which simple affects relations are implied by higher-order and conditional affects relations, we now focus on the question how higher-order affects relations are implied by different types of general affects relations. We begin with a generalization of \cref{thm:ZO-first}.

\begin{lemma}
    \label{thm:ZO}
    Let $S$ be a set of RVs in a causal model and $X, Y, Z_1, Z_2, W \subset S$ disjoint and $Z := Z_1 Z_2$. Then
    \begin{equation}
        X \affects Y \given \{ \doo(Z), W \} \implies Z_1 \affects Y \given \{ \doo(Z_2), W \} \ \lor \ X Z_1 \affects Y \given \{ \doo(Z_2), W \} \, .
    \end{equation}
    \begin{proof}
        The proof is directly analogous to (and a more general case of) the original proof for \cref{thm:ZO-first}. We prove this via contraposition:
        \begin{equation}
            \label{eq:ZO-contraposite}
            Z_1 \naffects Y \given \{ \doo(Z_2), W \} \ \land \ X Z_1 \naffects Y \given \{ \doo(Z_2), W \}
            \stackrel{!}{\implies} X \naffects Y \given \{ \doo(Z_1 Z_2), W \} \, .
        \end{equation}
        By definition, the left side gives
        \begin{align}
            P_{\mathcal{G}_{\doo(Z)}} (Y | Z_1=z_1, Z_2=z_2, W=w) &= P_{\mathcal{G}_{\doo(Z_2)}} (Y | Z_2=z_2, W=w) \quad and \\
            P_{\mathcal{G}_{\doo(XZ)}} (Y | X=x, Z_1=z_1, Z_2=z_2, W=w) &= P_{\mathcal{G}_{\doo(Z_2)}} (Y | Z_2=z_2, W=w) \, .
        \end{align}
        Equating these to each other yields the right-hand side of \cref{eq:ZO-contraposite}, and hence, yields the claim.
    \end{proof}
\end{lemma}

\begin{lemma}
    \label{thm:affects-to-HO}
    Let $S$ be a set of RVs in a causal model and $X, Y, Z_1, Z_2, W \subset S$ disjoint and $Z := Z_1 Z_2$. Then
    \begin{equation}
        X Z_1 \affects Y \given \{ \doo(Z_2), W \} \implies X \affects Y \given \{ \doo(Z), W \} \ \lor \ Z_1 \affects Y \given \{ \doo(Z_2), W \} \, .
    \end{equation}
    \begin{proof}
        We prove this via contraposition:\\
        \begin{equation}
            \label{eq:affects-to-HO-contraposite}
            X \naffects Y \given \{ \doo(Z), W \} \ \land \ Z_1 \naffects Y \given \{ \doo(Z_2), W \} \stackrel{!}{\implies} X Z_1 \naffects Y \given \{ \doo(Z_2), W \} \, .
        \end{equation}
        By definition, the left side gives
        \begin{align}
            P_{\mathcal{G}_{\doo(X Z)}} (Y | X=x, Z_1=z_1, Z_2=z_2, W=w) &= P_{\mathcal{G}_{\doo(Z)}} (Y | Z_1=z_1, Z_2=z_2, W=w) \quad and \\
            P_{\mathcal{G}_{\doo(Z_2)}} (Y, Z_2=z_2, W=w)  &= P_{\mathcal{G}_{\doo(Z)}} (Y | Z_1=z_1, Z_2=z_2, W=w) \, .
        \end{align}
        Equating these to each other yields the right-hand side of \cref{eq:affects-to-HO-contraposite}, and hence, yields the claim.
    \end{proof}
\end{lemma}

For the special case of $Z_2 = W = \emptyset$, we get the simplified expression
\begin{equation}
    \label{eq:affects-to-HO-simple}
    XZ \affects Y \implies X \affects Y \given \doo(Z) \ \lor \ Z \affects Y \, .
\end{equation}

\begin{corollary}
    \label{thm:HO-transfer}
    Let $S$ be a set of RVs in a causal model and $X, Y, Z_1, Z_2, W \subset S$ disjoint and $Z := Z_1 Z_2$. Then
    \begin{equation}
        \begin{split}
            & X \affects Y \given \{ \doo(Z), W \} \\
            \implies
            & X \affects Y \given \{ \doo(Z_2), W \} \ \lor \ Z_1 \affects Y \given \{ \doo(Z_2), W \} \ \lor \ Z_1 \affects Y \given \{ \doo(X Z_2), W \} \, .
        \end{split}
    \end{equation}
    \begin{proof}
        Consider that \cref{thm:affects-to-HO} can be restated by swapping $X$ and $Z_1$, arriving at
        \begin{equation}
            X Z_1 \affects Y \given \{ \doo(Z_2), W \} \implies Z_1 \affects Y \given \{ \doo(X Z_2), W \} \ \lor \ X \affects Y \given \{ \doo(Z_2), W \} \, .
        \end{equation}
        Plugging this into \cref{thm:ZO} gives the claim.
    \end{proof}
\end{corollary}

For the special case of $Z_2 = W = \emptyset$, we get the simplified expression
\begin{equation}
    \label{eq:HO-transfer-simple}
    X \affects Y \given \doo(Z) \implies X \affects Y \ \lor \ Z \affects Y \ \lor \ Z \affects Y \given \doo(X) \, .
\end{equation}

By this statement, we can basically flip the role of $X$ and $Z$ within higher-order affects relations. However, this flip does not preserve irreducibility, since irreducibility generates no statements regarding subsets of $Z$. Further, we see that the phenomenon of superficial nodes leading to \cref{def:reducible} is present in a similar way for HO affects relations:
For any $X, Y \in S$ with $X \affects Y$ irreducible, we usually have $X \affects Y \given \doo(Z)$ irreducible. In particular, this is the case for all $Z$ with no outgoing arrows, since intervening does not change the distributions of any other nodes in this case.

\begin{definition}
    \label{def:indecreasable}
    Let $S$ be a set of RVs in a causal model and $X, Y, Z, W \subset S$ disjoint.
    We call an affects relation \emph{indecreasable} if $X \affects Y \given \{ \doo(Z), W \}$, but
    $X \naffects Y \given \{ \doo(Z \setminus e_Z), W \} \ \forall e_Z \in Z$.
    Otherwise, we call it \emph{decreasable}.
    In particular, every 0th-order affects relation is indecreasable.
\end{definition}

\begin{lemma}
    \label{thm:HO-switch}
    Consider a causal model over a set of RVs $S$ where $X, Y, Z, W \subset S$ disjoint, $e_Z \in Z$ and $\tilde{Z} = Z \setminus e_Z$. Then
    \begin{equation}
        \begin{split}
            & X \affects Y \given \{ \doo(Z), W \} \ \land \ X \ \naffects Y \given \{ \doo(\tilde{Z}), W \} \\
            \implies \quad &
            e_Z \affects Y \given \{ \doo(\tilde{Z}), W \} \ \lor \
            e_Z \affects Y \given \{ \doo(\tilde{Z} X), W \}
        \end{split}
    \end{equation}
    with the (alternative) affects relations from $e_Z$ (i.e. with $e_Z$ in the first argument) being irreducible, if present.
    \begin{proof}
        Consider $Z = \tilde{Z} e_Z$. By \cref{thm:HO-transfer}, then
        \begin{equation}
            \begin{split}
                X \affects Y \given \{ \doo(Z), W \} \implies &
                    X \affects Y \given \{ \doo(\tilde{Z}), W \} \,\ \ \lor \\ &
                    e_Z \affects Y \given \{ \doo(\tilde{Z}), W \} \ \lor \
                    e_Z \affects Y \given \{ \doo(\tilde{Z} X), W \} \, .
            \end{split}
        \end{equation}
        However, by assumption,
        $X \naffects Y \given \{ \doo(\tilde{Z}), W \}$. Hence
        \begin{equation}
            e_Z \affects Y \given \{ \doo(\tilde{Z}), W \} \ \lor \
            e_Z \affects Y \given \{ \doo(\tilde{Z} X), W \} \, .
        \end{equation}
        Since $e_Z$ is a singleton, the respective affects relation is irreducible.
    \end{proof}
\end{lemma}

\begin{corollary}
    \label{thm:affects-to-cause-HO}
    Consider a causal model over a set of RVs $S$ where $X, Y, Z, W \subset S$ disjoint and $e_Z \in Z$. Then
    \begin{enumerate}
        \item $X \affects Y \given \doo(Z) \ \land \ X \naffects Y \given \doo(Z \setminus e_Z) \, \implies$
            $e_Z$ is a cause of at least one element $e_Y \in Y$.
        \item $X \affects Y \given \{ \doo(Z), W \} \ \land \ X \naffects Y \given \doo(Z \setminus e_Z) \, \implies$
            $e_Z$ is a cause of at least one element $e_{YW} \in YW$.
    \end{enumerate}
    \begin{proof}
        As the first case is a special case of the second with $W = \emptyset$, we only prove the latter.
        By applying \cref{thm:affects-to-cause} to \cref{thm:HO-switch}, both alternatives give that each $e_Z$ is a cause of at least one element $e_{YW} \in YW$.
        Therefore, we get the claim.
    \end{proof}
\end{corollary}

With this corollary, we have obtained a remarkably precise tool to extract individual causal relations using information about the \textit{absence} of affects relations, using information about the \enquote{critical} elements of higher-order affects relations. This method to read off information about causal relations can then be applied and simplified to indecreasable affects relations.
However, the indecreasability version of this statement is actually weaker and less universally applicable than the original one.\footnote{
    This raises the question if a similarly precise version exists for \cref{thm:affects-to-cause}.
    Indeed, a direct analogue exists by modifying the proof of that corollary, but looks much less attractive:
    $X \affects Y \given \doo(Z) \ \land \ e_X \affects Y \given \doo(ZX \setminus e_X)$ implies that $e_X$ is a cause of at least one element $e_Y \in Y$.
    Here, the original affects relation becomes completely irrelevant for performing causal inference.
    The information required to be more precise could just as well be given by the second affects relation.
}

\begin{corollary}
    \label{thm:HO-switch-indecreasable}
    Consider a causal model over a set of RVs $S$ where $X, Y, Z, W \subset S$ disjoint and $\tilde{Z} = Z \setminus e_Z$. Then
    \begin{equation}
        \begin{split}
            & X \affects Y \given \{ \doo(Z), W \} \ \text{indecreasable} \\
            \implies \ \ &
            e_Z \affects Y \given \{ \doo(\tilde{Z}), W \} \ \lor \
            e_Z \affects Y \given \{ \doo(\tilde{Z} X), W \} \quad
            \forall e_Z \in Z
        \end{split}
    \end{equation}
    with the alternative affects relations from $e_Z$ being irreducible, if present.
    \begin{proof}
        Follows directly from combining \cref{thm:HO-switch} for all $e_Z \in Z$.
    \end{proof}
\end{corollary}

\begin{corollary}
    \label{thm:affects-to-cause-HO-indecreasable}
    Consider a causal model over a set of RVs $S$ where $X, Y, Z, W \subset S$ disjoint. Then
    \begin{enumerate}
        \item $X \affects Y \given \doo(Z)$ indecreasable $\implies$ each $e_Z \in Z$ is a cause of at least one element $e_Y \in Y$.
        \item $X \affects Y \given \{ \doo(Z), W \}$ indecreasable $\implies$ each $e_Z \in Z$ is a cause of at least one element $e_{YW} \in YW$.
    \end{enumerate}
    \begin{proof}
        Directly analogous to \cref{thm:affects-to-cause-HO}, but with \cref{thm:HO-switch-indecreasable} instead of \cref{thm:HO-switch}.
    \end{proof}
\end{corollary}

Therefore, indecreasability complements irreducibility, which allows to perform causal inference for individual nodes from the elements of the first argument, with an analogue for the third argument.
For affects relations which are both irreducible and indecreasable, we can summarize this to a single expression.

\begin{corollary}
    \label{thm:irr-indec}
    Consider a causal model over a set of RVs $S$ where $X, Y, Z, W \subset S$ disjoint. Then
    \begin{equation}
        \label{eq:irr-indec}
        \begin{split}
            X \affects Y \given \{ \doo(Z), W \} \ \text{irreducible and indecreasable} \\
            \implies \forall \, e_{XZ} \in XZ \quad \exists \, e_{YW} \in YW \ \colon \quad e_{XZ} \cause e_{YW} \, .
        \end{split}
    \end{equation}
    \begin{proof}
        Follows directly by conjunction of \cref{thm:affects-to-cause} and \cref{thm:affects-to-cause-HO-indecreasable}.
    \end{proof}
\end{corollary}

Wondering if a similar complement for the second argument of an affects relation exists, we observe that here, a similar notion provides no real simplification:
$X, Y, W \subset S, \ X \affects YW \ \land \ X \affects Y$ already gives that each $e_X \in X$ is a cause of at least one element $e_Y \in Y$, which is precisely what we would want from reducing in the second argument. Therefore, introducing an analogue does not seem necessary.
Considering the presence or absence of further affects relations which are not directly related is not required here.

\subsection{Affects Causal Loops} \label{sec:acl}

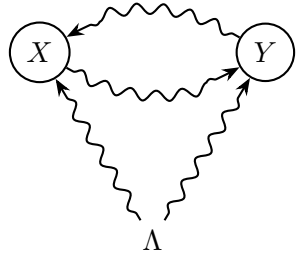
\begin{figure}[t]
	\centering
    \begin{tikzpicture}
        \begin{scope}[every node/.style={circle,thick,draw}]
            \node (X) at (0,0) {$X$};
            \node (Y) at (3,0) {$Y$};
        \end{scope}
        \node (L) at (1.5,-2.5) {$\Lambda$};

        \begin{scope}[>={Stealth[black]},
                      every node/.style={fill=white,circle},
                      every edge/.style=causalarrow]
            \path [->] (X) edge[bend right] (Y);
            \path [->] (Y) edge[bend right] (X);
            \path [->] (L) edge (X);
            \path [->] (L) edge (Y);
        \end{scope}
    \end{tikzpicture}
	\caption[An operationally undetectable causal loop between $X$ and $Y$.]{
        An operationally undetectable causal loop between $X$ and $Y$, as discussed in \cref{ex:HCL}. The common cause $\Lambda$ is unobserved, as is indicated here by omitting the circle.}
	\label{fig:HCL}
\end{figure}

Having introduced the concept and properties of affects relations and their relation to causal structures, we can now study under which circumstances we can use affects relations to detect loops in the causal structure.

Generally, a loop in the causal structure might not be operationally verifiable, in which case we call it a hidden causal loop (HCL). However, such a scenario could never be uncovered experimentally. Furthermore, a causal loop detectable only by studying the correlations in the causal model might be possible.

\begin{example}[{\cite[Ex.~VI.1]{VVC}}]
    \label{ex:HCL}
    Consider a causal model of \cref{fig:HCL} over the binary RVs $X, Y, \Lambda$, where $X, Y$ are observed and $\Lambda$ is unobserved.
    Further, $\Lambda$ is uniform and $X \cong \Lambda \oplus Y$ and $Y \cong \Lambda \oplus X$.
    Then, $X$ and $Y$ constitute a causal loop even though, due to uniformity of $\Lambda$, interventions on $X$ or $Y$ do not change the probability distribution of the other observed variable.
    Hence, $X \naffects Y$ and $Y \naffects X$, excluding all affects relations to be possible for this causal structure.
\end{example}

We will revisit and exclude these cases for graphs without unobserved nodes in \cref{rem:no-corr-loops}, though. For the scope of this thesis, we will not consider this scenario further.

In the following, we give a short outline of those types of affects causal loops (ACLs) which will be relevant for further discussion.
These are derived from \cite{VVC}, Section VI, but partially modified to be more general.
As already this leads to a wide variety of types of causal loops, we will restrict only to simple affects relations here.
For ease of presentation, we will first give the definitions of some causal loop types before providing proofs that they indeed constitute causal loops.

To begin, we present a revised selection of the types 1–5, which constitute subtypes of so called complete affects chains. As ACL5 constitutes the most general case here and encompasses the others, we only refer to a proof there.

\begin{definition}[ACL2a {\cite[Def. VI.4]{VVC}}]
    \label{def:ACL2a}
    Let $I$ be an index set, potentially empty.
    Then a set of affects relations $\mathscr{A}$ contains an \emph{ACL2a} if there exist RVs $X, Y, Z_i$ (for all $i \in I$) for which the affects relations
    $X \affects Z_1 \affects \dots \affects Z_i \affects Z_{i+1} \affects \dots \affects Z_k \affects Y \affects X$ are
    irreducible and in $\mathscr{A}$, respectively.
\end{definition}

An ACL2a gives the simple-most case for a causal loop: Only individual RVs are involved, which we can infer to form a circle in the causal structure.
To also encompass the case of ACL1, which is given by $X \affects Y \affects X$, we have adjusted the definition slightly.
Increasing complexity, we can also consider sets of random variables, as is the case for ACL3.

\begin{definition}[ACL3 {\cite[Def. VI.5]{VVC}}]
    \label{def:ACL3}
    Let $S_1, S_2$ be disjoint sets of RVs.
    Then a set of affects relations $\mathscr{A}$ contains an \emph{ACL3} if there exist RVs with
    $S_1 \affects e_2$ and $S_2 \affects e_1$ in $\mathscr{A}$ irreducible, where $e_1 \in S_1, e_2 \in S_2$.
\end{definition}

To consider more complex types, we will introduce a notion of a complete affects chain. Thereby, we have a chain of affects relations which implies a causal link from its beginning to its end at our disposal. Eventually, we can composite these to more complex causal loops.

\begin{notation}[Subset Relations]
    To emphasize that a given subset relation $A \subset B$ includes equality, we will write $A \subseteq B$. If on the contrary, we want to highlight that equality is excluded, we will write $A \subsetneq B$.
    This is not to be confused with $A \not\subseteq B$, which points out that $A$ is neither a subset nor equal to $B$.
\end{notation}

\begin{definition}[Complete Affects Chain]
    \label{def:cac}
    Let $I$ be an index set and $S_i \subseteq \hat S_i$ be sets of RVs $\forall i \in I$, with $n := \abs{I}$.
    Then a set of affects relations $\mathscr{A}$ contains a \emph{complete affects chain} $\mathscr{C}_{S_1}$ from $S_1$ to $S_n$ if there exist sets of RVs with
    $\hat S_i \affects S_{i+1} \in \mathscr{A}$ irreducible for all $1 \leq i < n$.
\end{definition}

\begin{definition}[ACL5 {\cite[Def VI.7]{VVC}} and Complete Affects Chains]
    \label{def:ACL5}
    A set of affects relations contains an \emph{ACL5} if it contains a complete affects chain from $S_1$ to $S_1$.
\end{definition}

The definition of ACL4, which is introduced in \cite{VVC}, has been skipped here since it constitutes a special case of ACL5 with $S_i = \hat{S_i} \ \forall i \in I$.

\begin{example}
    Let $A, B, C, D, E, F, G, H \in S$. An example of an ACL5 (which is no ACL4) is given by $\mathscr{A} = \{ AB \affects CD, CDE \affects F, F \affects GH, GH \affects A \}$, with all affects relations are presumed to be irreducible.
\end{example}

\begin{definition}[ACL6 {\cite[Def VI.8]{VVC}}]
    \label{def:ACL6}
    A set of affects relations $\mathscr{A}$ contains an \emph{ACL6} if there exist sets of RVs $S_i$ such that
    \begin{enumerate}
        \item $S_1 \affects S_2$ irreducible in $\mathscr{A}$.
        \item For each $e_2 \in S_2$, there exists $s_1 \subset S_1$ such that there is a complete affects chain from $e_2$ to $s_1$, as given in \cref{def:ACL5}.
    \end{enumerate}
\end{definition}

This is the most general kind of ACLs based on complete affects chains which has been put forward within \cite{VVC}.
With it, we have \enquote{split} an ACL for the first time, using an affects relation to a set of RVs.
Before generalizing, we will confirm that the types introduced so far do indeed imply the presence of a cyclic causal structure.

\begin{theorem}[{\cite[Theorem~VI.1]{VVC}}]
    Any set of affects relations $\mathscr{A}$ containing an ACL of type 1, 2, 3, 4, 5 or 6 can only arise from a cyclic causal structure.
\end{theorem}

However, we can generalize this class ACL6 further, by including recursive cases, thus leading to ACL6a.

\begin{definition}[ACL6a]
    \label{def:ACL6a}
    A set of affects relations $\mathscr{A}$ contains an \emph{ACL6a} if there exist sets of RVs $S_i$ such that
    \begin{enumerate}
        \item There exists a complete affects chain from $S_1$ to $S_2$ in $\mathscr{A}$.
        \item For each $e_2 \in S_2$, there exists a complete affects chain from $e_2$ to some $S_k$ disjoint from $S_2$.
        \item For each $e_j \in S_k$, there exists a complete affects chain from $e_j$ to either $s_1 \subseteq S_1$ or to some disjoint $S_k'$, for which the same holds.
        \item The third clause is only used a finite number of times.
    \end{enumerate}
\end{definition}

\begin{theorem}
    Any set of affects relations $\mathscr{A}$ containing an ACL of type 6a can only arise from a cyclic causal structure.
    \begin{proof}
        Applying \cref{thm:affects-to-cause}, we get that each $e^1 \in s_1 \subseteq S_1$ is a cause of some $e_2 \in S_2$,
        and that $e_2 \in S_2$ is a cause of some $e_k \in S_k$.
        Chaining \cref{thm:affects-to-cause}, each element $e_i \in S_k$ is a cause of an element $e^j \in S_k'$.
        Recursing point 3, we get that each element $e_i \in S_k$ is a cause of an element $e^2 \in s_1$.
        If $e^1 = e^2$, we are finished. Otherwise, since both $S_1$ and the number of paths (due to point 4) are finite, we can repeat until we reach an element of $S_1$ already visited.
        Therefore, $\mathscr{A}$ implies the presence of a causal loop.
    \end{proof}
\end{theorem}

As an ACL6 is a special case for an ACL6a, this proof also shows that any ACL6 can arise only from a cyclic causal structure.

\begin{example}
    \label{ex:ACL6a}
    An example of an ACL6a which is no ACL6 is given by
    \begin{equation}
        \mathscr{A} = \{ AB \affects CD , C \affects A , D \affects EF, E \affects B, F \affects A \} \, ,
    \end{equation}
    with all affects relations are presumed to be irreducible.
\end{example}

In \cref{sec:graph}, we will introduce a graphical representation of examples like this one, and show its representation in \cref{fig:ACL6a-loop}.

All examples from ACLs considered so far are built from complete affects chains.
Intuitively it is apparent that the causal structure spanned by the affects relations leading to an ACL6a has a central core or \enquote{trunk}, from $S_1$ to $S_2$, while the remainder of the ACL can be represented as a tree, whose nodes are given by sets of RVs.
In this picture, both the root and all leaves of this tree are given by $S_1$.

While it is possible to generalize this type of causal loop further by allowing branches to rejoin and mix recursively, we will refrain from working the respective details out specifically and develop a more general framework in the following section instead.

In addition to the complete affects chains considered here, incomplete affects chains exist. There, we replace the condition $S_i \subseteq \hat S_i$ from \cref{def:cac} with $S_i \cap \hat S_i \neq \emptyset$. While this alone does not suffice to certify cyclicity, multiple of such chains can complement each other to do so, as will be apparent from \cref{ex:ACL7} later on. However, since even including these types does not lead to an exhaustive list, as exhibited in \cite[App.~B]{VVC}, we will refrain from considering these classes further, and turn to developing and studying a general representation instead.

\section{New Results on Causal Inference for Affects Causal Loops} \label{sec:heavy}

After reviewing types of causal loops introduced in Section IV.A of \cite{VVC} and generalizing some of them, we will develop a general graphical representation for causal loops. To the best of the author's knowledge, this constitutes the first procedure which is able to exhaustively construct all causal loops operationally detectable using affects relations (i.e. affects causal loops).

\subsection{Completeness of Causal Inference} \label{sec:completeness}

In this section, we will change our point of view to study the interplay between affects relations and the causal structure from the opposite perspective.
If just the graph of a causal structure is given, which kind of affects relations can we read off it, and which sets of affects relations contain the full amount of causal information which may be extracted?
Here, we will focus on causal structures with free interventions, i.e. interventions are possible on each observed nodes. First, we will study the case without unobserved nodes more thoroughly, before afterwards making some statements which hold even in the case of unobserved nodes.
This will endow us with the necessary ingredients to constructively define the notion of an affects causal loop in the following sections.

\subsubsection{Completeness for Causal Structures without Unobserved Nodes}
In the case of no unobserved nodes, we can always use a functional approach to describe the causal model. Such a model can be completely determined by classical channels, constituting the structural equations of a causal model \cite[Rem.~IV.3]{VVC}.\footnote{
    Sometimes, an exogenous so-called error or noise variable $E_Y$ is included explicitly as a parameter of $f_Y$, in addition to $\parent(Y)$.
    This allows to transform any classical channel $f_Y$, which is implemented in the causal model, to a deterministic function of the exogenous variables of the model \cite[p.~21]{VVC}\cite{Barrett2021}.
    This poses a certain analogy to purification, which recovers a unitary transformation from a generic cptp map by extending the system under consideration.
    Some authors require this property for their notion of classical causality \cite[p.~16]{PhysRevA.81.062348}\cite{arxiv.1906.10726}.
    Since this is clearly a special case of the formalism where $E_Y$ is not explicitly introduced, but considered as a potential element of $\parent(Y)$, we will not regard it explicitly here.
}

\begin{definition}[Structural Equations]
    \label{def:structural}
    The \emph{structural equations} of a classical causal model are given by stochastic maps
    \begin{equation}
        Y = f_Y ( \parent(Y))
    \end{equation}
    for any $Y$ which are not exogenous. Here, we presume that no subset $S \subsetneq \parent(Y)$ exists such that $f_Y (S)$ leads to the same probability distribution.
    In particular, this excludes the case of $f_Y$ being constant.
\end{definition}

If the causal structure of this causal model is cyclic, we additionally require that different structural equations must be consistent with each other, as has been pointed out before in \cref{sec:cyclic-structures}.
More generally, in a causal structure $X \leftsquigarrow\!\!\!\!\!\dircause Y$, it must hold that $f_X \circ f_Y = f_Y \circ f_X = \text{id}$.
In the acyclic case, inconsistencies are not possible.

In the presence of unobserved nodes, a direct functional approach is no longer possible, since the respective nodes in $\parent(Y)$ are not available as arguments to $f_Y$.

\begin{mdframed}
    \centering
    For the remainder of this section, we will assume all RVs $A$ associated with observed nodes to be given by $A = f_A (\parent(A))$.
\end{mdframed}

\begin{notation}
    For $e_X, e_Y \in S$, $e_X$ being a direct cause of $e_Y$ will be denoted as $e_X \dircause e_Y$.
    For $X, Y \subset S$, if $X$ is a cause of $Y$ (cf. \cref{def:cause}) and this relation can be inferred via affects relations (using \cref{thm:affects-to-cause} or \cref{thm:affects-to-cause-HO}),
    we will denote this as $X \cause Y$.
\end{notation}

For the purpose of the statements of this section, we will sometimes assume that we have complete knowledge regarding which affects relations are present and absent, to give some basic statements which such information is redundant for purposes of causal inference.
To actually obtain such a set, we require the possibility of free interventions on all observable nodes.

\begin{definition}[Present and Absent Affects Relations]
    In addition to a set of affects relations $\mathscr{A}$, which contains some affects relations which are known to be present for a causal model, we may also consider a set of \enquote{does not affect}-affects relations $\mathscr{B}$, which is composed of affects relations which are known to be absent.
\end{definition}

Hence, we can also represent that from one node in our causal structure to another, no signalling occurs. As the affects relations in $\mathscr{A}$ may generally imply further affects relations by the lemmata of \cref{sec:affects-rules}, it is possible for $\mathscr{A}$ and $\mathscr{B}$ to be inconsistent with one another.
As a special case, we consider a complete set of affects relations.

\begin{definition}[Complete Set of Affects Relations]
    \label{def:affects-complete}
    A \emph{complete set of affects relations} $\mathscr{A}_\text{full}$ contains all generalized affects relations which are satisfied by a certain causal model.
    In this case the set of absent affects relations $\mathscr{B}$ is given by all valid potential affects relations which are not contained in $\mathscr{A}_\text{full}$.
    This definition naturally generalizes to a complete set under certain conditions, e.g. a complete set of affects relations of a certain type.
\end{definition}

Of course, it is possible for a complete set of affects relations $\mathscr{A}_\text{full}$ to be inconsistent, if it is not closed under affects relation transformations, i.e. if it does not contain all affects relations which are implied by its elements.

A complete set of affects relations can also be thought of as the list of affects relations an agent capable of free interventions on arbitrary subsets of a set of observed nodes $S$ can experimentally derive.
For each $X, Y, Z, W \subset S$ disjoint, this list specifies whether an affects relation is present. In particular, this also implies that the list of all affects relations which are absent is known.
An incomplete set of affects relations, on the other hand, would not capture all affects relations which might be theoretically detected by a causal model.
\Cref{sec:affects} and \cref{sec:affects-new} give rules which may allow to derive some of these missing affects relations.

\begin{lemma}
    \label{thm:cause-from-affects}
    Let $S$ be a set of RVs in a causal model without unobserved nodes and $X, Y \in S$ with $X \neq Y$. Then
    \begin{align}
        X \dircause Y
        &\iff X \affects Y \given \doo (\parent(Y) \setminus X) \label{eq:cause-from-affects-first} \\
        &\iff X \affects Y \given \doo (S \setminus XY) \, .  \label{eq:cause-from-affects-second}
    \end{align}
    \begin{proof}
        We have that $X \dircause Y$ is precisely stating that $f_Y$, and therefore $Y$, is having a non-trivial, direct dependence on $X$. We use this to prove the first claim:

        \hin If a dependence like this exists, there exist values of all (other) RVs $\parent(Y) \setminus X$ constituting its arguments, for which $X$ modifies the value of $Y$.
        This is precisely the definition of $X \affects Y \given \doo(\parent(Y) \setminus X)$.

        \rueck If $X \affects Y \given \doo(\parent(Y) \setminus X)$, there exists a combination of values of the RVs $\parent(Y) \setminus X$ such that the influence of $X$ is detectable on $Y$,
        implying a cause by \cref{thm:affects-to-cause}. As all other values $Y$ may depend on are fixed, this cause must be direct.

        Since all other nodes can not influence $Y$ if all direct influences are predetermined (via interventions), fixing them all does not change the argument in the proof. Therefore, it yields \cref{eq:cause-from-affects-second}.
    \end{proof}
\end{lemma}

Hence, this gives us an immediate way to express a causal arrow in an arbitrary causal structure without unobserved nodes through a single affects relation, even in the presence of fine-tuning.
Switching perspective, this allows to restrict the types of affects relations necessary to characterize causal structures without unobserved nodes significantly.

\begin{lemma}
    \label{thm:causal-discovery}
    Consider a causal model over a set of RVs $S$ and let $X, Y, Z \subset S$ disjoint.
    Then each associated causal structure $\mathcal{G}$ without unobserved nodes can be entirely reconstructed using the complete set of (irreducible) unconditional affects relations $\mathscr{A}_\text{full}$ (of the form $X \affects Y \given \doo(Z)$).
    Further, affects relations with $\abs{X} = \abs{Y} = 1$ (\textit{from} and \textit{to} individual RVs) suffice.
    \begin{proof}
        By \cref{thm:cause-from-affects}, for any pair of random variables $X$ and $Y$, we have
        \begin{equation}
            \label{eq:causal-discovery}
            X \dircause Y \quad \implies \quad X \affects Y \given \doo( S \setminus XY ) \, .
        \end{equation}
        By evaluating the higher-order part of this affects relation, we perform the total intervention given by $\doo (S \setminus Y)$.
        Since therefore all nodes except for $Y$ are exogenous in the post-intervention graph, indirect causes of $Y$ can not exist there.
        Therefore, only considering affects relations of this type, which particularly are irreducible and fulfill $\abs{X} = \abs{Y} = 1$, suffices to discover the full causal structure.
     \end{proof}
\end{lemma}

\begin{lemma}
    \label{thm:causal-discovery-simple}
    Consider a causal model over a set of RVs $S$ and let $X, Y \subset S$ disjoint.
    Then the existence of every directed path in an associated causal structure $\mathcal{G}$ without unobserved nodes can be confirmed or excluded using the complete set of \textit{irreducible} unconditional 0th-order affects relations $\mathscr{A}_\text{full}$ (of the form $X \affects Y$) with $\abs{Y} = 1$ (\textit{to} individual RVs).
    \begin{proof}
        Again, we first consider a property which yields \textit{all} direct causal relations, and then deduce that it yields \textit{only} direct causal relations.

        Due to the functional approach given in \cref{def:structural}, we have $\parent (Y) \affects Y$. Otherwise, $Y$ would not change for any change in its parents, and would therefore be constant, which is in contrast to the definition.
        This affects relation is irreducible, since applying the definition of irreducibility (\cref{def:reducible}) yields
        \begin{equation}
            X \affects Y \given \doo(\parent(Y) \setminus X) \, .
        \end{equation}
        These are precisely all associated affects relations of the form of \cref{thm:cause-from-affects}.

    \end{proof}
\end{lemma}

While this version retains all information which RVs are a cause of which, it does not preserve the information which of these causes are direct and which indirect.
This is due to the fact that generally, further irreducible affects relations which are neither from $\parent(Y)$ nor from its subsets may exist.
While all affects relations of the form $A \parent(Y) \affects Y$ are reducible since $A \naffects Y \given \doo(\parent(Y))$,
this is not necessarily the case for $X \subsetneq \parent(Y)$ and $A \subset S$ disjoint from $\parent(Y)$ with $AX \affects Y$.
For example, this may be the case if $A$ is an indirect cause of $Y$, hence, a cause of $\parent(Y) \setminus X$.

Nonetheless, we point out that while in \cref{thm:causal-discovery}, irreducibility has been added by hand, it is required here for the statement to hold.
This is linked to the fact that checking for reducibility involves checking for the presence of higher-order affects relations.

This emphasizes how irreducibility is naturally distilling higher-order affects relations from a single RV out of an unconditional 0th-order affects relation, rendering the higher-order information implicit.
Also, the freedom to not consider higher-order affects relations explicitly -- to retain general causal links, at least -- in the absence of unobserved nodes suits the observation of \cref{sec:affects-intro}, that the post-intervention relation is completely specified by the pre-intervention distribution in a purely classical model \cite[p.~42ff.]{VVphd}.

\begin{remark}
    \label{rem:no-corr-loops}
    This implies that if we have a causal structure $\mathcal{G}$
    \begin{itemize}
        \item without unobserved nodes,
        \item which can be freely intervened upon,
        \item for which we have a sufficiently complete set of affects relations $\mathscr{A}$, containing at least
        \begin{itemize}
            \item either all irreducible 0th-order unconditional affects relations with $\abs{Y} = 1$
            \item or all irreducible unconditional affects relations with $\abs{X} = \abs{Y} = 1$,
        \end{itemize}
    \end{itemize}
    \textbf{each} causal loop would be detectable using affects relations. Hidden causal loops or causal loops which might be detectable only by studying correlations, as discussed at the beginning of \cref{sec:acl}, are not possible there.
\end{remark}

Eventually, we have two complementary perspectives on irreducible affects relations with \cref{thm:causal-discovery} and \cref{thm:causal-discovery-simple}.
Incidentally, this also validates that at least in absence of unobserved nodes, only unconditional 0th-order affects relations have to be considered explicitly to study arbitrary examples of definite causal structures. However, this still requires encapsulating higher-order information into irreducibility.
Further, some, but not all information on the directness of causation is lost in the process.

\begin{example}
    Consider a causal model over a set $S = \{ A, B, C \}$ and a set of affects relations given by $\mathscr{A} = \{ A \affects B, B \affects C \}$. Then $A$ is an \textit{indirect} cause of $C$, as $A \cause B \cause C$. However, we can not exclude that there is a direct cause from $A$ to $C$ as well.
    For the causal relations $A \cause B$ and $B \cause C$, we can not infer whether they are direct or indirect from the given set of affects relations.
    However, if there is a direct cause from (exemplarily) $A$ to $B$, by \cref{thm:cause-from-affects}, $A \affects B \given \doo(C)$ must hold, if the causal structure has no additional unobserved nodes.
\end{example}

As a final remark, we point out that it is not possible to alternatively use only indecreasability properties to characterize a causal structure, as unconditional 0th-order affects relations with $\abs{X} = 1$ can not be rephrased as such.

Having all of this in place, we can use \cref{thm:affects-to-cause} (and \cref{thm:affects-to-cause-HO}, if indecreasability is given) to deconstruct complex affects relations into causal relations.
Beware that the implied causes may be indirect. Therefore, it is not possible to apply \cref{thm:cause-from-affects} to construct an actual mapping between causal structure and affects relations.

\begin{remark}
    In causal structures \emph{without unobserved nodes},
    affects relations with $\abs{Y} \neq 1$ in $\mathscr{A}$ (\emph{to} multiple RVs) indicate some ignorance regarding the target node of a causal relation.
    With respect to causal discovery, they are redundant to any affects relation $X \affects s_Y : s_Y \subset Y$, as conditioning additional RVs on $X$ can not break dependence.
    Similarly, \enquote{does not affect}-relations can supplement this information using \cref{thm:affects-to-HO}.
    However in causal structures without unobserved nodes, these are not required to reconstruct the causal structure.
\end{remark}

\subsubsection{Completeness for General Causal Structures} \label{sec:completeness-general}

When generalizing to the case of a causal structure with unobserved nodes, we notice that these can not be part of any affects relation, since they can not be intervened upon.\footnote{
    It is conceivable that under certain conditions, it may be possible to generalize the concept of interventions to unobserved nodes.
    For example, in the sense of (classical) experimental preparation, one might impose the condition that there are no exogenous hidden nodes. In this case, an unobserved node which has only classical parents could be intervened on: by removing the parents and replacing them with an intervention corresponding to a (theory-dependent) maximally general preparation of the state.
    If the respective theory has a notion of replacing (unknown) hidden information with known information, this approach generalizes even to all hidden nodes.
    Potentially, a theory only allows for a restricted set of interventions, leading to additional novel features.
    Exemplarily, \cite{VVR}\cite[Sec.~6]{arxiv.1906.10726} give arguments in this direction for quantum theory, however further research is needed, particularly incorporating the notion of generalized affects relations.
}
Hence, we must resort to affects relations of the form
\begin{equation}
    X \affects Y \given \doo (\parent_\text{obs} (Y) \setminus X) \, ,
\end{equation}
intervening only on the observable parents of $Y$, or, without knowledge of the causal structure,
\begin{equation}
    X \affects Y \given \doo (S \setminus XY) \, .
\end{equation}
As always, $S$ represents the set of observed nodes here.
Therefore, we can no longer do full causal discovery, as was possible through \cref{thm:causal-discovery} in the observed case.

Further, we can no longer exclude affects relations $X \affects Y \given  \{ \doo(Z) \}$ with $\abs{Y} > 1$ from a complete set of affects relations $\mathscr{A}_\text{full}$ to be able to do maximal causal discovery, as those imply additional causal relations.

\begin{example}
    Consider the jamming scenario introduced in \cref{ex:IV.6}, where we have no unconditional affects relations between two individual RVs, even including higher-order ones. However, $B \affects AC$ nevertheless holds.
\end{example}

Also, we can not maintain some of the causal inference results from last section in this case. In particular, this holds for \cref{thm:causal-discovery}: Even after intervening on all observed parents $\parent_\text{obs} (Y)$ of a node $Y$ or even on all observed nodes $S \setminus XY$, there may still be additional causes intervening on other nodes, mediated via unobserved nodes.
Further, affects relations of the form $X \affects Y \given \doo(S \setminus XY)$ do no longer imply the presence of a direct cause, as they might originate from a causal structure contaning $X \dircause \Lambda \dircause Y$, with an unobserved node $\Lambda$.
As its precondition of being able to isolate all parents of $Y$ is missing, we can find no immediate generalization of \cref{thm:causal-discovery-simple}.

In fact, we can show that the best causal inference one can perform in this case, using only affects relations, is the inference given in \cref{thm:affects-to-cause}. For stronger statements, we would require restrictions on the space of causal models, hence, on the theory or the set of causal structures.

\begin{figure}[t]
	\centering
    \begin{subfigure}[b]{0.3\textwidth}
        \centering
        \begin{tikzpicture}
            \begin{scope}[every node/.style={circle,thick,draw,inner sep=0pt,minimum size=0.8cm}]
                \node (X) at (0,1) {$X$};
                \node (Y) at (2,1) {$Y$};
                \node (Z) at (4,1) {$Z$};
            \end{scope}
            \node (P) at (0,0) {};

            \begin{scope}[>={Stealth[black]},
                          every node/.style={fill=white,circle},
                          every edge/.style=causalarrow]
                \path [->] (X) edge[bend right] (Y);
                \path [->] (Y) edge[bend right] (X);
                \path [->] (Y) edge (Z);
            \end{scope}
        \end{tikzpicture}
        \caption{Original causal structure $\mathcal{G}$}
        \label{fig:fine-grainining-original}
    \end{subfigure}%
    \hspace{0.1\textwidth}
    \begin{subfigure}[b]{0.3\textwidth}
        \centering
        \begin{tikzpicture}
            \begin{scope}[every node/.style={circle,thick,draw,inner sep=0pt,minimum size=0.8cm}]
                \node (X_1) at (0,0) {$X_1$};
                \node (X_2) at (0,2) {$X_2$};
                \node (Y_1) at (2,0) {$Y_1$};
                \node (Y_2) at (2,2) {$Y_2$};
                \node (Z_1) at (4,0) {$Z_1$};
                \node (Z_2) at (4,2) {$Z_2$};
            \end{scope}

            \begin{scope}[>={Stealth[black]},
                          every edge/.style=causalarrow]
                \path [->] (X_2) edge (Y_1);
                \path [->] (Y_2) edge (X_1);
                \path [->] (Y_2) edge (Z_1);
            \end{scope}
        \end{tikzpicture}
        \caption{Fine-grained causal structure $\mathcal{G}'$}
        \label{fig:fine-grainining-after}
    \end{subfigure}%
	\caption[Fine-graining of a cyclic causal structure to an acyclic causal structure.]{
        Example for a fine-graining of a cyclic causal structure to an acyclic causal structure, as is done as part of the proof of \cref{thm:affects-completeness}.}
	\label{fig:fine-graining}
\end{figure}
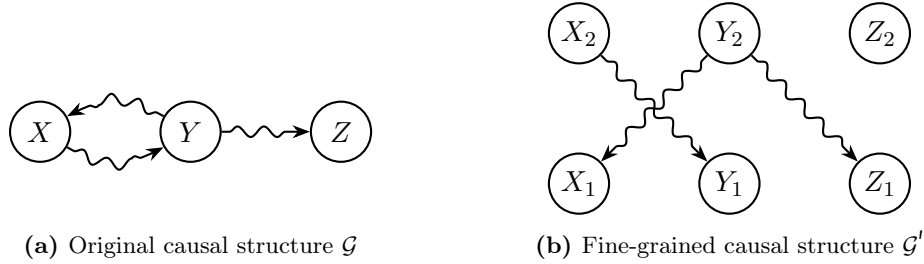

\begin{lemma}[Completeness of Causal Inference]
    \label{thm:affects-completeness}
    Consider a set of present (irreducible) affects relations $\mathscr{A}$.
    Then, if no further information is given, the complete set of causal relations (potentially) implied by $\mathscr{A}$ can be extracted using only \cref{thm:affects-to-cause}.
    \begin{proof}
        For some sets of RVs $X, Y, Z, W \subset S$, a single generic affects relation $X \affects Y \given \{ \doo(Z), W \}$
        only gives the causal relations between its RVs which are given by \cref{thm:affects-to-cause-first}.
        This can be seen from the example that when not imposing additional conditions, just having $\exists e_X \in X, e_Y \in Y$ with $e_X \dircause e_Y$ and $e_X \cong e_Y$ provides an example of a causal model where just causal relations potentially implied by this lemma exist.
        Notice that this is not in contradiction to the transformation theorems of \cref{sec:affects}, as even for special cases none of these can be used to tighten the statements of \cref{thm:affects-to-cause-first}.

        As a special case, we consider that for some affects relations in $\mathscr{A}$, we have given that they are irreducible.
        This does nothing more than provide a short-hand for a larger set of affects relations.
        By \cref{def:reducible}, an irreducible affects relation $X \affects Y \given \{ \doo(Z), W \}$ implies $e_X \affects Y \given \{ \doo(ZX \setminus e_X), W \} \ \forall e_X \in X$.
        This precisely corresponds to the implications of \cref{thm:affects-to-cause}, which therefore are no stronger than applying \cref{thm:affects-to-cause-first} to each of these affects relation individually.

        Now, we consider a set of generic affects relations and their joint implications of the causal structure.
        Here, we show that generally, for each choice of $\mathscr{A}$, a compatible faithful model exists which contains only causal relations potentially implied by using \cref{thm:affects-to-cause-first} on each affects relation. Therefore, we show that the conjunction of multiple affects relations does not allow for additional causal inference, if no further information about the causal model is known.

        Given $\mathscr{A}$, we first construct a possible causal structure $\mathcal{G}$ by adding an arbitrary relation $e_X \dircause e_Y$ for each $(X \affects Y \given \{ \doo(Z), W \}) \in \mathscr{A}$.
        Generally, this causal structure may be cyclic. Then, we can transform to another causal structure $\mathcal{G}'$, as depicted in \cref{fig:fine-graining}, where we replace each node $N \in \mathcal{G}$ with two nodes $N_1$ and $N_2$, satisfying
        \begin{equation}
            \parent(N_1) = \parent(N) \,, \quad \child(N_2) = \child(N) \,, \quad \child(N_1) = \parent(N_2) = \emptyset \, .
        \end{equation}
        The respective graph is always acyclic and for any nodes $M, N \in \mathcal{G}'$ satisfies
        \begin{equation}
            N \dircause M \ \iff \ N_1 N_2 \dircause M_1 M_2 \ \iff \ N_2 \dircause M_1 \, .
        \end{equation}
        As this graph is acyclic and has no unobserved nodes, any functional approach as given by \cref{def:structural} is possible, yielding a faithful causal model over $\mathcal{G}'$.

        Now, considering the original graph $\mathcal{G}$, we set any node $N$ to be isomorphic to $N_1 \times N_2$. For instance, this is possible by presuming $N_1$ and $N_2$ to be binary, and $N$ to be quaternary:
        \begin{equation}
            (N_1, N_2) \in \{ (0,0), (1,0), (0,1), (1,1) \} \ \text{is isomorphic to} \ N \in \{ 0, 1, 2, 3 \} \, .
        \end{equation}
        Then, we can presume each non-exogenous node $M$ to be given by bitwise addition of all its parents:
        \begin{equation}
            M \cong \bigoplus_{N \in \parent(M)} N \, .
        \end{equation}

        Thereby, we specify a (rather atypical) causal model which is compatible with the causal structure:
        Due to the distribution associated with $N$, incoming and outgoing causal influences are independent from each other. Therefore, all nodes which are not directly connected by $\dircause$ are independent, rendering this probability distribution compatible with $\mathcal{G}$ as per \cref{def:compat-stoch}.

        As this shows that for each choice of $\mathscr{A}$, the conjunction of multiple affects relation does not generally lead to any additional causal information, this concludes the proof.
    \end{proof}
\end{lemma}

The technique used for the second part of this proof is similar to the technique used for fine-graining cyclic causal structures that is introduced in Section 2.4.1 of \cite{VVR}.

As its proof is performed by composing a specific causal model, this statement does generally only apply if a set of present affects relations $\mathscr{A}$ is everything we know about the probability distribution of a system. If additional information about the causal model is present, it is possible that the proof method used to conjoin the affects relations fails.
For example, this is the case if the domain of the random variables is known to be binary or ternary, as in this case, it is not isomorphic to any cartesian product of two sets.
Otherwise, further knowledge of the probability distribution $P$, or of a set of absent affects relations $\mathscr{B}$, may similarly lead to additional conditions which are incompatible with this causal model.

Specifically, we have seen in \cref{sec:affects-new} that in particular for indecreasability, the absence of affects relations can indeed generally be used to uncover additional causal relations.

Nonetheless, even affects relations which are not required to discover the causal structure may be helpful to gain further insight into the properties of the model, even though they do not add any information for causal inference.
This is particularly true for conditional affects relations, which are, to the author's best knowledge, generally not equivalent to some logical combination of unconditional affects relations.

Using the result on the completeness of causal inference, we arrive at the following characterization of ACLs:

\begin{lemma}[Characterization of Affects Causal Loops]
    \label{thm:ACL}
    Let $\mathscr{A}$ be a set of 0th-order unconditional affects relations. Then the existence of an ACL in $\mathscr{A}$ implies
    $\exists (S_1 \affects S_2) \in \mathscr{A}$ irreducible such that the set of affects relations
    $\mathscr{A}$ implies that $\exists e_1 \in S_1, e_2 \in S_2$ with $e_2$ is a cause of $e_1$.
    \begin{proof}
        We show by contradiction that the presence of an ACL implies the existence of such a $(S_1 \affects S_2) \in \mathscr{A}$.

        If we have an ACL, the presence of a causal loop is implied by the set of affects relations $\mathscr{A}$.
        These imply precisely the causal relations given by \cref{thm:affects-to-cause}, as discussed in \cref{thm:affects-completeness}.
        For $S_1 \affects S_2$ in particular, we arrive at $\forall e_1 \in S_1 \ \exists e_2 \in S_2 : e_1 \cause e_2$.

        Now we assume the negation of the outer existence claim: $\forall (S_1 \affects S_2) \in \mathscr{A}$ irreducible the set of affects relations $\mathscr{A}$ does not imply the existence of
        $e_1 \in S_1, e_2 \in S_2$ non-empty with $e_2$ is a cause of $e_1$.
        Hence, there is no chain of causal arrows from any element of $S_2$ to $S_1$ for any affects relation.
        However, if this is the case, no affects relation can contribute a causal relation which is part of a causal loop, which is in contradiction to the presence of an ACL. $\lightning$
    \end{proof}
\end{lemma}

Beware that this statement is not constructive regarding for which irreducible affects relation in $\mathscr{A}$ this is the case.
Since any conditional affects relation implies the presence of an unconditional one with identical causal implications, the statement can be generalized accordingly to conditional affects relations.

\subsection{Graphical Representation of Causal Loops} \label{sec:graph}

With the completeness results at hand, we can now find a graphical representation for the causal relations implied by affects relations.
To do so, we use all causal relations which are implied by \cref{thm:affects-to-cause} for irreducible affects relations to construct a \textit{potential cause graph}.
This graph captures the whole set of causal structures that could give rise to the affects relations, and can be imagined to represent an equivalence class of causal structures.
Once the potential cause graph is constructed, we define a variation which gives an immediate and exhaustive way to detect the presence of cycles in the causal structure implied by a set of affects relations.

\begin{definition}[Potential Cause Graph]
    \label{def:causal-graph}
    Consider a causal model over a set of RVs $S$ with $X, Y_1, Y_2, Z \subset S$.
    Let $\mathscr{A}$ be the set of affects relations which are present on $S$.
    A \emph{(Presence) Potential Cause Graph} $\mathcal{G}^{\potcause}$ is constructed by providing the elements of $S$ as nodes and adding arrows
    $e_X \potcause_Y e_Y$ for each $e_X \in X, e_Y \in Y_1 Y_2$ for each
    $X \affects Y_1 \given \{\doo(Z), Y_2 \}$ in $\mathscr{A}$ irreducible.
\end{definition}

Therefore, the respective potential cause graph contains all causal relations implied by \cref{thm:affects-to-cause}, with the arrows being indexed by the set $Y_1 Y_2$ from the second and fourth argument, which enter on the existence side of that lemma. In particular, it is possible for two nodes to be connected by multiple arrows with different indices.
Only if from a certain node $X$ there are sets of arrows $\potcause_A$ and $\potcause_B$ with $A \subsetneq B \subset S$, deduplication is possible by removing \textit{all} arrows $\potcause_B$ originating from $X$.

\begin{corollary}
    The potential cause graph contains the complete set of (potential) causal relations that can be inferred from a set of irreducible affects relations.
    \begin{proof}
        This is a direct application of \cref{thm:affects-completeness}.
    \end{proof}
\end{corollary}

\begin{notation}
    \label{not:cause-graph}
    If a $\potcause_Y$ arrow in a potential cause graph has $\abs{Y} = 1$, we write $\cause$ instead, as the arrow corresponds to an actual cause (which may be indirect) in this case. This precisely matches to the notion of \cref{not:affects-cause}, and the information of the index is redundant here. If $\abs{Y} \neq 1$, we will also write $\actpotcause_Y$ instead of $\potcause_Y$.

    Further, we will drop the index of $\potcause_Y$ (and $\actpotcause_Y$, respectively) if for a given node, all such outgoing arrows share their index $Y$.
    For the presence causal graph, this is the case if for a given RV $X$, there is only one affects relation $S_1 \affects S_2 \given \{ \doo(S_3), S_4 \}$ with $X \in S_1$ and $\abs{S_2 S_4} \neq 1$.
    In this case, the respective information of the index is redundant.
\end{notation}

As we have considered in \cref{sec:affects-new}, there are additional causal relations which can be inferred from higher-order affects relations using information about the absence of affects relations, as given by \cref{thm:affects-to-cause-HO}. As a special case, this includes the information inferable from indecreasable affects relations (cf. \cref{def:indecreasable}).
As this information is not available in the setup chosen for \cref{def:causal-graph}, we consider an enhanced version, using an additional set $\mathscr{B}$ of absent affects relation complementary to the set of present affects relations.

\begin{definition}[Extended Potential Cause Graph]
    \label{def:causal-graph-HO}
    Consider a causal model over a set of RVs $S$ with $X, Y_1, Y_2, Z \in S$.
    Let $\mathscr{A}$ the set of affects relations which are present and $\mathscr{B}$ the set of affects relations which are known to be absent on $S$.
    An \emph{Extended Potential Cause Graph} $\mathcal{G}^{\potcause}$ is constructed from a presence potential cause Graph by adding further arrows
    $e_Z \potcause_Y e_Y$ for each $e_Z \in Z, e_Y \in Y_1 Y_2$ with
    $X \affects Y_1 \given \{\doo(Z), Y_2 \}$ in $\mathscr{A}$ and $X \naffects Y_1 \given \{\doo(Z \setminus e_Z), Y_2 \}$ in $\mathscr{B}$.
\end{definition}

The same notational simplification can be performed for this graph as well.

\begin{definition}[Loop Graph]
    \label{def:loop-graph}
    A \emph{(Presence/Extended) Loop Graph} $\mathcal{G}^\circ$ is constructed by taking a (presence or extended) potential cause graph $\mathcal{G}^{\potcause}$ and repeat the following procedure until every node has children:
    \begin{enumerate}
        \item Remove any node $e_Y$ which has no children (but potentially, parents), as well as all incoming arrows $e_X \potcause_A e_Y$ for any $e_X$ (\textit{to} $e_Y$).
        \item For each arrow $e_X \potcause_A e_Y$ removed this way, remove all other arrows $e_X \potcause_A e_W$ for any $e_W$ (\textit{from} $e_X$ with the same $A$).
    \end{enumerate}
    Optionally, we may also recursively remove any node with no parents.
\end{definition}

This procedure has the goal of removing causal relations which can not be part of a causal loop, and successfully does so, as we will show in \cref{thm:graph}.
While the last step is not necessary for the proof, it will remove nodes which are not actually involved into any causal loop, and thereby improve the visual representation. Affects relations imply no (potential) causal influence to a node which has no parent in the potential cause graph.
For the examples of loop graphs depicted in this work, we will assume this step to be performed.

We introduce some examples, depicted in \cref{fig:ACL-pot} and \cref{fig:ACL-loop}, to familiarize ourselves with the concepts of potential cause graph and loop graph before continuing.
A further example of intermediate complexity can be seen in \cref{fig:ACL7-clg}.
We will not explicitly show presence or absence of a causal loop for these examples, since this will be covered by the general proof to follow in \cref{thm:graph}.

\begin{figure}[t]
	\centering
    \begin{subfigure}[b]{0.3\textwidth}
        \centering
        \begin{tikzpicture}
            \node (A) at (0,2) {$e_1$};
            \node (B) at (2,2) {$e_2$};
            \node (C) at (2,4) {$e_3$};
            \node (D) at (0,4) {$e_4$};
            \node (E) at (0,0) {$e_5$};
            \node (F) at (2,0) {$e_6$};

            \begin{scope}[>={to[black]},
                          every edge/.style=affectsarrow]
                \path [->>] (A) edge[bend right] (B);
                \path [->>] (B) edge[bend right] (A);
                \path [->>] (C) edge (B);
                \path [->>] (D) edge (A);
                \path [->>] (D) edge (B);
                \path [->>] (E) edge (A);
                \path [->>] (B) edge (F);
            \end{scope}
        \end{tikzpicture}
        \caption{an ACL3 (\cref{ex:ACL3})}
        \label{fig:ACL3-pot}
    \end{subfigure}%
    \begin{subfigure}[b]{0.3\textwidth}
        \centering
        \begin{tikzpicture}
            \node (A) at (0,0) {$A$};
            \node (B) at (0,2) {$B$};
            \node (C) at (2,0) {$C$};
            \node (D) at (2,2) {$D$};

            \begin{scope}[>={to[black]},
                          every edge/.style=affectsarrow]
                \path [dashed,->>] (A) edge (B);
                \path [dashed,->>] (A) edge (C);
                \path [->>] (B) edge (D);
                \path [->>] (D) edge (A);
            \end{scope}
        \end{tikzpicture}
        \caption{\cref{ex:noACL}}
        \label{fig:noACL-pot}
    \end{subfigure}%
    \begin{subfigure}[b]{0.3\textwidth}
        \centering
        \begin{tikzpicture}
            \node (A) at (0,2) {$A$};
            \node (X) at (2,2) {$X$};
            \node (C) at (0,0) {$C$};
            \node (B) at (2,0) {$B$};
            \node (D) at (0,-2) {$D$};
            \node (E) at (2,-2) {$E$};

            \begin{scope}[>={to[black]},
                          every node/.style={fill=white,circle,inner sep=0pt},
                          every edge/.style=affectsarrow]
                \path [->>] (A) edge[bend left] (X);
                \path [dashed,->>] (X) edge (A);
                \path [dashed,->>] (X) edge (B);
                \path [dashed,->>] (B) edge[bend left] node {$\scriptscriptstyle{CD}$} (C);
                \path [dashed,->>] (B) edge node {$\scriptscriptstyle{CD}$} (D);
                \path [dashed,->>] (B) edge node {$\scriptscriptstyle{AC}$} (A);
                \path [dashed,->>] (B) edge[bend right] node {$\scriptscriptstyle{AC}$} (C);
                \path [dashed,->>] (C) edge (A);
                \path [dashed,->>] (C) edge (B);
                \path [dashed,->>] (D) edge[bend left] (A);
                \path [dashed,->>] (D) edge (C);
                \path [dashed,->>] (D) edge[bend right] node {$\scriptscriptstyle{BE}$} (B);
                \path [dashed,->>] (D) edge node {$\scriptscriptstyle{BE}$} (E);
                \path [->>] (B) edge (E);
            \end{scope}

        \end{tikzpicture}
        \caption{\cref{ex:ACL11}}
        \label{fig:ACL11-pot}
    \end{subfigure}%
	\caption[Examples of potential cause graphs for different ACLs.]{
        These graphs depict multiple potential cause graphs, as defined by \cref{def:causal-graph}, for different examples.
        As stated by \cref{not:cause-graph}, causes which are guaranteed to exist through the affects relations are denoted as $\cause$, while causes which alternatively exist are denoted as $\actpotcause$.
        If there are different families of potential causes, where it is necessary to pick at least one cause from each family, the respective family is given on top of the arrow.
    }
	\label{fig:ACL-pot}
\end{figure}

\begin{figure}[t]
	\centering
    \begin{subfigure}[b]{0.3\textwidth}
        \centering
        \begin{tikzpicture}
            \node (A) at (0,2) {$e_1$};
            \node (B) at (2,2) {$e_2$};
            \node (C) at (2,4) {};
            \node (D) at (0,4) {};
            \node (E) at (0,0) {};
            \node (F) at (2,0) {};

            \begin{scope}[>={to[black]},
                          every edge/.style=affectsarrow]
                \path [->>] (A) edge[bend right] (B);
                \path [->>] (B) edge[bend right] (A);
            \end{scope}
        \end{tikzpicture}
        \caption{any ACL3 (\cref{ex:ACL3})}
        \label{fig:ACL3-loop}
    \end{subfigure}%
    \begin{subfigure}[b]{0.3\textwidth}
        \centering
        \begin{tikzpicture}
            \node (A) at (0,0) {$A$};
            \node (B) at (0,2) {$B$};
            \node (C) at (2,0) {$C$};
            \node (D) at (2,2) {$D$};
            \node (E) at (2,4) {$E$};
            \node (F) at (0,4) {$F$};

            \begin{scope}[>={to[black]},
                          every edge/.style=affectsarrow]
                \path [dashed,->>] (A) edge (C);
                \path [dashed,->>] (A) edge (D);
                \path [dashed,->>] (B) edge (C);
                \path [dashed,->>] (B) edge (D);
                \path [->>] (C) edge[bend left] (A);
                \path [dashed,->>] (D) edge (E);
                \path [dashed,->>] (D) edge (F);
                \path [->>] (E) edge (B);
                \path [->>] (F) edge[bend right] (A);
            \end{scope}
        \end{tikzpicture}
        \caption{an ACL6a (\cref{ex:ACL6a})}
        \label{fig:ACL6a-loop}
    \end{subfigure}%
    \begin{subfigure}[b]{0.3\textwidth}
        \centering
        \begin{tikzpicture}
            \node (A) at (0,2) {$A$};
            \node (X) at (2,2) {$X$};
            \node (C) at (0,0) {$C$};
            \node (B) at (2,0) {$B$};
            \node (D) at (0,-2) {$D$};
            \node (E) at (2,-2) {};

            \begin{scope}[>={to[black]},
                          every node/.style={fill=white,circle,inner sep=0pt},
                          every edge/.style=affectsarrow]
                \path [->>] (A) edge[bend left] (X);
                \path [dashed,->>] (X) edge (A);
                \path [dashed,->>] (X) edge (B);
                \path [dashed,->>] (B) edge[bend left] node {$\scriptscriptstyle{CD}$} (C);
                \path [dashed,->>] (B) edge node {$\scriptscriptstyle{CD}$} (D);
                \path [dashed,->>] (B) edge node {$\scriptscriptstyle{AC}$} (A);
                \path [dashed,->>] (B) edge[bend right] node {$\scriptscriptstyle{AC}$} (C);
                \path [dashed,->>] (C) edge (A);
                \path [dashed,->>] (C) edge (B);
                \path [dashed,->>] (D) edge[bend left] (A);
                \path [dashed,->>] (D) edge (C);
            \end{scope}

            \path (D) edge[bend right,draw=none] (E);
        \end{tikzpicture}
        \caption{\cref{ex:ACL11}}
        \label{fig:ACL11-loop}
    \end{subfigure}%
	\caption[Examples of Loop Graphs for different ACLs.]{
        These graphs depict multiple Loop Graphs, as defined by \cref{def:loop-graph}, for different examples.
        The meaning of the arrows is identical to \cref{fig:ACL-pot}, as given by \cref{not:cause-graph}.
    }
	\label{fig:ACL-loop}
\end{figure}
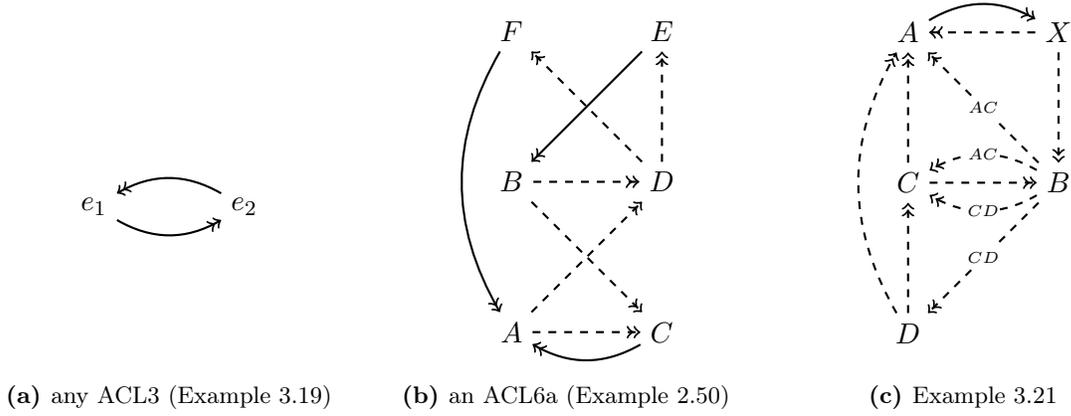

\begin{example}
    \label{ex:ACL3}
    Let $e_1, e_2, e_3, e_4, e_5, e_6 \in S$.
    Consider a set of irreducible affects relations given by
    \begin{equation}
        \mathscr{A} = \{ e_1 e_3 e_4 \affects e_2, \, e_2 e_5 \affects e_1, \, e_4 \affects e_1, \, e_2 \affects e_6 \} \, ,
    \end{equation}
    containing an ACL3 as given by \cref{def:ACL3}, with $S_1 = e_1 e_3 e_4$ and $S_2 = e_2 e_5$. The respective potential cause graph is shown in \cref{fig:ACL3-pot}, while the respective causal loop graph, which looks identical for any causal structure containing an ACL3 as only causal loop, is shown in \cref{fig:ACL3-loop}.
\end{example}

\begin{example}
    \label{ex:noACL}
    Let $A, B, C \in S$ and consider a set of irreducible affects relations given by
    \begin{equation}
        \mathscr{A} = \{ A \affects BC, B \affects D, D \affects A \} \,
    \end{equation}
    with the respective potential cause graph depicted in \cref{fig:noACL-pot}. Even though $A, B$ and $D$ are apparently forming a circular dependence, this set of affects relations does not imply the presence of a causal loop.
    Since $C$ is a childless node, both $A \potcause_{BC} B$ and $A \potcause_{BC} C$ are removed by construction of the loop graph. This again renders $A$ a childless node.
    Continuing, we arrive at an empty loop graph, that implies that there exists an acyclic causal structure which give rise to these affects relations.
    One example for a causal structure which may yield these affects relations is given by $B \dircause D \dircause A \dircause C$.
\end{example}

To show the full power of the potential cause graph, we introduce another, more complex example for an affects causal loop.

\begin{example}{\cite[Ex.~B.3]{VVC}}
    \label{ex:ACL11}
    Let $A, B, C, D, E, X \in S$. Consider a set of irreducible affects relations given by
    \begin{equation}
        \mathscr{A} = \{ X \affects AB, A \affects X, C \affects AB, B \affects CD, BD \affects AC, D \affects BE, B \affects E \} \, ,
    \end{equation}
    which contains an affects causal loop. The last two affects relations are actually irrelevant for the causal loop and are therefore not contained in the loop graph shown in \cref{fig:ACL11-loop}.
    The respective potential cause graph is shown in \cref{fig:ACL11-pot}.
\end{example}

We continue by giving some final, technical definitions before stating the theorem.

\begin{notation}
    We denote the number of nodes in a graph $\mathcal{G}$ by $\abs{\mathcal{G}}$.
\end{notation}

\begin{definition}[Strongly Connected Component of a Directed Graph]
    A \emph{strongly connected component} $\mathfrak{g}$ of a directed graph $\mathcal{G}$ is a subgraph $\mathfrak{g}$ where every node $A \in \mathfrak{g}$ satisfies $\mathfrak{g} = \desc(A) = \anc(A)$.
\end{definition}

Therefore, in this case, from any node in a strongly connected component $\mathfrak{g}$, there exists a path both to the node itself and to any other node in $\mathfrak{g}$.

Hence, if a graph consists of multiple strongly connected components, it has multiple subgraphs which are not connected with each other.

\begin{theorem}
    \label{thm:graph}
    A set of affects relations $\mathscr{A}$ implies the presence of an ACL if and only if the corresponding presence loop graph $\mathcal{G}^\circ$ is not empty.
    \begin{proof}
        \rueck First, we will show, that if the graph is not empty, the presence of an ACL is implied.

        By construction of the loop graph, any node has children. This implies that for any node $A$, the induced subgraph obtained from removing some $B \not\in \desc(A)$ is still cyclic, as from each $A' \in \desc(A)$ there is an arrow either to $A$ or to another element in $\desc(A)$.
        We choose $A_1 \in \mathcal{G}^\circ$. If $A_1 \in \desc(B) \ \forall B \in \mathcal{G}^\circ$, we are done.
        Otherwise, $\exists A_2 \in \mathcal{G}^\circ : A_1 \not\in \desc(A_2)$. We therefore remove $A_1$ and the associated arrows from the graph, obtaining an induced subgraph $\mathfrak{g}^{(1)}$, and recursively repeat the procedure with $A_2$ instead of $A_1$.
        Since the number of nodes in the graph is finite, we will eventually reach a strongly connected component $\mathfrak{g}^{(n)}$, where there exists a path to some node $A_n$ from each node. Otherwise, we would obtain an empty graph after $\abs{\mathcal{G}}$ repetitions. However, with the empty graph being acyclic, this poses a contradiction.

        If $\mathscr{A}$ would imply no causal loop, we would be able to choose the $\potcause_Y$ relations to be realized (at least one for each combination of parent node and $Y$) such that there is no cycle consisting of causal arrows.
        However, as $\mathfrak{g}^{(n)} = \desc(A) = \anc(A)$, each node is both a descendant and an ancestor of $A$. As there is at least one $\potcause_Y$ relation from each node (for some $Y$), it is impossible to choose the causal relations without the path visiting at least one node twice. Therefore, $\mathscr{A}$ implies a causal loop if the graph is not empty.

        \hin Second, we will show that if the graph is empty, there can not be an ACL implied.

        By \cref{thm:affects-completeness}, the potential cause graph $\mathcal{G}^{\potcause}$ contains the entire information on the causal structure which can be extracted from a set of affects relations.
        Therefore, all causal relations and in particular all ACLs within are included there.
        If the loop graph $\mathcal{G}^\circ$ is empty, from each node in the original potential cause graph $\mathcal{G}^{\potcause}$, there exists a path built from a path of $\potcause$ relations -- and therefore, a choice of causal relations -- leading to a node without children, compatible with an acyclic causal structure.
        This holds since the rule of \cref{def:loop-graph} precisely removes $\potcause$ arrows where a choice can be made to arrive at a node without children.
        Since this causal structure is compatible with acyclicity, it implies no causal loop.
    \end{proof}
\end{theorem}

The proof is essentially an abstraction of the method used by \cite{VVC} to prove the presence of ACLs for specific sets $\mathscr{A}$.
Analogous, we can define:

\begin{theorem}
    \label{thm:graph-HO}
    A set of affects relations $\mathscr{A}$ and a set of absent affects relations $\mathscr{B}$ imply the presence of an ACL if the corresponding Extended Loop Graph $\mathcal{G}^\circ$ is not empty.
    \begin{proof}
        The proof is analogous to \enquote{$\Rightarrow$} of \cref{thm:graph}.
    \end{proof}
\end{theorem}

To study whether an only-if-condition is fulfilled here, one would need to study the causal implications of the absence of certain affects relations. We have not aimed to exhaustively do this in this work, but it remains an area of further study.

To visualize examples, it is useful to \enquote{unwind} this graph by duplicating a node for each encounter, and terminate the respective branch when reencountering a RV visited before. In doing so, we retrieve the heralded tree with $X$ as a base.

\begin{remark}
    One may identify the strongly connected component $\mathfrak{g}^{(n)}$ of the loop graph $\mathcal{G}^\circ$ with a causal loop.
    However, there may be other causal loops like this within $\mathcal{G}^\circ$, as is for example the case for
    \begin{equation}
        \mathfrak{g} = \{X \cause Y, Y \cause Z, Z \cause X, X \cause A, A \cause B, B \cause A \} \, ,
    \end{equation}
    which contains two distinct subgraphs which form an ACL by itself.
\end{remark}

\newpage
\part{Compatibility of Information-Theoretic and Spacetime Notions of Causality}
\section{Spacetime and Compatibility}
After treating causality from a information-theoretic point of view in the last two sections, we now finally turn to the better-known concept of relativistic causality.
In the following, we will introduce a minimal model of the causal structure of spacetime. Then, we will embed random variables into it to relate both perspectives on causality to each other.
Finally, we will postulate a set of properties for these embeddings which are required to gain a consistent picture from both approaches.
This will complete the toolset we require to study the potential physicality of causal loops in a acyclic spacetime.

\subsection{Relativistic Causality} \label{sec:poset}
To model the causal properties of spacetime in this framework, we aim to be as general as possible. Therefore, we will model the causal structure of spacetime as a partially ordered set (poset) $\TT$ of its points, as suggested by \cite{Kronheimer1967}.
The fundamental definitions of this section are a synthesis of the information on related concepts from nLab \cite{nlab}, as well as German and English Wikipedia.

\begin{definition}[Partial Order]
    A \emph{(strict) partial order} is a binary relation $\prec$ on a set $T$ which satisfies
    \begin{itemize}
        \item Irreflexivity: $a \not\prec a$.
        \item Asymmetry: $a \prec b \implies b \not\prec a$.
        \item Transitivity: $a \prec b$ and $b \prec c \implies a \prec c$.
    \end{itemize}
    for all $a, b, c \in T$.
\end{definition}

\begin{definition}[Poset]
    A poset $\TT$ is given by a set $T$ endowed with a partial order: $\TT := (T, \prec)$.
\end{definition}

\begin{figure}[t]
    \centering
    \tikzset{surface/.style={draw=blue!70!black, fill=blue!20!white, fill opacity=.6}}

    \newcommand{\coneback}[4][]{
    \draw[canvas is xy plane at z=#2, #1] (45-#4:#3) arc (45-#4:225+#4:#3) -- (O) --cycle;
    }
    \newcommand{\conefront}[4][]{
    \draw[canvas is xy plane at z=#2, #1] (45-#4:#3) arc (45-#4:-135+#4:#3) -- (O) --cycle;
    }
    \scalebox{0.9}{
    \begin{tikzpicture}[tdplot_main_coords, grid/.style={help lines,violet!40!white,opacity=0.5},scale=1]
        \coordinate (O) at (0,0,0);

        \coneback[surface]{-3}{2}{-12}
        \conefront[surface]{-3}{2}{-12}

        \fill[violet!40!white,opacity=0.5] (-4,-4,0) -- (-4,4,0) -- (4,4,0) -- (4,-4,0) -- cycle;

        \foreach \x in {-4,...,4}
        \foreach \y in {-4,...,4}
        {
         \draw[grid] (\x,-4) -- (\x,4);
         \draw[grid] (-4,\y) -- (4,\y);
         \draw[violet] (-4,4)--(-4,-4)--(4,-4)--(4,4)--cycle;
        }

        \draw[->] (-4,0,0) -- (4,0,0) {};
        \draw[->] (0,-4,0) -- (0,4,0) {};
        \coneback[surface]{3}{2}{12}
        \draw[-,dashed] (0,0,-2.65) -- (0,0,2.65) node[above] {};
        \draw[-,dashed] (0,0,-4) -- (0,0,-3.35) node[above] {};
        \draw[->,dashed] (0,0,3.35) -- (0,0,4) node[above] {$time$};
        \conefront[surface]{3}{2}{12}
        \fill (4,0,2) circle (2pt) node[above right] {$C$};
        \fill (0,0,0) circle (2pt) {};
        \fill (-0.5,-0.85,2.2) circle (2pt) node[above left] {$A$};
        \fill (1.3,0.5,2) circle (2pt) node[above left] {$B$};
        \draw[->,red] (0,0,0) -- (4,0,2) node[below, pos=0.6, rotate=26.5651,scale=0.70,black] {$\textbf{spacelike vector}$};
        \draw[->,red] (0,0,0) -- (1.3,0.5,2) node[below, pos=0.65, rotate=55.1459,scale=0.70,black] {$\textbf{lightlike vector}$};
        \draw[->,red] (0,0,0) -- (-0.5,-0.85,2.2) node[above, pos=0.57, rotate=-65.8557,scale=0.70,black] {$\textbf{timelike vector}$};
        \node[black] at (0,0,3) {$Future\,\,Light\,\,Cone$};
        \node[black] at (0,0,-3) {$Past\,\,Light\,\,Cone$};
        \node[black] at (0,0.05,0.3) {$O$};
        \node[black] at (0,4.7,0) {$space$};
        \node[black] at (5,-0.3,0) {$space$};
    \end{tikzpicture}
    }
	\caption[A light cone in 2+1-Minkowski spacetime with space-like and time-like region.]{
        A light cone in 2+1-Minkowski spacetime. For each point $O$, there exist time-like separated points $A$, in this case with $A \succ O$, space-like separated points $C$ with $C \unord O$, which are unordered with regard to $O$, and light-like separated points, which are again ordered with respect to $O$: Here, $B \succ O$.
        Therefore, relative to $O$, there exist a region of space-like separated points and two regions of time-like separated points, which are separated by a surface of codimension 1 of points which are separated in a light-like way.
        Picture published by SandyG and Nick at \href{https://tex.stackexchange.com/questions/640441/}{TeX Stack Exchange} and licensed under \href{https://creativecommons.org/licenses/by-sa/4.0/}{CC BY-SA 4.0}.}
	\label{fig:minkowski}
\end{figure}
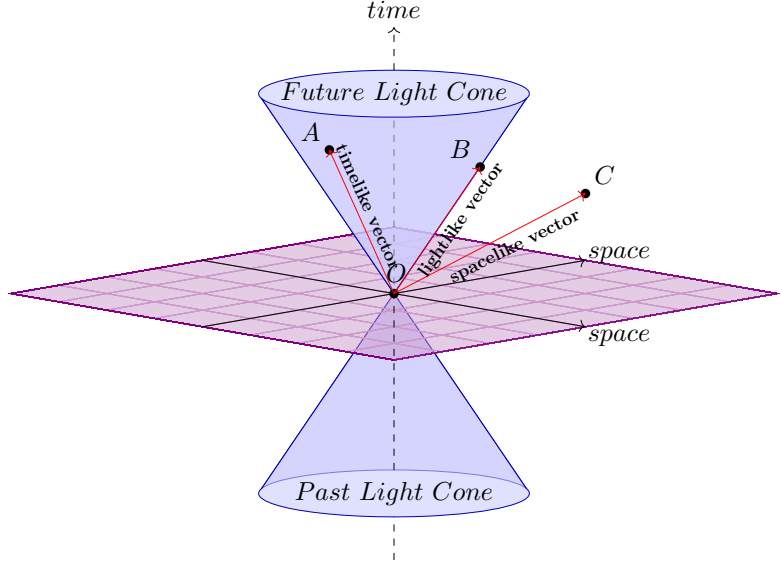

Hence, this approach removes almost the entire structure which is usually considered when studying spacetime: All information can be expressed by the pairwise ordering of points $a, b \in \TT$, which can be exactly one of the following:
\begin{equation}
    a = b \, , \quad
    a \prec b \, , \quad
    a \succ b \, , \quad
    a \unord b \, .
\end{equation}
Hence, if $a$ and $b$ are not identical, they can either be in the causal future/past and therefore timelike or lightlike, or causally unordered and therefore spacelike with regard to each other.

By allowing to order events in spacetime in a transitive way, this approach is generic enough to model causality for an arbitrary spacetime manifold without closed timelike curves (CTC), as for example Minkowski spacetime depicted in \cref{fig:minkowski}.
Allowing for CTCs in the manifold would allow for causal loops of the form $b \prec a \prec b$, which would violate both the asymmetry property of the partial order and the conjunction of irreflexitivity and transitivity.
Therefore, such spacetimes can be modelled as a \emph{(weak) preorder}, which does not demand asymmetry and relaxes irreflexivity to reflexivity: $a \preceq a \ \forall a \in T$.
However, this generality is undesirable here since it allows the causal future and the causal past of a point to be non-disjoint, and would thereby allow the embedding of arbitrary causal loops later on.

\begin{figure}[t]
	\centering
    \begin{subfigure}[b]{0.45\textwidth}
        \centering
        \begin{tikzpicture}
            \begin{scope}[every node/.style={rectangle,thick,draw}]
                \node (A) at (2,1) {$A$};
                \node (B) at (0,2) {$B$};
                \node (C) at (4,2) {$C$};
                \node (D) at (0,4) {$D$};
                \node (E) at (4,4) {$E$};
                \node (F) at (2,5) {$F$};
            \end{scope}

            \begin{scope}[every edge/.style={draw=black,thick}]
                \path [-] (A) edge (B);
                \path [-] (B) edge (D);
                \path [-] (A) edge (C);
                \path [-] (C) edge (E);
                \path [-] (E) edge (F);

                \path [-] (A) edge (C);
                \path [-] (B) edge (E);
                \path [-] (C) edge (D);
                \path [-] (D) edge (F);
            \end{scope}
        \end{tikzpicture}
        \caption{Poset which is no lattice}
        \label{fig:hasse-poset}
    \end{subfigure}%
    \begin{subfigure}[b]{0.45\textwidth}
        \centering
        \begin{tikzpicture}
            \begin{scope}[every node/.style={rectangle,thick,draw}]
                \node (A) at (2,1) {$A$};
                \node (B) at (0,2) {$B$};
                \node (C) at (4,2) {$C$};
                \node (D) at (0,4) {$D$};
                \node (E) at (4,4) {$E$};
                \node (F) at (2,5) {$F$};
                \node (G) at (-2,3) {$G$};
                \node (H) at (2,3) {$H$};
                \node (I) at (6,3) {$I$};
            \end{scope}

            \begin{scope}[every edge/.style={draw=black,thick}]
                \path [-] (A) edge (B);
                \path [-] (B) edge (G);
                \path [-] (G) edge (D);
                \path [-] (A) edge (C);
                \path [-] (C) edge (I);
                \path [-] (I) edge (E);
                \path [-] (H) edge (D);
                \path [-] (H) edge (E);
                \path [-] (E) edge (F);

                \path [-] (A) edge (C);
                \path [-] (B) edge (H);
                \path [-] (C) edge (H);
                \path [-] (G) edge (D);
                \path [-] (D) edge (F);
            \end{scope}
        \end{tikzpicture}
        \caption{Lattice}
        \label{fig:hasse-lattice}
    \end{subfigure}%
	\caption[Hasse diagrams, representing finite posets: A non-lattice and a lattice.]{
        Finite posets can be depicted using \textit{Hasse diagrams}. Here, two nodes $A$ and $B$ are connected and $B$ is shown above $A$ if $B$ covers $A$ (cf. \cref{def:cover}).
        In (a), we see a poset which is no lattice, since $B \prec D \succ C$ and $B \prec E \succ C$. Therefore, since $D \unord E$, $B$ and $C$ have no least upper bound.
        In (b), we see a poset which forms a lattice. Here, $H = B \lor C$.
    }
	\label{fig:hasse}
\end{figure}
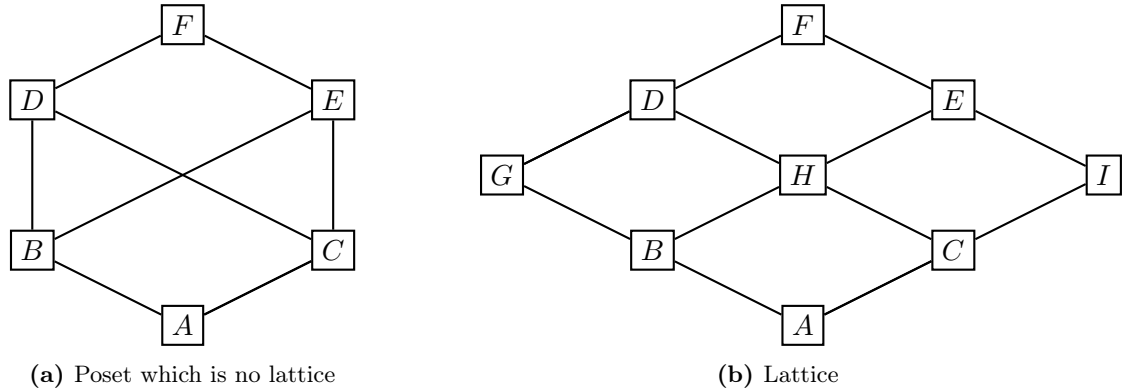

Further, it allows to study discrete generalizations of spacetime \cite{PhysRevLett.59.521}\cite{PhysRevD.87.064022}, since we make no further assumptions on the properties of $\TT$. Here, the most notable representative of the latter group are \emph{causal sets} (causets), which are locally finite in addition to being posets. In such sets, we can find a notion of immediate neighbors:
\begin{definition}
    \label{def:cover}
    Let $\TT$ be a poset and $x, y \in \TT$. We say $y$ \emph{covers} $x$ if $x \prec y$ and there is no $z \in \TT$ such that $x \prec z \prec y$.
\end{definition}
We will use this definition to give discrete posets a graphical representation in \cref{fig:hasse}.

Finally, we point out that for spacetime manifolds, the notion of causal future of $a \in \TT$ given by $\{ b \in \TT : b \succeq a \}$ matches literature conventions from general relativity \cite{Penrose1972}.

For the remainder of this section, we will study additional properties posets may show as well as their relation to the properties of relevant physical examples for $\TT$.

\begin{definition}[Join]
    \label{def:join}
    Let $\TT$ be a poset. For two elements $x, y \in \TT$, their \emph{join} $x \lor y$ is an element of $\TT$ such that:
    \begin{itemize}
        \item $x \preceq x \lor y$ and $y \preceq x \lor y$
        \item if $x \preceq a$ and $y \preceq a$, then $x \lor y \preceq a \ \forall a \in \TT$
    \end{itemize}
    or equivalently: $\forall a \in \TT, \ x \lor y \preceq a \iff x \preceq a \text{ and } y \preceq a$.
\end{definition}

\begin{definition}[Meet]
    Let $\TT$ be a poset. For two elements $x, y \in \TT$, their \emph{meet} $x \land y$ is an element of $\TT$ such that:
    \begin{itemize}
        \item $x \succeq x \land y$ and $y \succeq x \land y$
        \item if $x \succeq a$ and $y \succeq a$, then $x \land y \succeq a \ \forall a \in \TT$
    \end{itemize}
    or equivalently: $\forall a \in \TT, \ x \land y \succeq a \iff x \succeq a \text{ and } y \succeq a$.
\end{definition}

Given two points in a poset, the existence of join and meet is not guaranteed. However if they exist, they are respectively unique. Both fulfill the following properties, where $\odot$ stands either for $\land$ or $\lor$:
\begin{itemize}
    \item Associativity: $(a \odot b) \odot c = a \odot (b \odot c)$
    \item Commutativity: $a \odot b = b \odot a$
    \item Idempotency: $a \odot a = a$
\end{itemize}

Generalizing the concept of the join, we will define the notion of the minimal elements of a poset. Analogously, we could also define the maximal elements for the meet.
This, however, will not be required going forward, as we will focus on the causal future to model the availability of information.

\begin{definition}[Minimal Elements of a Poset]
    \label{def:min}
    Let $M \subseteq \TT$. Then
    $\min M := \{ x \in M \,|\, \not\exists \, y \in M : y \prec x \}$.
\end{definition}

\begin{lemma}
    \label{thm:join-min}
    Let $\TT$ be a poset.
    If and only if two elements $x, y \in \TT$ have a join $x \lor y$, then $\min \{ a \in \TT | x \preceq a \succeq y \} = \{ x \lor y \}$.
\end{lemma}

\begin{definition}[Semilattice]
    \label{def:semilattice}
    A poset $\TT$ is a \emph{join-semilattice} if $\forall x, y \in \TT$ the join $x \lor y$ exists. Dually, $\TT$ is a \emph{meet-semilattice} if $\forall x, y \in \TT$ the meet $x \land y$ exists.
\end{definition}

\begin{definition}[Join- and Meet-free Poset]
    \label{def:join-free}
    Let $\TT$ be a poset.
    We call $\TT$ a \emph{join-free poset} if for all $x, y \in \TT$, the existence of a join $x \lor y$ implies $x \lor y = x$ or $x \lor y = y$.
    Dually, $\TT$ is a \emph{meet-free poset} if no meet $x \land y$ exists for analogous choices of $x$ and $y$.
\end{definition}

Hence, a join-free poset can be considered the \enquote{opposite} of a join-semilattice, since in the former, for a pair of points $x, y$, joins exist only for the trivial case of $x \prec y$ or $x \succ y$.
All other posets are in-between these two extremes: Some non-trivial combinations of two points have a join there, while others do not.

\begin{definition}[Lattice]
    \label{def:lattice}
    A \emph{lattice} is a poset $\TT$ which is both a join- and a meet-semilattice.
\end{definition}

An example of a lattice, in contrast to a poset which forms no lattice, is depicted in \cref{fig:hasse}.

The theory of (order) lattices constitutes a well-studied subbranch of mathematics, which can be studied both from an order-theoretic (as done here) and from an algebraic perspective. More precisely, one can show that instead of demanding both semilattices to originate from the same partial order, one can equivalently demand join and meet to satisfy the absorption law
\begin{equation}
    a \lor (a \land b) = a \land (a \lor b) = a \quad \forall a,b \in \TT \, .
\end{equation}
Accordingly, a huge amount of further properties a lattice may fulfill are known \cite{Davey2002}. However, most of these have not proven directly relevant to this work and will therefore not be reviewed here.

\begin{proposition}[1+1-Minkowski Spacetime]
    \label{thm:1-plus-1}
    The \emph{light cone structure of 1+1-Minkowski spacetime}, having one spatial and one temporal dimension, is given by a lattice of its points. The meet of two points is given by the latest point where their past light cones intersect, while their join is given by the earliest point where their future light cones intersect \cite{Mattern2002}.
\end{proposition}

\begin{proposition}[Higher-dim. Minkowski Spacetime]
    \label{thm:no-jsl}
    In the case of $n$ spatial dimensions, we speak of $n$+1-Minkowski spacetime.
    For $n \geq 2$, Minkowski spacetime (as depicted in \cref{fig:minkowski}) is a join- and meet-free poset.
    \begin{proof}[Proof sketch]
        This is due to the fact that there are no unique minimal and maximal points, respectively. Specifically, the intersection of the future of two light cones is larger than any light cone that can be found within the intersection \cite{Mattern2002}.
    \end{proof}
\end{proposition}

Nonetheless, causally closed subsets of Minkowski spacetime actually assemble a so-called \emph{complete orthomodular lattice} for any number of dimensions. The respective sets are sometimes refered to as causal diamonds. Remarkably, the same mathematical structure is also exhibited by the set of projectors in Hilbert space, as used in quantum mechanics \cite{Casini2002}\cite[p.~109--115]{Minkowski2010}.
However, we are not aware of any deeper results which have originated from this observation.

We will return to more complex properties of higher-dimensional Minkowski spacetime in \cref{sec:minkowski-prop}.

\subsection{Ordered Random Variables (ORVs)} \label{sec:orvs}

To gain a link between the set of RVs $S$, which are part of a causal model as studied in \cref{sec:causality-info}, and a spacetime given by a poset $\mathcal{T}$, in which the respective physical experiments are ultimately performed, we introduce the concept of an embedding. Each random variable will be embedded into a single location in spacetime. Moreover, each random variable will be assigned a region of spacetime $\RC$ where its information is accessible.

This section summarizes the ideas introduced in sections V.A and V.B of \cite{VVC}. However, we will expand to look into the induced properties of sets of ORVs explicitly.

\begin{definition}[Ordering]
    An \emph{Ordering} $O$ of a set of random variables (RVs) $S$ on a poset $\TT$ is given by a map
    \begin{align}
        O : S &\to \TT \\
            X &\mapsto O(X) \, .
    \end{align}
\end{definition}

\begin{definition}[Ordered Random Variable (ORV)]
    \label{def:orv}
    Given an ordering $O$, an \emph{ORV} is defined as the pair $\XX := (X, O(X))$, where $X$ is a random variable (RV) and $O(X) \in \TT$ its assigned location.
    Given a set of RVs $S$, the associated set of ORVs is denoted by $\SC$.
\end{definition}

\begin{remark}
    In total, there are three different preorders relevant for ORVs: In addition to the partial order on the points of spacetime given by $\TT$, these are the preorder given by the set of 0th-order affects relations $\mathscr{A}$, as studied in \cref{sec:graph}, and the causal structure $\mathcal{G}$ itself, which also forms a preorder.
    While these preorders will often agree with each other, it is all the more important to look for cases where they disagree.
    Notice that a preorder corresponds to a directed graph, while a strict partial order can be represented by a DAG. Therefore, if the causal structure is acyclic, it is given by a strict partial order.
\end{remark}

\begin{notation}
    We will carry over the notion of a partial order from $\TT$ to $\SC$. Hence, $O(X) \prec O(Y) \to \XX \prec \YY$, however with $\XX = \YY$ if and only if $X = Y$, which, as $X$ and $Y$ refer to the same node, implies $O(X) = O(Y)$.
    Furthermore, we will denote $O(\XX) := O(X)$.
\end{notation}

\begin{remark}
    The partial order of locations of ORVs forms a subset of the poset $\TT$. However, it does not necessarily carry over all additional structure from $\TT$. For instance, since there does not need to be a RV embedded at $O(X) \, \lor \, O(Y)$ for arbitrary $X, Y \in S$, it does not carry over the structure of a join-semilattice.
\end{remark}

\begin{definition}[Future]
    \label{def:future}
    The \emph{inclusive} and \emph{exclusive future} of an ORV $\XX$ are defined as the sets
    \begin{align}
        \Fut (\XX) &:= \{ a \in \TT : a \succeq O(\XX) \} \, , \\
        \mathcal{F} (\XX) &:= \{ a \in \TT : a \succ O(\XX) \}
    \end{align}
    respectively.
\end{definition}

Returning to the example of $\TT$ being a spacetime manifold in addition to being a poset, beware that both the exclusive and the inclusive future include the lightlike surface of the light cone. Therefore, this concept is unrelated to topological closed- and openness.

\begin{definition}[Copy of a RV]
    \label{def:copy}
    In a causal model over a set of observed variables $S$, $X' \in S$ is a \emph{copy of} $X \in S$ if $\parent(X') = \{ X \}$ and $X' \cong X$.
\end{definition}

It is convenient to imagine an original causal model which does contain $X$, but not $X'$, which is then augmented with an additional node $X'$ which is a copy of $X$.
This serves as a potential model for transport of the information encoded in a certain RV over spacetime, since $X'$ may be embedded into a different location $O(X') \neq O(X)$. However, in practice, we will keep this construct implicit by referring to an accessible region of a RV $X$ instead.

\begin{definition}[Accessible Region of a RV]
    The \emph{accessible region} of a RV $X \in S$ is defined as the subset $\RC_X \subseteq \TT$ where it is possible to have a copy $X'$ of a random variable $X$.
    By contrast, the \emph{inaccessible region} of $X$ is given by $\RC_X^\perp = \TT \setminus \RC_X$.

    Let $\mathcal{P} (\TT)$ denote the power set of $\TT$. We then define a map $R$, mapping each RV to this accessible region:
    \begin{align}
        R : S &\to \mathcal{P} (\TT) \\
            X &\mapsto \RC_X \, .
    \end{align}
\end{definition}

Notably, this concept does not yet specify any particular mapping $R$ that shall be used to determine the accessible region for each RV. In \cref{sec:compat}, we will follow up on this to specify a physically reasonable condition for $R$. In particular, this concept does not take the structure of the spacetime $\TT$ into account and is therefore a priori independent of the notion of ORVs or the concept of future introduced in \cref{def:future}.
It strictly reflects the information-theoretic aspects of accessibility.

Now, we will widen the concepts introduced so far to encompass the causal model as a whole.

\begin{definition}[Embedding]
    \label{def:embedding}

    An ordering $O$ and a mapping $R: X \mapsto \RC_X$ together canonically induce an \emph{embedding} $\mathcal{E}$ of a set of RVs $S$ in $\TT$, yielding a set of ORVs $\SC = (S, O(S))$.
    \begin{align}
        \varepsilon&: S \to \SC \\
        \varepsilon&: X \mapsto \XX = (X,O(X)) \\
        \mathcal{E}&: S \mapsto (\SC, R(S)) :=  \{ (X,O(X),\RC_X) | X \in S \}
    \end{align}
    Here, $\mathcal{E}$ refers to the embedding in its entirety, while $\varepsilon$ represents only the part assigning a location to each RV.
    If $O$ is injective, its embedding is considered \emph{non-degenerate} ($O(X) \neq O(Y)$ for all distinct $X,Y \in S$).
    If this condition at least holds for all $X,Y$ with $X \affects Y \in \mathscr{A}$, the embedding is called \emph{non-trivial}.
    Whenever we augment the original causal model with additional ORVs by adding copies or intervention nodes, we will not include these nodes for discussing degeneracy and triviality for sake of simplicity.
\end{definition}

\begin{mdframed}
    \centering
    From now on, let the (sets of) ORVs $\AA, \BB, \CC, \XX, \YY, \ZZ, \SC_i, e_\XX$ etc.\ always be associated to the (sets of) RVs $A, B, C, X, Y, Z, S_i, e_X$ etc.\ with respect to some embedding $\mathcal{E}$, and vice versa.
\end{mdframed}

Having assigned a location to each RV in $S$, we will expand on the concepts explicitly introduced in \cite{VVC}, by generalizing from individual ORVs to sets of ORVs.

\begin{definition}[Accessible Region of Sets of RVs]
    \label{def:corv}
    The accessible region of a set of RVs $X$ on a poset $\TT$ is given by
    \begin{equation}
        \RC_X := \bigcap_{X_i \in X} \RC_{X_i} \, .
    \end{equation}
    In particular, this may be empty.
\end{definition}

\begin{definition}[Support Future]
    \label{def:supp-future}
    Let $\XX$ be a subset of a set of ORVs $\SC$. We define
    \begin{equation}
        \Fut_s (\XX) := \bigcap_{\XX_i \in \XX} \Fut(\XX_i)
    \end{equation}
    and call it the \emph{support future}.
\end{definition}

Having these definitions in place, it is natural to wonder whether it might be possible to generalize the basic concept of location to sets of random variables as well.
For this purpose, we will use the notion of minimal elements of a poset introduced in \cref{def:min}, which trivially satisfies
\begin{equation}
    \{ O(X) \} = \min \Fut_s ((X, O(X))) \ \forall X \in S \, .
\end{equation}
While it generally does no longer correspond to a single point, one can consider the set $\min \Fut_s (\XX)$ as the natural generalization of location to a set of ORVs $\XX \subset \SC$. This will be underpinned by the following Lemma.

\begin{lemma}
    \label{thm:supp-min}
    Let $\XX \subset \SC$, $A \subset \TT$ and $M = \{ y \in \TT \, | \, \exists x \in A \, \colon \, y \succeq x \}$.
    Then the following statements hold:
    \begin{alignat}{3}
        \min M \subseteq A& \subseteq M \, , \\
        \min \Fut_s (\XX) \subseteq A &\implies&& \Fut_s (\XX) \subseteq M \, , \\
        \min \Fut_s (\XX) = A &\implies&& \Fut_s (\XX) = M \, , \\
        \min \Fut_s (\XX) \subseteq A \subseteq \Fut_s (\XX)
        &\Longleftarrow&& \Fut_s (\XX) = M \, .
    \end{alignat}
    \begin{proof}
        The first statement follows from construction of $M$, since it is precisely built from $A$ and all elements in $\TT$ which are larger than at least one element of $A$.

        By definition, $y \in \Fut_s (\XX)$ implies that either
        $y \in \min \Fut_s (\XX)$ or $\exists x \in \min \Fut_s (\XX) : y \succ x$.
        As $M$ contains all elements which are in the future of at least one element of $A$, in particular $\min \Fut_s (\XX)$ being contained in $A$ precisely leads to the presence of $\Fut_s (\XX)$ in $M$, yielding the second and third claim.

        To prove the last claim, notice that by the first claim $A \subset M$, which immediately gives $A \subseteq \Fut_s (\XX)$.

        The relation $\min \Fut_s (\XX) \subseteq A$ can be shown by contradiction.
        If $A \not\ni y \in \min \Fut_s (\XX)$, by definition of the $\min$ this implies both that $y \in \Fut_s (\XX)$
        and that there is no $x \prec y$ in $\Fut_s (\XX)$.
        However, by assumption $M = \Fut_s (\XX)$ and therefore
        $\exists x \in A \, \colon \, y \succeq x$.
        Together, this gives $y = x \in A$, which poses a contradiction. $\lightning$
    \end{proof}
\end{lemma}

\begin{lemma}
    \label{thm:jsl}
    Let $\SC$ be a set of ORVs induced by an ordering $O$. Then, if $\TT$ is a join-semilattice,
    $\min \Fut_s (\XX) = \left\{ \bigvee_{\XX_i \in \XX} O(\XX_i) \right\} \ \forall \XX \subset \SC$.
    Hence, in this case there is a natural extension of $O$ to sets of RVs which gives
    $O(\XX) := \bigvee_{\XX_i \in \XX} O(\XX_i)$ and $\Fut (\XX) := \Fut_s (\XX)$.
    \begin{proof}
        Follows directly from \cref{def:min} applied to the locations of sets of ORVs.
    \end{proof}
\end{lemma}

\begin{figure}[t]
	\centering
	\includegraphics[width=0.65\textwidth]{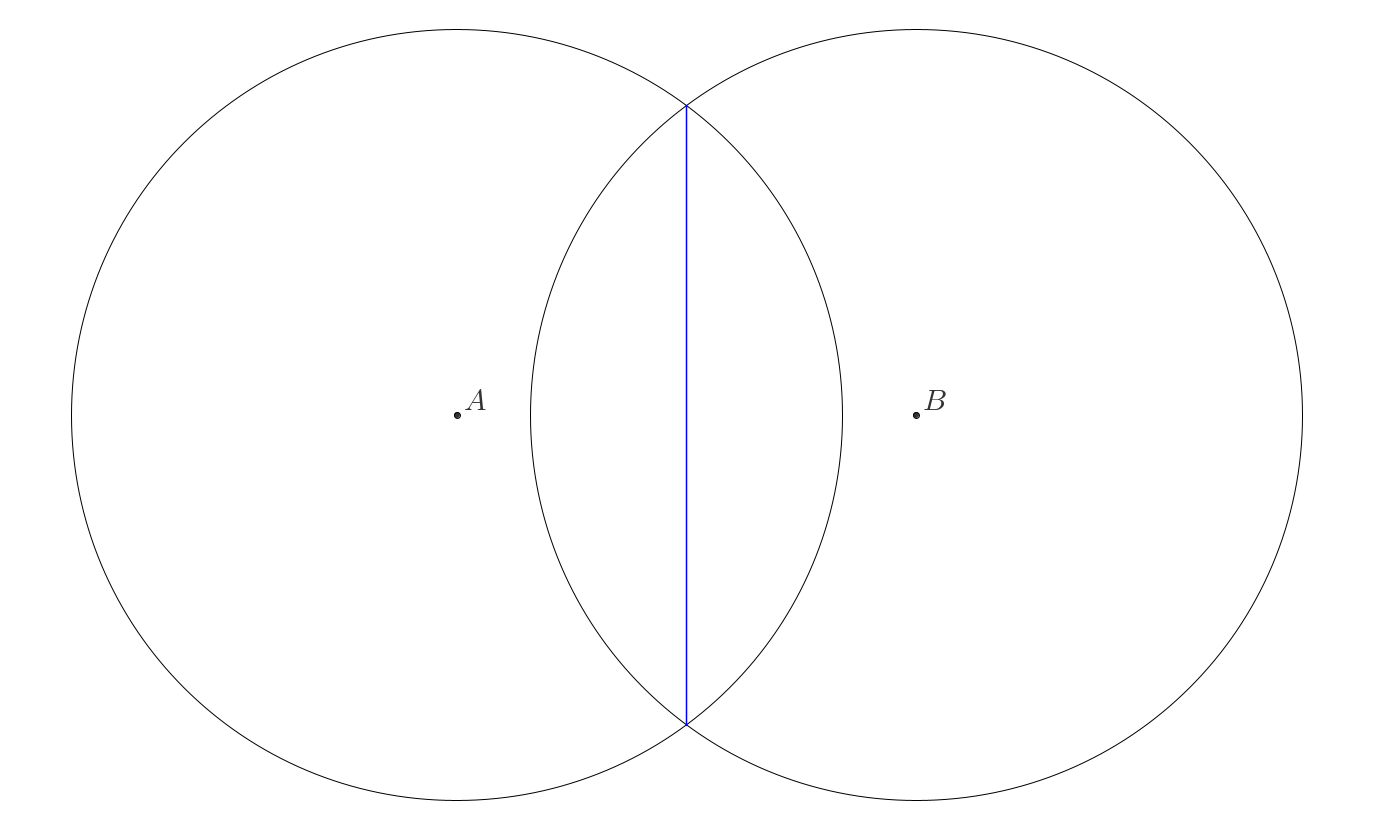}
	\caption[Time slice of the future light cones of $A$ and $B$ in 2+1-Minkowski spacetime.]{
        Time slice of the future light cones of $A$ and $B$ in 2+1-Minkowski spacetime.
        In particular along the perpendicular bisector, which constitutes the minimum of the joint future in top view, $\Fut_s (\AA\BB)$ grows faster than the future light cone from any point on the minimum surface.
        If only two space-like separated points are involved, we can always find an inertial system where both points share the same time coordinate, as depicted here.
        For more points, this is no longer the case. The farther the time-slice is in the future, the more similar to the future of a single ORV the intersection becomes in shape. The same holds the more additional points are added nearby, since that leads to smoother points of intersection.}
	\label{fig:joint-future}
\end{figure}

By contrast, if $\TT$ is no join-semilattice, we may have $\XX \subset \SC$ such that $\abs{\min \Fut_s (\XX) } > 1$. In this case, there is no single earliest point where the associated information becomes accessible. Physically, this corresponds to the fact that this information is seemingly transmitted faster than the speed of light.
However, since this information is not actually originating at $\min \Fut_s (\XX)$, but at the original locations $O(\XX_i)$, this is not problematic from a perspective of relativistic causality.
Additionally, if there is any $\YY \in \SC$ with $\YY \preceq \XX_i \ \forall \XX_i \in \XX$, $\Fut_s (\XX) \subseteq \Fut (\YY)$ still holds.

\subsection{Compatibility of Embeddings} \label{sec:compat}

In this section, we will postulate compatibility conditions for the information-theoretic concept of signalling as encoded into the affects relations with relativistic causality, as given by the spacetime poset $\TT$. While demanding all causation to go into the future seems to be a natural assumption, it is unnecessarily strict: Generally, special relativity does only forbid superluminal \textit{signalling}, as this would allow for transmitting information, encoded into RVs, into the past. Further, as discussed in \cref{sec:causal-structures}, there are correlations in nature which seem to originate in space-like separation to each other, further strengthening this point of view.

Now, we will introduce and justify the compatibility condition as given in Section V.C of \cite{VVC}, although slightly renamed. This serves as distinction to a stronger compatibility condition which we will derive and study in \cref{sec:compat-indec}. Afterwards, we will study its general implications as well as its properties under transformations in detail.

\begin{definition}[\textbf{compat-irreducible}]
    \label{def:compat}
    Let $\SC$ be a set of ORVs from a set of RVs $S$ and a poset $\TT$ with an embedding $\mathcal{E}$. Then a set of affects relations $\mathscr{A}$ is said to be \emph{compatible} with $\mathcal{E}$ if the following conditions hold:
    \begin{itemize}
        \item \textbf{compat1-irreducible}: Let $X, Y \subset S$ be disjoint non-empty sets of RVs, $W, Z \subset S$ two more disjoint sets of RVs, potentially empty.
        If ($X \affects Y \given \{ \doo(Z), W \}$) $\in \mathscr{A}$ and is irreducible, then $\RC_{YWZ} = \RC_Y \cap \RC_W \cap \RC_Z \subseteq \RC_X$.
        \item \textbf{compat2}: With respect to $\mathcal{E}$,
        $\RC_X = \Fut(\XX) \ \forall \XX \in \SC$.
    \end{itemize}
    We will also say that \textbf{compat-irreducible} is satisfied in this case.
\end{definition}

With this compatibility condition we link the concept of accessible regions of random variables with the light cone structure of spacetime for the first time.

With the first condition, we recognize generic affects relations as a comprehensive, mathematically rigid model for information-theoretic signalling between agents, as has been discussed in \cref{sec:affects}.
This can be seen as a generalization of the fact that each $e_X \in X$ affects each of its copies $e_X'$, which are required to be located in the accessible region:
We ensure that general signals encoded in $YZW$, originating from $X$, can only be sent to locations in $\TT$ where $X' = \bigcup_{e_X \in X} e_X'$ could also be obtained, hence, precisely that $R_{YZW} \subseteq \RC_X$.
This further ensures that signalling is directed: Any location, where all information required to detect an affects relation is accessible, needs to be a location where the original information the signal is derived from may also be accessed.

The second condition establishes the (possibility of) broadcasting of information given by a classical RV, such that it is accessible in the full relativistic future.
In principle, this possibility may be restricted by only requiring $\RC_X \subseteq \Fut(\XX) \ \forall \XX \in \SC$, yielding a broader notion of compatibility, which implies $\RC_Y \subset \Fut_s (\XX)$ for $X \affects Y$. However, already presuming that $O(Y) \in \RC_Y$ recovers $\Fut (\YY) \subset \Fut_s (\XX)$.
Another alternative approach is considered in \cite[Sec.~V.D]{VVC}, by additionally formulating an alternative to \textbf{compat1} in terms of the relativistic futures $\Fut_s$. Here, by modifying the embedding, an equivalency to the original formulation is shown.
Hence, we do not consider such variations further.

\begin{lemma}
    \label{thm:compat2-rephr}
    \hyperref[def:compat]{\textbf{compat2}} $\iff \RC_X = \Fut_s (\XX)
    \ \forall \XX \subset \SC$
    \begin{proof}
        \hin By \cref{def:corv}, any set of ORVs $\XX$ fulfills
        $\RC_X = \bigcap_{X_i \in X} \RC_{X_i}$.\\
        By \hyperref[def:compat]{\textbf{compat2}}, this implies
        $\RC_X = \Fut_s (\XX)$.

        \rueck $\RC_X = \Fut_s (\XX)
        \ \forall \XX \subset \SC$ for the case that $\XX \in \SC$ is precisely \hyperref[def:compat]{\textbf{compat2}},
        as $\Fut$ and $\Fut_s$ are equivalent for individual ORVs.
    \end{proof}
\end{lemma}

This delivers the expected result that \hyperref[def:compat]{\textbf{compat2}} can be generalized to sets of random variables, when replacing the relativistic future of an individual RV with the support future.
Hence, it is possible to obtain a copy of a composite ORV in the joint causal future of the locations of the random variables in its support.

\begin{remark}
    \label{thm:emptyset}
    Let $\AA$ be a set of ORVs. If $\Fut_s (\AA) = \emptyset$, there is no location in $\TT$ which is in the future of all $O(\AA_i)$.
    With \cref{thm:compat2-rephr}, this means that $\RC_A = \emptyset$.

    For $X,Y,Z \in S$, consider an affects relation $X \affects Y \given \{ \doo(Z), W \}$.
    If $\RC_{YZW} = \emptyset$, there is no agent who can verify the respective affects relation, rendering it operationally meaningless.
    If $\RC_X = \emptyset$, compatibility implies that $\RC_{YZW} = \emptyset$.
    Therefore, we will usually disregard the case where either of these sets is empty.
\end{remark}

A physical example for this would be given by two ORVs $\mathcal{A}_1$ and $\mathcal{A}_2$ located in two distinct classical black holes or outside their respective cosmological event horizons, as these would have no joint future. Therefore, they can be signalling to each other arbitrarily without breaking compatibility.

To gain a better understanding of the compatibility conditions and aim at their application, we observe how under certain transformations of the set of affects relations $\mathscr{A}$, both compatibility and causal inference are impacted.

\begin{notation}[Affects Relations]
    \label{not:affects-tuple}
    When referring to $(X \affects Y \given \{ \doo(Z), W \}) \in \mathscr{A}$, for the remainder of this section, we will use $(X, Y, Z, W)$ as a short-hand.
\end{notation}

\begin{lemma}[Conditionality Transformation {\cite[Rem.~V.2]{VVC}}]
    \label{thm:conditional-equiv}
    Let $\mathscr{A}$ be a set of 0th-order affects relations with an embedding $\mathcal{E}$, and $X, Y, Z, W$ as in \hyperref[def:compat]{\textbf{compat-irreducible}}.
    Then $Z = \emptyset$ for all affects relations in $\mathscr{A}$ and the transformation
    \begin{equation}
        (X, Y, \emptyset, W) \mapsto (X, YW, \emptyset, \emptyset) \quad \forall (X, Y, \emptyset, W) \in \mathscr{A} \, .
    \end{equation}
    to a set $\mathscr{A}'$ of unconditional 0th-order affects relations,
    \begin{enumerate}
        \item preserves (irreducible) compatibility
        \item preserves inferable information about the causal structure.
    \end{enumerate}
    \begin{proof}
        We consider the transformation $(X, Y, \emptyset, W) \to (X, YW, \emptyset, \emptyset)$ acting on all elements of $\mathscr{A}$.
        Then for any element of $\mathscr{A}$, the first statement follows directly from \hyperref[def:compat]{\textbf{compat1-irreducible}} being invariant under this transformation.
        Explicitly, $X \affects Y \given W$ implies $\RC_Y \cap \RC_W \subseteq \RC_X$, and is transformed to $X \affects YW$, which implies $\RC_{YW} \subseteq \RC_X$.
        By \cref{def:corv}, these subset relations are equivalent. See also \cite[Rem.~V.2]{VVC}.

        For the second part, the causal relations implied by \cref{thm:affects-to-cause} are invariant, while \cref{thm:affects-completeness} gives their completeness.
    \end{proof}
\end{lemma}

This shows that w.l.o.g., we can focus on unconditional affects relations to study the interplay of causal structures with spacetime compatibility.
Unfortunately, no simple analogue does exist for higher-order affects relations. We have the following candidate transformations to be performed on all affects relations on $\mathscr{A}$, all raising significant issues:
\begin{enumerate}
    \item $(X, Y, Z, W) \mapsto (X, ZYW, \emptyset, \emptyset)$ preserves compatibility, but reduces causal information,
    \item $(X, Y, Z, W) \mapsto (X, YW, \emptyset, \emptyset)$ preserves causal information, but tightens compatibility,
    \item $(X, Y, Z, W) \mapsto \{ (X, YW, \emptyset, \emptyset), (X, ZYW, \emptyset, \emptyset) \}$ preserves causal information, but tightens compatibility.
\end{enumerate}
For approaches splitting $Z \subset S$ into subsets or individual elements, similar issues will occur, since removing $Z$ even partially tightens compatibility, while moving subsets of it to the second argument reduces causal information.

This shows that any proof forbidding embeddability of causal structures associated with certain 0th-order affects relations does not necessarily generalize to higher-order affects relations. For the instance of causal loops detectable via affects relations, while reducing causal information may lead to causal loops to be removed, tightening compatibility may lead to them being no longer compatible.

Therefore, we will focus on the 0th-order case for now and will return to the HO-case for a more elaborate treatment in \cref{sec:compat-indec}.
Furthermore, by \cref{thm:causal-discovery-simple}, if all nodes are observed and the set of 0th-order affects relations $\mathscr{A}$ is complete, this is sufficient for full causal discovery.

\begin{remark}
    Consider a causal model without unobserved nodes, as well as the equivalency between observed affects relations and causal relations given by \cref{thm:cause-from-affects}.
    Let $\mathcal{E}$ be an embedding of a set of RVs $S$ compatible with a \emph{complete} set of affects relations $\mathscr{A}_\text{full}$ (cf. \cref{def:affects-complete}), and let $X, Y \in S$ with $X \neq Y$ be observed. Then we obtain that
    \begin{equation}
        X \dircause Y \implies \RC_Y \cap \RC_{\parent(Y) \setminus X} \subseteq \RC_X \, .
    \end{equation}
    Exemplarily, consider a causal structure given by $A \dircause C \diresuac B$. Then, we get
    \begin{align}
        A \dircause C &\implies \RC_C \cap \RC_B \subseteq \RC_A \, , \\
        B \dircause C &\implies \RC_C \cap \RC_A \subseteq \RC_B \, , \\
        A \dircause C \ \land \ B \dircause C & \implies \RC_C \cap \RC_B = \RC_C \cap \RC_A \, .
    \end{align}
    Hence $C$ may only be accessible where either both or none of $A$ and $B$ are accessible.

    For the example of higher-dimensional Minkowski spacetime, we will consider in \cref{thm:location-symmetry} that there are no embeddings where both conditions may occur in parallel.
    This means we must choose between either
    $\Fut(\CC) \subseteq \Fut_s (\AA \BB)$ or $\Fut_s (\AA \BB) = \Fut (\BB) = \Fut (\AA)$, with $\Fut (\CC)$ arbitrary for the latter case.

    As 1+1-Minkowski spacetime is a join-semilattice, we additionally have the option there that $\Fut (\AA) = \Fut (\AA \BB)$, corresponding to $\Fut(\AA) \subsetneq \Fut(\BB)$, as well as the symmetric option obtained by permuting $\AA$ and $\BB$.

    Hence, already without explicitly knowing the probability distribution of a set of RVs $S$ or a set of affects relations $\mathscr{A}$ over this set, the knowledge of an embedding that is compatible is sufficient to gain some information about the causal structure.
\end{remark}

\subsection{Stability of Embeddings} \label{sec:stable-aff}

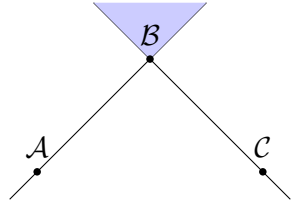
\begin{figure}[t]
	\centering
    \begin{tikzpicture}[dot/.style={circle,inner sep=1pt,fill,name=#1},
                        extended line/.style={shorten >=-#1,shorten <=-#1},
                        extended line/.default=1cm]

        \node [dot=A,label=$\AA$] at (0,0) {};
        \node [dot=C,label=$\CC$] at (3,0) {};

        \node (left) at (-0.5,-0.5) {};
        \node (right) at (3.5,-0.5) {};
        \draw (left) -- ++(2.75,2.75);
        \draw (right) -- ++(-2.75,2.75);
        \fill[fill=blue!20] (1.5,1.5) -- (2.25,2.25) -- (0.75,2.25) -- (1.5,1.5);

        \node [dot=B,label=$\BB$] at (1.5,1.5) {};



    \end{tikzpicture}
	\caption[A compatible embedding of $\mathscr{A} = \{ B \affects AC, AC \affects B \}$ into 1+1- Minkowski spacetime.]{
        Sketch of a compatible, non-degenerate embedding of the ACL $\mathscr{A} = \{ B \affects AC, AC \affects B \}$ into 1+1-Minkowski spacetime.
        Space and time are given along the horizontal and vertical axis, respectively.
        The black lines correspond to light-like surfaces, while the blue region is $\Fut_s (\AA\CC) = \Fut (\BB)$, where detection of the causal loop is principally possible.}
	\label{fig:measure-zero-embedding}
\end{figure}

When applying the concept of compatibility to causal loops, we observe that already for simple examples, compatibility is possible, as has been pointed out by \cite{VVC_Letter}.
For example, we can consider $\mathscr{A} = \{ B \affects AC, AC \affects B \}$, containing an ACL5, which admits a non-degenerate compatible embedding for $\Fut (\BB) = \Fut (\AA\CC)$.
This condition can actually be realized in 1+1-Minkowski spacetime, as depicted in \cref{fig:measure-zero-embedding}, but is enforcing $A$ and $C$ to be lightlike to $B$, thereby posing quite a strong restriction.

In light of this example, in \cite[p.~32f.]{VVC} the authors introduce the notion of \textit{measure-zero type embeddings}, as the lightlike surfaces in $n$+1-Minkowski spacetime are $n$-dimensional and therefore of measure zero. On the contrary, they describe all other embeddings as \textit{stable}, and wonder if such embeddings are compatible with the presence of causal loops. We will return to this question in \cref{sec:general-loops}.

However, this argument does not carry over to general posets $\TT$, which do not admit a unique notion of a measure. Therefore, we choose an adjusted definition which does not refer to any additional structure.
Hence, we choose the subset relations given by the respective compatibility conditions to be strict, which likewise rules out the example above.
Afterwards, we will discuss two additional notions of stability and briefly relate to the notions of (non-)trivial embeddings, as reviewed in \cref{def:embedding}.
While both \mbox{(non-)triviality} and (non-)degenerateness carry a similar intuition, they are hard to compare since they operate on a level of individual ORVs instead of full sets.
Therefore, they are only marginally addressed here.

First, we introduce the concept of support stability, which we will predominately use over the remainder of this work.

\begin{definition}[(Support) Stability]
    \label{def:stable-supp}
    An embedding $\mathcal{E}$ is considered \emph{(support-)stable}, if for all $\XX, \YY, \ZZ, \WW \subset \SC$ disjoint
    with $X \affects Y \given \{ \doo(Z), W \}$:
    \begin{align}
        \Fut_s (\YY) \cap \Fut_s (\ZZ) \cap \Fut_s (\WW) &\subsetneq \Fut_s (\XX) \, , \\
    \intertext{explicitly excluding equality. Equivalently, we can state}
        \Fut_s (\YY \ZZ \WW) &\subsetneq \Fut_s (\XX) \, .
    \end{align}
\end{definition}

This introduces a strict partial time-ordering for sets of ORVs, induced from the subset relations of the accessible regions $\RC$.
In particular, this notion eliminates affects relations where the RVs on both sides are only accessible in empty sets, as discussed in \cref{thm:emptyset}.
Due to the closeness of this definition to the definition of compatibility itself, we can immediately state a lemma for 0th-order affects relations.

\begin{lemma}
    \label{thm:conditional-equiv-stable}
    The conditionality transformation of \cref{thm:conditional-equiv} preserves support stability.
    \begin{proof}
        This proof is analogous to the proof of \cref{thm:conditional-equiv} preserving compatibility, replacing general with strict subset relations.
    \end{proof}
\end{lemma}

Moreover, there is an alternative notion which is interesting from a physical perspective on information processing:

\begin{definition}[Minimum Stability]
    \label{def:stable-min}
    An embedding is considered \emph{minimum-stable}, if for all $\XX, \YY, \ZZ, \WW \subset \SC$ disjoint
    with $X \affects Y \given \{ \doo(\ZZ), \WW \}$:
    \begin{equation}
        \Fut_s (\YY \ZZ \WW) \subseteq \Fut_s (\XX) \setminus \min \Fut_s (\XX) \, .
    \end{equation}
    Here, for any $M \subseteq \TT$, $\min M \subset \TT$ is given as in \cref{def:min}.
\end{definition}

The intuition behind this definition becomes apparent by recalling $\min \Fut_s$, as discussed in \cref{thm:supp-min}, which gives the points of earliest future where a set of ORVs is available. Hence, it can be seen as the set of locations where the joint probability distribution is mended together.
Therefore, minimum stability captures the physical idea that non-trivially deriving some information $Y$ from some data given by the set of RVs $X$ can not happen instantaneously with respect to the partial order of $\TT$.

If not only the partial order, but further structure is present, further stronger notions of stability may become apparent.
For the example of pseudo-Riemannian spacetime of dimension $d$, a possible notion is that $\YY$ may not already be constructed on the $(d-1)$-dimensional surface of the light cones given by the elements of $\XX$.
However, it could be constructed $\varepsilon$-close to this surface (for some small $\varepsilon > 0$). In this example, demanding minimum stability transforms from topologically closed light cones to topologically open light cones (in spatial direction, respectively).
To arrive at this notion, an additional partial order, indicating chronology \cite{Penrose1972}, or topological properties may be used.
We will refer to this notion as \emph{chronological stability}, but not study it in more detail.

\begin{lemma}
    \label{thm:stable}
    Let $\XX \subseteq \SC$ and $\TT$ be a discrete poset or a spacetime manifold with closed light cones coinciding with the inclusive future $\Fut (\XX)$. Then minimum stability implies support stability.
    \begin{proof}
        Consider that w.l.o.g., $\Fut_s (\XX) \neq \emptyset$, since it could not affect anything otherwise due to \cref{thm:emptyset}.
        Then it remains to show that $\min \Fut_s (\XX) \neq \emptyset$. This is sufficient, as then minimum stability implies that
        \begin{equation}
            \Fut_s (\YY\ZZ\WW) \subseteq \Fut_s (\XX) \setminus \min \Fut_s (\mathcal{X}) \subsetneq \Fut_s (\XX) \, ,
        \end{equation}
        which again implies support stability.

        If $\TT$ is a discrete set, this holds since in this case minimal sets always exist.
        If $\TT$ is a spacetime with closed light cone structure, the conical surface of the light cone is included in $\Fut_s (\XX)$ and (topologically) closed (while open into the future).
        As intersections of closed sets are closed, values for $\min \Fut_s (\XX)$ are attained.
    \end{proof}
\end{lemma}

\begin{example}
    Let $\TT$ be a poset with $a, b \in \TT$ and let $\XX, \YY, \ZZ \in \SC$. Then the set $\mathscr{A} = \{ XY \affects Z \}$ embedded in $\TT$ such that
    \begin{align}
        a, b &\in \min \Fut_s (\XX\YY) \\
        \ZZ &\unord b \unord a \\
        a &= O (Z)
    \end{align}
    is support-stable, but not minimum-stable: $O(Z) \in \min \Fut_s (\XX\YY)$,
    while $b \not\in \Fut_s (\ZZ)$ and therefore $\Fut(\ZZ) \subsetneq \Fut_s (\XX\YY)$.
\end{example}

Finally, we will give a connection between stability and triviality, albeit a rather loose one.

\begin{lemma}
    \label{thm:stable-trivial}
    Every trivial embedding of a set of affects relations $\mathscr{A}$ satisfying \hyperref[def:compat]{\textbf{compat-irreducible}} is neither minimum-stable nor support-stable.
    \begin{proof}
        Due to triviality
        $\exists X, Y \in S : \ X \affects Y \in \mathscr{A}$ such that $O(X) = O(Y)$ and therefore $\Fut_s (\XX) = \Fut_s (\YY)$.
        This is in contradiction to support stability, and since $\min \Fut_s (\XX) = \min \Fut_s (\YY) \subset \Fut_s (\YY)$, also to minimum stability. $\lightning$
    \end{proof}
\end{lemma}

Hence, both notions of stability offer a suitable generalization of non-triviality, which acts only on individual RVs, for arbitrary affects relations.

A direct connection to degeneracy, however, can not be found.
As degenerateness is not concerned with the presence and absence of affects relations, stability does generally not exclude degenerateness.
Similarly, non-degenerateness does not imply stability, as we have seen for \cref{fig:measure-zero-embedding}.

\subsection{Special Properties of Minkowski and Similar Spacetimes} \label{sec:minkowski-prop}

In this brief section, we will quickly motivate and introduce special order-theoretic properties for Minkowski spacetime, as those can be used to gain additional insights from compatibility properties. While we have characterized 1+1-Minkowski spacetime as a lattice in \cref{thm:1-plus-1}, there are different properties in the higher-dimensional case. In particular, it is a join- and meet-free poset, as pointed out in \cref{thm:no-jsl}. In this section, we introduce three further properties for posets which we conjecture to be satisfied for higher-dimensional Minkowski spacetime.
Further research is needed to determine and characterize the full space of posets satisfying these properties.

\begin{definition}[Spanning Elements]
    \label{def:span}
    Let $\TT$ be a poset and $\XX \subset \SC$ a set of ORVs on this poset.
    Then, the set of \emph{spanning elements}, denoted by $\spann(\XX)$, is given by the union of all sets $s_\XX \subseteq \XX$ which satisfy
    \begin{equation}
        \Fut_s (s_\XX) = \Fut_s (\XX) \ \land \ \Fut_s (t_\XX) \neq \Fut_s(\XX) \ \forall t_\XX \subsetneq s_\XX \, .
    \end{equation}
\end{definition}

Hence it forms the union of all possible minimal subsets of $\XX$ which span the same light cone.
In particular, this is a subset of the \enquote{latest} elements of $\XX$ in the poset, given by
\begin{equation}
    \{ \XX_i \in \XX \,|\, \not\exists \XX_k \in \XX \ \colon \ \XX_k \succ \XX_i \} \, .
\end{equation}
Generally, and in particular, for $n$+1-Minkowski spacetime with $n \geq 2$, both sets are not identical, as can be seen by considering the joint future of three ORVs $\AA, \BB, \CC$ equidistantly embedded on a space-like line next to each other. These will be space-like separated, but still satisfy $\Fut_s (\AA\BB\CC) = \Fut_s (\AA\CC)$. If $\abs{\XX} \leq 2$ however, both sets are indeed the same.

\begin{definition}[Union Property]
    \label{def:union}
    Let $\TT$ be a poset and $\XX \subset \SC$ be a set of ORVs on this poset. Let $\YY \in \SC$ be another ORV with $\Fut(\YY) \neq \Fut(\XX_i) \ \forall \XX_i \in \XX$.
    Then $\TT$ satisfies the \emph{union property} if
    \begin{equation}
        \Fut(\YY) \supseteq \Fut_s (\XX) = \Fut_s (\spann(\XX)) \quad \implies \quad \neg \left( \Fut (\YY) \subsetneq \bigcup_{\XX_i \in \spann(\XX)} \Fut(\XX_i) \right) \, .
    \end{equation}
\end{definition}

Here, excluding equality on the right side is mostly a technicality, as equality could only occur in the degenerate case of $\Fut (\YY) = \Fut (\XX_i) \ \forall \XX_i \in \spann(\XX)$.

\begin{conjecture}
    \label{thm:union}
    For $n \geq 2$, $n$+1-Minkowski spacetime satisfies the union property.
    \begin{proof}[Proof sketch]
        We show this by contradiction.
        Assume there is $\Fut (\YY) \subsetneq \bigcup_{\XX_i \in \spann(\XX)} \Fut(\XX_i)$.
        This implies that $\exists \XX_i : \YY \succ \XX_i$ (as equality of locations is ruled out by assumption), since otherwise $O(Y)$ would not be contained within this union.
        We therefore get that everywhere the boundary of the light cone of $\YY$ will be inside the light cone of $\XX_i$.
        In particular, this implies that $\YY$ replaces the spherical sections of the boundary of the joint future of $\XX$ which have been contributed by $\XX_i$ when considering $\Fut(\XX\YY)$ instead of $\Fut(\XX)$.
        This yields $\Fut_s (\XX\YY) \subsetneq \Fut (\YY)$.
        However, $\Fut(\YY) \supseteq \Fut_s (\XX)$ implies that $\Fut_s (\XX) = \Fut_s (\XX\YY)$.
        This poses a contradiction and thereby proves the claim. $\lightning$
    \end{proof}
\end{conjecture}

\begin{definition}[Conicality]
    \label{def:conical}
    Let $\TT$ be a poset and $\XX \subset \SC$ a set of ORVs on this poset.
    Then $\TT$ is a \emph{conical} poset if the knowledge of $\Fut_s (\XX)$ implies the locations $O (\XX_i)$ for all its spanning elements $\XX_i \in \spann (\XX)$.
\end{definition}

As $\Fut_s (\XX)$ itself is completely determined by the locations of all ORVs $\XX_i \in \spann(\XX)$, this gives an equivalency between the knowledge of these locations and the knowledge of the joint future.

\begin{conjecture}
    \label{thm:conical}
    For $n \geq 2$, $n$+1-Minkowski spacetime is a conical poset.
    \begin{proof}[Proof sketch]
        Let $\XX = \bigcup_i \XX_i$ a set of ORVs.
        Considering a time-slice of a future light cone in $n$+1-Minkowski spacetime given by $\Fut (\XX_i)$, we observe that its radius allows to infer the position of the origin of the light cone, given by $O (\XX_i)$.
        As any non-empty time-slice of the intersection of light cones is compiled from spherical pieces of the boundaries of the respective light cones (circular arcs for 2-dim. spacetime), this argument generalizes to $\Fut_s (\XX)$.
        However, since by definition, precisely the $\XX_i \in \spann(\XX)$ contribute to this boundary, we can only infer the locations for these $\XX_i$.
    \end{proof}
\end{conjecture}

\begin{definition}[Location Symmetry]
    \label{def:location-symmetry}
    Let $\TT$ be a poset and $I$ be an index set. Then $\TT$ shows \emph{location symmetry} if for all $\XX, \YY_i \subset \SC$ with $i \in I$ it satisfies
    \begin{align}
        & \Fut_s (\XX \YY_i) = \Fut_s (\XX \YY_k) \quad \forall i, k \in I \\
        \ \implies \
        & \Fut_s (\XX) \subseteq \Fut_s \left( \textstyle{\bigcup_{i \in I}} \YY_i \right) \ \lor \
            \forall \YY_i \ \exists s_{\YY_i} \subseteq \YY_i \ \colon \
            \forall \YY_k \ \exists s_{\YY_k} \subseteq \YY_k \ \colon \
            \Fut_s (s_{\YY_i}) = \Fut_s (s_{\YY_k}) \, ,
    \end{align}
    where all subsets are presumed to be non-empty.
\end{definition}

This property is easiest to understand for the special case where $\abs{I} = 2$, with $\YY_1 = \YY$ and $\YY_2 = \emptyset$. Here, it reduces to
\begin{equation}
    \Fut_s (\XX\YY) = \Fut_s (\XX) \implies \Fut_s (\XX) \subseteq \Fut_s (\YY) \, .
\end{equation}
Generally, for $\abs{I} = 2$ we get
\begin{align}
    & \Fut_s (\XX\YY_1) = \Fut_s (\XX\YY_2) \\
    \ \implies \
    & \Fut_s (\XX) \subseteq \Fut_s (\YY_1 \YY_2) \ \lor \
        \exists s_{\YY_1} \subseteq \YY_1, \, s_{\YY_2} \subseteq \YY_2  \ \text{non-empty} \  \colon \ \Fut_s (s_{\YY_1}) = \Fut_s (s_{\YY_2}) \, .
\end{align}

\begin{conjecture}
    \label{thm:location-symmetry}
    For $n \geq 2$, $n$+1-Minkowski spacetime shows location symmetry.
    \begin{proof}[Proof sketch]
        Since it would require more complex geometric tools to derive this property formally, we only argue the conjecture informally.
        As argued in the proof sketch for \cref{thm:conical}, for each non-empty time-slice, the boundary of the joint future of a set of ORVs is given by spherical pieces of the boundaries of the light cones of some of the spanning ORVs in the set.
        We will now argue the implications of this for the case of $\abs{I} = 2$, but the argument generalizes to an arbitrary amount of sets involved.

        Hence, if for some $\XX, \YY_1, \YY_2 \subset \SC$, $\XX\YY_1$ and $\XX\YY_2$ share the same joint future, this can have multiple reasons.
        First, the joint future of the elements in $\YY_1$ or $\YY_2$ is later than the joint future of the elements in $\XX$, and therefore, $\Fut_s (\YY_1) \supseteq \Fut_s (\XX)$ and $\Fut_s (\YY_2) \supseteq \Fut_s (\XX)$. Conjoining these points gives precisely the first option that $\Fut_s (\XX) \subseteq \Fut_s (\YY_1\YY_2)$.

        Otherwise, we must have that $\YY_1$ and $\YY_2$ contribute the same spherical pieces to the boundary of $\XX \YY_1 \YY_2$, and hence, by the argument of \cref{thm:conical}, satisfy
        \begin{equation}
            \exists s_{\YY_1} \subseteq \spann(\YY_1), \, s_{\YY_2} \subseteq \spann(\YY_2) \, \colon \ \Fut_s (s_{\YY_1}) = \Fut_s (s_{\YY_2}) \, .
        \end{equation}
        This point in particular generalizes to a larger number of sets as all of which then must contribute the same spherical pieces to the boundary.
        As $\spann(\AA) \subseteq \AA \ \forall \AA \in \SC$, the statement follows immediately.
    \end{proof}
\end{conjecture}

From these proof sketches, it is strongly hinted that conicality implies location symmetry, and that there is a relation to join-/meet-free posets. However, as we are mostly interested in making statements for Minkowski spacetime, we will not study this point further here.

In summary, we have claimed four properties to hold for higher-dimensional Minkowski spacetime: Being a join- and meet-free poset, the union property, conicality, and location symmetry.
Finally, we will conjecture that these properties generalize to other spacetimes than Minkowski. Therefore, all results we show which require them would generalize to this class of spacetimes.

\begin{conjecture}
    Any $n$-dimensional spacetime manifold with time orientation and no CTCs, which is homotopic to Minkowski spacetime, is join- and meet-free, conical, and shows both the union property and location symmetry.
    \begin{proof}[Proof sketch]
        This follows from the fact that the consideration above only refers to qualitative features of the light cone structure of $n$+1-Minkowski spacetime, which is locally invariant for spacetime manifolds.
        However, it does not necessarily carry over if the spacetime has singularities or is differently not simply-connected, since light cones do change their fundamental geometry in such situations.
    \end{proof}
\end{conjecture}

Here, we refer to closed timelike curves as used general relativity, which we have ruled out by choosing a partial order model for spacetime in \cref{sec:poset}.
As we do not presume all information transfer within an information-theoretic causal structure to be directed to the future, this does not entirely rule out the possibility of cyclic information transfer.
However, as we do not associate the arrows in a causal structure with time, we will not refer to this scenario as featuring CTCs.

Further, we presume it plausible that this conjecture may require additional technical conditions on the spacetime, as for example given by global hyperbolicity \cite{Wald1984}.

\section{Causal Loops in Spacetime} \label{sec:loops}

\subsection{Generic Spacetimes} \label{sec:general-loops}
In this section, we will study how ACLs can be compatibly embedded into spacetime. In doing so, we will focus on the possibility of a stable embedding for different types of causal loops, as introduced in \cref{sec:stable-aff}, and remain mostly generic in the properties of spacetime.
While we rule out the existence of stable embeddings for a wide class of causal loops, we also show that for general posets $\TT$, it is impossible to rule out their existence entirely.

While all results will be phrased using unconditional 0th-order affects relations only, our results from \cref{sec:compat} show that they hold equally for conditional affects relations, as compatibility and causal inference change symmetrically under transformation. The case of higher-order affects relations will be covered individually in \cref{sec:compat-indec}.

As introduced in \cref{sec:orvs}, we will continue to always associate RVs $S_i$ with ORVs $\SC_i$, without pointing this out explicitly.

\begin{lemma}
    \label{thm:cac-embedding}
    Consider a complete affects chain $\mathscr{C}$ from $S_1$ to $S_n$, with $S_i, \hat S_i$ as in \cref{def:cac}.
    Then any embedding $\mathcal{E}$ into a poset $\TT$ fulfills $\Fut_s (\hat \SC_n) \subseteq \Fut_s (\SC_n) \subset \Fut_s (\SC_1)$.
    If the embedding is stable, the latter subset relation must be strict.
    \begin{proof}
        Let $S_i, \hat S_i$ be as in \cref{def:cac}. Then $S_i \subset \hat S_i$ and $\hat S_i \affects S_{i+1} \forall 1 \leq i < n$.
        Further, consider any embedding $\mathcal{E}$.
        Then, by \cref{def:corv} and \hyperref[def:compat]{\textbf{compat-irreducible}}, we get $\Fut_s (\hat \SC_i) \subseteq \Fut_s (\SC_i)$ and $\Fut_s (\SC_{i+1}) \subset \Fut_s (\hat \SC_i)$.
        By \cref{def:stable-supp}, the latter subset relation is strict for a stable embedding. Chaining those, we get
        $\Fut_s (\hat \SC_n) \subseteq \Fut_s (\SC_n) \subset \Fut_s (\SC_1)$.
    \end{proof}
\end{lemma}

\begin{corollary}
    \label{thm:ACL5}
    Consider an embedding $\mathcal{E}$ of a complete affects chain $\mathscr{C}_1$ from $S_1$ to itself, forming an ACL5 (cf. \cref{def:ACL5}). Then $\Fut_s (\hat \SC_i) = \Fut_s (\SC_i) = \Fut_s (\SC_1)$.
    In particular, this implies that any embedding of an ACL5 is not stable.
\end{corollary}

This heavily constricts the space of compatible embeddings for ACL5. In addition to being unstable, depending on the properties of $\TT$, we gain additional conditions by forcing all futures of ORVs to coincide.
In particular, if there is at least one $S_k$ in a complete affects chain $\mathscr{C}$ with $\abs{S_k} = 1$, all sets of random variables in $\mathscr{C}$ must be accessible in a region shaped like a (generalized) light cone.\footnote{
In this sense, a light cone is an allowed shape for the future of a single point in the respective poset $\TT$.}
Therefore, in a join-semilattice, we get no restrictions (cf. \cref{thm:jsl}), while in a join-free poset, we are heavily restricted.

\begin{lemma}
    \label{thm:corv-collapse}
    Consider an embedding $\mathcal{E}$ of a complete affects chain $\mathscr{C}$ into a join-free poset $\mathcal{T}$, where there exists a $S_j \in \mathscr{C}$ such that $\abs{S_j} = 1$.
    Then, any set of ORVs $S_i \in \mathscr{C}$ fulfills the condition
    \begin{equation}
        \exists e_k \in S_i \ \forall e_i \in S_i \ : O(e_k) \succeq O(e_i) \, .
    \end{equation}
    \begin{proof}
        This follows directly from the fact that $\Fut_s (\SC_i)$ must be shaped like an individual light cone, which in a join-free poset is only possible if there exists an $e_k$ with $\Fut(\varepsilon(e_k)) = \Fut_s (\SC_i)$.
    \end{proof}
\end{lemma}

In the language of \cref{sec:minkowski-prop}, this is equivalent to $\abs{\spann(S_i)} = 1 \ \forall S_i \in \mathscr{C}$, if the embedding is non-degenerate.

Moreover, we can use this result to attain further corollaries for more special types of causal loops. If these also include individual RVs in their affects chains, the respective embeddings are not only unstable, but also degenerate.

\begin{corollary}[{\cite[Lem.~VI.2]{VVC}}]
    Any embedding of an ACL2a (cf. \cref{def:ACL2a}) is degenerate and unstable.
    \begin{proof}
        This is a special case of \cref{thm:ACL5}.
        For individual RVs $X, Y, Z_i$ as given in \cref{def:ACL2a}, identical futures $\Fut(\XX) = \Fut(\YY) = \Fut(\ZZ_i)$ imply identical locations $O(X) = O(Y) = O(Z_i)$ for all $i$, yielding a degenerate embedding.
    \end{proof}
\end{corollary}

\begin{corollary}[{\cite[Lem.~VI.3]{VVC}}]
    Any embedding of an ACL3 (cf. \cref{def:ACL3}) is degenerate and unstable.
    \begin{proof}
        This is a special case of \cref{thm:ACL5}.
        For individual RVs $e_1, e_2$ as given in \cref{def:ACL3}, identical futures $\Fut(\varepsilon(e_1)) = \Fut(\varepsilon(e_2))$ imply identical locations $O(e_1) = O(e_2)$, yielding a degenerate embedding.
    \end{proof}
\end{corollary}

Using these results, we take another look at the example for a non-degenerate compatible embedding for an ACL5 which was exemplarily given in \cite[Ex.~VI.2]{VVC}.

\begin{example}
    \label{ex:ACL5}
    Let $S = \{A, B, C \}$ and consider the ACL5 given by $\mathscr{A} = \{ B \affects AC, AC \affects B \}$, which we have already discussed before in the beginning of \cref{sec:stable-aff}.
    Then for any compatible embedding $\mathcal{E}$, we get $\Fut_s (\AA\CC) = \Fut (\BB)$ by \cref{thm:cac-embedding}.
    In a join-semilattice, a non-degenerate embedding exists by embedding $\BB$ into the point of earliest joint future of $\AA$ and $\CC$. However, this embedding is not stable,
    as the condition on the futures violates \cref{def:stable-supp}.
    On the contrary, in a join-free poset, every compatible embedding is degenerate, since \cref{thm:corv-collapse} would imply
    $\Fut (\CC) = \Fut (\BB) \ \lor \ \Fut (\AA) = \Fut (\BB)$.
\end{example}

Therefore, this example can not be embedded into 3+1-Minkowski spacetime.

\begin{lemma}
    \label{thm:merge-branches}
    Let $I$ be an index set. Consider sets of RVs $X, S_i \subset S \ \forall i \in I$ with a set of complete affects chains $\mathscr{C}_{i}$ from every $e_i \in X$ to some $S_i$.
    Then with respect to any compatible embedding $\mathcal{E}$,
    $\Fut_s \left( \bigcup_{i \leq \abs{S}} \SC_i \right) \subseteq \Fut_s (\XX)$.
    \begin{proof}
        We have $\Fut_s (\SC) = \bigcap_{e_i \in S} \Fut (\varepsilon(e_i))$ and, by \cref{thm:cac-embedding}, $\Fut_s ( \SC_i ) \subset \Fut( \varepsilon(e_i) )$, where $\varepsilon(e_i)$ denotes the ORV associated to the RV $e_i$.
        Since an intersection of subsets is a subset (or equal) to the intersection of original sets, the claim follows.
    \end{proof}
\end{lemma}

While intuitively, one expects that demanding support stability would carry over to a strict subset relation here, this is not necessarily the case for a general poset.

\begin{example}
    Consider a poset $\TT = \{ a, b, c, d \}$, where the complete set of order relations is given by $a \prec b \prec c \succ d$ and embed RVs $A, B, C, D \in S$ into the respective locations.
    Further, assume $\mathscr{A} = \{ A \affects B \}$ and therefore the presence of a complete affects chain from $A$ to $B$. Note that for this choice of $\mathscr{A}$, the embedding is stable and compatible.
    Then $\Fut_s (\AA \DD) = \Fut_s (\BB \DD)$ even though $\Fut (\BB) \subsetneq \Fut (\AA)$.
\end{example}

In higher-dimensional Minkowski spacetime, we see that this scenario is not possible due to \cref{thm:conical}.

\begin{corollary}
    \label{thm:ACL6a}
    There exists no stable embedding for an ACL6a. (cf. \cref{def:ACL6a})
    \begin{proof}
        Let all $S_i, S_i', e_i$ be as in \cref{def:ACL6a} and assume the presence of a stable embedding.
        Then, by the first property of this definition, we get $\Fut_s (\SC_2) \subsetneq \Fut_s (\SC_1)$.
        By \cref{thm:merge-branches} the second property allows to further deduce that $\Fut_s (\SC_k) \subseteq \Fut_s (\SC_2) \subsetneq \Fut_s (\SC_1)$.
        With the third and fourth property of the definition of an ACL6a, we can repeat this process until we reach $\Fut_s (\SC_1) \subsetneq \Fut_s (\SC_1)$. $\lightning$
    \end{proof}
\end{corollary}

This basically captures the idea that if an ACL has at least one central nexus $S_1 \affects S_2$ in a causal loop, from which $S_1$ is always reached again somehow, it can not have a stable embedding.

\begin{theorem}
    \label{thm:no-stable-loops}
    Stable embeddings of ACL1–ACL6a are impossible in any spacetime $\TT$.
    \begin{proof}
        This follows directly by conjoining \cref{thm:ACL5} for ACLs of type 1–5, which all constitute special cases of type 5,
        and \cref{thm:ACL6a} for ACLs of type 6a, which include ACLs of type 6 as a special case.
    \end{proof}
\end{theorem}
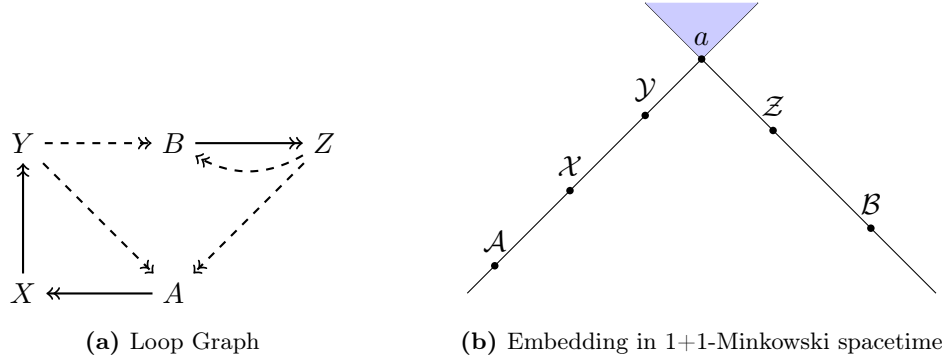
\begin{figure}[t]
	\centering
    \begin{subfigure}[b]{0.4\textwidth}
        \centering
        \begin{tikzpicture}
            \node (X) at (0,0) {$X$};
            \node (Y) at (0,2) {$Y$};
            \node (A) at (2,0) {$A$};
            \node (B) at (2,2) {$B$};
            \node (Z) at (4,2) {$Z$};

            \begin{scope}[>={to[black]},
                          every edge/.style=affectsarrow]
                \path [->>] (A) edge (X);
                \path [->>] (X) edge (Y);
                \path [dashed,->>] (Y) edge (A);
                \path [dashed,->>] (Y) edge (B);
                \path [->>] (B) edge (Z);
                \path [dashed,->>] (Z) edge (A);
                \path [dashed,->>] (Z) edge[bend left] (B);
            \end{scope}
        \end{tikzpicture}
        \caption{Loop Graph}
        \label{fig:ACL7-clg}
    \end{subfigure}%
    \begin{subfigure}[b]{0.4\textwidth}
        \centering
        \begin{tikzpicture}[dot/.style={circle,inner sep=1pt,fill,name=#1}]

            \node [dot=A,label=$\AA$] at (0,0) {};
            \node [dot=X,label=$\XX$] at (1,1) {};
            \node [dot=Y,label=$\YY$] at (2,2) {};
            \node [dot=Z,label=$\ZZ$] at (3.7,1.8) {};
            \node [dot=B,label=$\BB$] at (5,0.5) {};

            \node (left) at (-0.5,-0.5) {};
            \node (right) at (6,-0.5) {};
            \draw (left) -- ++(4,4);
            \draw (right) -- ++(-4,4);
            \fill[fill=blue!20] (2.75,2.75) -- (3.5,3.5) -- (2,3.5) -- (2.75,2.75);

            \node [dot=join,label=$a$] at (2.75,2.75) {};



        \end{tikzpicture}
        \caption{Embedding in 1+1-Minkowski spacetime}
        \label{fig:ACL7-compat}
    \end{subfigure}%
	\caption[Loop graph and a stable embedding of \cref{ex:ACL7} into 1+1-Minkowski spacetime.]{
        Loop graph (a) and sketch of a compatible, stable, non-degenerate embedding of \cref{ex:ACL7} into 1+1-Minkowski spacetime (b).
        In the latter, all ORVs which are part of the causal loop are located on a light-like surface, and therefore on the boundary of the light cone of the respecive earlier RVs.
        Indicated in blue is $\Fut_s (\AA\BB)$, which is the region the causal loop can be operationally detected in, and equal to the future of $a$.
    }
	\label{fig:ACL7}
\end{figure}%
\begin{example}
    \label{ex:ACL7}
    Let $X, Y, Z, A, B \in S$. Consider $\mathscr{A} = \{ X \affects Y, Y \affects AB, A \affects X, Z \affects AB, B \affects Z \}$.
    As shown in \cite[Ex.~B.2]{VVC} and apparent from the non-empty loop graph shown in \cref{fig:ACL7-clg}, this set of affects relations constitutes an affects causal loop (of type 7), which is not built from complete affects chains.
    Embedding this into a poset spanned by the conditions
    \begin{align}
        \AA \prec \XX \prec \YY &\prec a \, , \\
        \BB \prec \ZZ &\prec a \, ,
    \end{align}
    we obtain an embedding which is both stable and minimum-stable.
    Note that we have $\min \Fut_s (\AA\BB) = \{ a \}$ here.
    This embedding can be realized in 1+1-Minkowski spacetime, as shown in \cref{fig:ACL7-compat}, but not in the higher-dimensional case.
    However, this realization does not satisfy chronological stability.
\end{example}

\begin{figure}[t]
	\centering
    \begin{subfigure}[b]{0.4\textwidth}
        \centering
        \begin{tikzpicture}
            \node (A) at (0,0) {$A$};
            \node (B) at (2,0) {$B$};
            \node (C) at (1,1.732) {$C$};

            \begin{scope}[>={to[black]},
                          every edge/.style=affectsarrow]
                \path [dashed,->>] (A) edge[bend right] (B);
                \path [dashed,->>] (B) edge[bend right] (C);
                \path [dashed,->>] (C) edge[bend right] (A);
                \path [dashed,<<-] (A) edge (B);
                \path [dashed,<<-] (B) edge (C);
                \path [dashed,<<-] (C) edge (A);
            \end{scope}
        \end{tikzpicture}
        \caption{Loop Graph}
        \label{fig:ACL12-clg}
    \end{subfigure}%
    \begin{subfigure}[b]{0.4\textwidth}
        \centering
        \begin{tikzpicture}
            \begin{scope}[every node/.style={rectangle,thick,draw}]
                \node (A) at (0,0) {$O(A)$};
                \node (B) at (2,0) {$O(B)$};
                \node (C) at (4,0) {$O(C)$};
                \node (D) at (2,2) {$O(A) \lor O(B) \lor O(C)$};
            \end{scope}

            \begin{scope}[every edge/.style={draw=black,thick}]
                \path [-] (A) edge (D);
                \path [-] (B) edge (D);
                \path [-] (C) edge (D);
            \end{scope}
        \end{tikzpicture}
        \caption{Hasse diagram}
        \label{fig:ACL12-compat}
    \end{subfigure}%
	\caption[Loop graph and a stable embedding of \cref{ex:ACL12}.]{
        Loop graph (a) for \cref{ex:ACL12} and Hasse diagram (b, cf. \cref{fig:hasse}) for a compatible, stable, non-degenerate embedding into a discrete join-semilattice.
    }
	\label{fig:ACL12}
\end{figure}

\begin{example}
    \label{ex:ACL12}
    Let $A, B, C \in S$. Consider $\mathscr{A} = \{ A \affects BC, B \affects AC, C \affects AB \}$. Then, as can be seen in \cref{fig:ACL12-clg}, this set of affects relations implies the presence of a causal loop.
    For this loop, there exists a compatible embedding into a spacetime $\TT$ which is both stable and minimum-stable, as shown in \cref{fig:ACL12-compat}.
    Here, $\TT$ satisfies the union property (cf. \cref{def:union}), but still has $\Fut_s (\AA\BB) = \Fut(\AA\CC) = \Fut(\BB\CC)$.
\end{example}

\subsection{Higher-dimensional Minkowski Spacetime} \label{sec:minkowski-loops}

In this section, we will give some arguments for embedding affects causal loops into Minkowski spacetime.
To do so, we use geometrical properties of higher-dimensional Minkowski spacetime, as derived in \cref{sec:minkowski-prop}, as well as results regarding causal loops, compatibility and stability which have been derived over the course of this work.

In the previous section, we have found some examples for compatible embeddings for affects causal loops. However, in all of these cases we have found that some $\AA, \BB, \CC \in \SC$ with $\AA \unord \BB$ exist, which satisfy $\Fut_s (\AA\CC) = \Fut_s (\BB\CC)$.
In 1+1-Minkowski spacetime, this implies that $\BB$ and $\CC$ must be light-like with regard to each other, as otherwise $O(A) \lor O(B) = O(A) \lor O(C)$ can not be fulfilled here.
As in higher dimensions, we presume Minkowski spacetime to show conicality (cf. \cref{def:conical}) and the related location symmetry (cf. \cref{def:location-symmetry}), this class of embeddings can not be realized there.
Therefore, showing all compatible embeddings to be of this type would rule out all non-degenerate embeddings for higher-dimensional Minkowski spacetime.

Further, we can provide a link between non-degeneracy and stability, which can be made for any conical spacetimes like Minkowski spacetime.

\begin{lemma}
    \label{thm:stable-degenerate}
    Any non-degenerate compatible embedding $\mathcal{E}$ into a conical poset $\TT$ (cf. \cref{def:conical}) is stable.
    \begin{proof}
        We show the contraposite of this claim.
        For an embedding to be unstable, there must exist $\XX, \YY \subset \SC$ disjoint with $\Fut_s (\XX) = \Fut_s (\YY)$.
        However, by conicality, this implies that $\forall \XX_i \in \spann(\XX) \ \exists \YY_i \in \spann(\YY) : O(\XX_i) = O(\YY_i)$.
        Therefore, the embedding is degenerate. This concludes the proof for the contraposite.
    \end{proof}
\end{lemma}

Presuming \cref{thm:conical} already yields formally that all classes of ACLs built from complete affects chains, as we have given in \cref{sec:acl}, have no non-degenerate embedding into Minkowski spacetime.

\begin{corollary}
    Non-degenerate embeddings of ACL1–ACL6a into any conical poset $\TT$ are impossible.
    \begin{proof}
        By \cref{thm:no-stable-loops}, any compatible embedding of ACL1-ACL6a into $\TT$ is unstable. By \cref{thm:stable-degenerate}, for any conical poset every unstable embedding is degenerate.
        Therefore, we get that any such embedding is degenerate, yielding the claim.
    \end{proof}
\end{corollary}

\begin{corollary}
    Presume $n$+1-Minkowski spacetime to show conicality for $n \geq 2$, as given by \cref{thm:conical}. Then, non-degenerate embeddings of ACL1-ACL6a into Minkowski spacetime are impossible.
\end{corollary}

\section{Compatibility for Higher-Order Affects Relations: Role of Indecreasability} \label{sec:compat-indec}

Up until now, we have focused on the case of 0th-order affects relations when studying the compatibility of affects relations.
However, we can also study how the properties of HO affects relations, as established in \cref{sec:affects-new}, interact with compatibility.
In particular, this will allow us to judge how far our results regarding the embeddability of causal loops carry over to the higher-order case.

In \cref{sec:compat}, we have seen that there exists no obvious way to transform a set of HO affects relations to 0th order while leaving both the causal structure and its compatibility invariant.
This poses the question whether it may be possible to adjust the notion of compatibility in some way to address these problems. Indeed, the notion of indecreasability introduced in \cref{def:indecreasable} hints at a way out: Here, we have seen in \cref{thm:irr-indec} that in a causal model over a set of RVs $S$, for any affects relation which is both irreducible and indecreasable, for all $X, Y, Z, W \subset S$ disjoint we have
\begin{equation}
    \begin{split}
        X \affects Y \given \{ \doo(Z), W \} \ \text{irreducible and indecreasable} \\
        \implies \forall \, e_{XZ} \in XZ \quad \exists \, e_{YW} \in YW \ \colon \quad e_{XZ} \cause e_{YW} \, .
    \end{split}
\end{equation}
This shows that in addition to the symmetry between $Y$ and $W$ we have seen in \cref{thm:affects-to-cause} for irreducible affects relations, a certain symmetry between $X$ and $Z$ exists for affects relations that are additionally indecreasable. For any irreducible affects relation, the symmetry between $Y$ and $W$ carries over (cf. \cref{thm:conditional-equiv}) to the compatibility condition \hyperref[def:compat]{\textbf{compat-irreducible}}, demanding
\begin{equation}
    X \affects Y \given \{ \doo(Z), W \} \ \text{irreducible} \quad \implies \quad \RC_Y \cap \RC_W \cap \RC_Z \subseteq \RC_X \, .
\end{equation}
This naturally poses the question if something similar is possible by studying compatibility for indecreasable affects relations, even though the original compatibility condition does not exhibit the respective symmetry.

Therefore, we will suggest stronger versions of compatibility, and study their properties and how they relate to original compatibility.
These relations will provide strong justification why these versions of compatibility are natural choices.

\begin{definition}[Strong and Weak Indecreasability]
    \label{def:indecreasable-strong}
    An affects relation $X \affects Y \given \{ \doo(Z), W \}$ is \emph{strongly indecreasable} if it is indecreasable (cf. \cref{def:indecreasable}) and $X \naffects Y \given W$.
    An indecreasable affects relation that is not strongly indecreasable is called \emph{weakly indecreasable}.
\end{definition}

By comparing definitions, all first-order affects relations which are indecreasable are strongly indecreasable.

\begin{definition}[\textbf{compat-strong-indecreasable}]
    \label{def:compat-strong-indecreasable}
    Let $\SC$ be a set of ORVs from a set of RVs $S$ and a poset $\TT$ with an embedding $\mathcal{E}$. Then a set of affects relations $\mathscr{A}$ is said to satisfy \textbf{compat-strong-indecreasable} with regard to $\mathcal{E}$ if the following conditions hold:
    \begin{itemize}
        \item \textbf{compat1-strong-indecreasable}: Let $X, Y \subset S$ be disjoint non-empty sets of RVs, $W, Z \subset S$ two more sets of RVs, potentially empty.
        If ($X \affects Y \given \{ \doo(Z), W \}$) $\in \mathscr{A}$ is \textbf{strongly indecreasable}, then
        \begin{itemize}
            \item $\RC_{YWX} = \RC_Y \cap \RC_W \cap \RC_X \subseteq \RC_Z \,$.
        \end{itemize}
        \item \textbf{compat1-irreducible}: Let $X, Y \subset S$ be disjoint non-empty sets of RVs, $W, Z \subset S$ two more disjoint sets of RVs, potentially empty.
        If ($X \affects Y \given \{ \doo(Z), W \}$) $\in \mathscr{A}$ is \textbf{irreducible}, then
        \begin{itemize}
            \item $\RC_{YWZ} = \RC_Y \cap \RC_W \cap \RC_Z \subseteq \RC_X \,$.
        \end{itemize}
        \item \textbf{compat2}: With respect to $\mathcal{E}$,
        $\RC_X = \Fut(\XX) \ \forall \XX \in \SC$.
    \end{itemize}
\end{definition}

\begin{definition}[\textbf{compat-indecreasable}]
    \label{def:compat-indecreasable}
    The definition is identical to \textbf{compat-strong-indecreasable}, except for relaxing \hyperref[def:compat-strong-indecreasable]{\textbf{compat1-strong-indecreasable}} to \textbf{compat1-indecreasable} by including arbitrary indecreasable affects relations.
    Specifically, this condition reads:
    \begin{itemize}
        \item \textbf{compat1-indecreasable}: Let $X, Y \subset S$ be disjoint non-empty sets of RVs, $W, Z \subset S$ two more sets of RVs, potentially empty.
        If ($X \affects Y \given \{ \doo(Z), W \}$) $\in \mathscr{A}$ is \textbf{indecreasable}, then
        \begin{itemize}
            \item $\RC_{YWX} = \RC_Y \cap \RC_W \cap \RC_X \subseteq \RC_Z \,$.
        \end{itemize}
    \end{itemize}
\end{definition}

As discussed in \cref{sec:affects-new}, indecreasable affects relations can be defined by including specific information regarding the \textit{absence} of affects relations, requiring an additional set of affects relations known to be absent.
Contrasting the new notions of \textbf{compat-(strong-)indecreasable} with the original one of \hyperref[def:compat]{\textbf{compat-irreducible}} introduced in \cref{sec:compat}, we observe that they are precisely related by swapping $\RC_X$ and $\RC_Z$. The reason for this is rooted in the transformation properties of affects relations and will be explored further in the remainder of this section.

\begin{lemma}
    \hyperref[def:compat-indecreasable]{\textbf{compat-indecreasable}} $\implies$
    \hyperref[def:compat-strong-indecreasable]{\textbf{compat-strong-indecreasable}} $\implies$
    \hyperref[def:compat]{\textbf{compat-irreducible}}.
    \begin{proof}
        The first implication follows by relaxing the condition of \hyperref[def:compat-indecreasable]{\textbf{compat1-indecreasable}} to only hold for strongly indecreasable affects relations instead of for all indecreasable affects relations.
        The second implication follows by relaxing \hyperref[def:compat-strong-indecreasable]{\textbf{compat-strong-indecreasable}} by removing \hyperref[def:compat-strong-indecreasable]{\textbf{compat1-strong-indecreasable}}. The remaining set of conditions is identical to \hyperref[def:compat]{\textbf{compat-irreducible}}.
    \end{proof}
\end{lemma}

\begin{lemma}
    \label{thm:compat-irr-to-strong-first}
    Let $X, Y, Z, W \subset S$ as in \hyperref[def:compat]{\textbf{compat-irreducible}}, with the respective affects relation given by $X \affects Y \given \{ \doo(Z), W \}$.
    Further, let it be known that $X \naffects Y \given \{ \doo(Z \setminus e_Z), W \}$ for a given $e_Z \in Z$. Then for any compatible embedding $\mathcal{E}$,
    \begin{equation}
        \hyperref[def:compat]{\textbf{compat-irreducible}} \implies \Fut_s (\YY\WW\XX) \cap \Fut_s (\tilde{\ZZ}) \subset \Fut_s (e_\ZZ) \, .
    \end{equation}
    \begin{proof}
        By \cref{thm:HO-switch},
        \begin{equation}
            X \affects Y \given \{ \doo(Z), W \} \ \land \ X \ \naffects Y \given \{ \doo(\tilde{Z}), W \} \implies
               e_Z \affects Y \given \{ \doo(\tilde{Z}), W \} \ \lor \
               e_Z \affects Y \given \{ \doo(\tilde{Z} X), W \}
        \end{equation}
        either of which is irreducible, but not necessarily indecreasable. Therefore, by \hyperref[def:compat]{\textbf{compat1-irreducible}}, we get
        \begin{equation}
            \RC_{YW} \cap \RC_{\tilde{Z}} \subset \RC_{e_Z} \ \lor \
            \RC_{YW} \cap \RC_{X\tilde{Z}} \subset \RC_{e_Z} \implies
            \RC_{YWX} \cap \RC_{\tilde{Z}} \subset \RC_{e_Z} \, .
        \end{equation}
        Using \hyperref[def:compat-strong-indecreasable]{\textbf{compat2}}, we get the claim.
    \end{proof}
\end{lemma}

This statement can be used already to state a tighter result on compatibility, when information on the absence of affects relations of lower order is known.
It hints at a more general possibility to arrive at stronger compatibility conditions using the absence of affects relations, as we have similarly discussed for causal discovery in \cref{sec:graph}.
However, for this work, we will focus on the special case of indecreasability to arrive at a concise statement.

\begin{lemma}
    \label{thm:compat-irr-to-strong-second}
    Let $X, Y, Z, W \subset S$ as in \hyperref[def:compat]{\textbf{compat-irreducible}} and let their respective affects relation $X \affects Y \given \{ \doo(Z), W \}$ be indecreasable. Then for any compatible embedding $\mathcal{E}$,
    \begin{equation}
        \hyperref[def:compat]{\textbf{compat-irreducible}} \implies \Fut_s (\XX\YY\WW\ZZ) = \Fut_s (\XX\YY\WW\ZZ \setminus e_\ZZ) \quad \forall e_\ZZ \in \ZZ \, .
    \end{equation}
    \begin{proof}
        Analogous to \cref{thm:compat-irr-to-strong-first}, we can use \cref{thm:HO-switch-indecreasable} and transform
        \begin{align}
            \RC_{YWX} \cap \RC_{Z \setminus e_Z} \subset \RC_{e_Z} \implies
            \RC_{YWX} \cap \RC_{Z \setminus e_Z} &\subset \RC_{YWX} \cap \RC_{e_Z} &\forall e_Z \in Z \\
            \implies
            \RC_{YWX} \cap \RC_{Z \setminus e_Z^i} &\subset \RC_{YWX} \cap \RC_{e_Z^i} \cap \RC_{Z \setminus \{ e^i_Z, e^j_Z \}} &\forall e_Z^i, e_Z^j \in Z \\
            \iff
            \RC_{YWX} \cap \RC_{Z \setminus e_Z^i} &\subset \RC_{YWX} \cap \RC_{Z \setminus e_Z^j} &\forall e_Z^i, e_Z^j \in Z \, . \!\!\!
        \end{align}
        Combining these for all $e_Z$ and using \hyperref[def:compat-strong-indecreasable]{\textbf{compat2}}, we get
        \begin{align}
            \RC_{YWX} \cap \RC_{Z \setminus e^i_Z} &= \RC_{YWX} \cap \RC_{Z \setminus e^j_Z} \quad \forall e_Z^i, e_Z^j \in Z \\
            \implies \Fut_s (\XX\YY\WW\ZZ) &= \Fut_s (\XX\YY\WW\ZZ \setminus e_\ZZ) \quad \forall e_\ZZ \in \ZZ \, .
        \end{align}
        This is precisely the claim.
    \end{proof}
\end{lemma}

\begin{theorem}
    \label{thm:compat-weak-to-strong}
    Let $\TT$ show conicality (cf. \cref{def:conical}) and location symmetry (cf. \cref{def:location-symmetry}).
    Then for any non-degenerate embedding, \hyperref[def:compat]{\textbf{compat-irreducible}} $\implies$ \hyperref[def:compat-indecreasable]{\textbf{compat-indecreasable}}.
    Further, for any embedding, \hyperref[def:compat]{\textbf{compat-irreducible}} $\implies$ \hyperref[def:compat-strong-indecreasable]{\textbf{compat-strong-indecreasable}}.
    \begin{proof}
        By \cref{thm:compat-irr-to-strong-second}, for any indecreasable affects relation $X \affects Y \given \{ \doo(Z), W \}$, we get
        \begin{equation}
            \Fut_s (\XX\YY\WW\ZZ) = \Fut_s (\XX\YY\WW\ZZ \setminus e_\ZZ) \quad \forall e_\ZZ \in \ZZ \, .
        \end{equation}
        By using location symmetry on this chain of equalities we arrive at
        \begin{equation}
            \Fut_s (\XX\YY\WW) \subseteq \Fut_s (\ZZ) \quad \lor \quad
            \forall \ZZ_i \ \exists s_{\ZZ_i} \ \colon \
            \forall \ZZ_k \ \exists s_{\ZZ_k} \ \colon \
            \Fut_s (s_{\ZZ_i}) = \Fut_s (s_{\ZZ_k}) \, ,
        \end{equation}
        where $Z_i := Z \setminus e_i$.
        Translating from relativistic futures to accessible regions using \hyperref[def:compat-strong-indecreasable]{\textbf{compat2}}, the first alternative evaluates to $\RC_{YW} \cap \RC_X \subseteq \RC_Z$.

        For the second alternative, beware that the different $\ZZ_i$ are not disjoint.
        By conicality, the chain of equalities given by $\Fut_s (s_{\ZZ_i}) = \Fut_s (s_{\ZZ_k})$ implies that the locations of the ORVs spanning (cf. \cref{def:span}) the respecting joint futures must be identical.
        However, independent of the contents of $s_{\ZZ_i} \subseteq \ZZ_i$, for any $e_k \in s_{\ZZ_i}$ there exists a $s_{\ZZ_k}$ in this chain such that $e_k \not\in s_{\ZZ_k} \subseteq \ZZ_k$.
        Therefore, multiple ORVs must share the same location to yield these identical joint futures.
        As this corresponds to a degenerate embedding, we arrive at the first claim.

        For the second claim, we can carry over the argument for the first alternative from the first claim. To conclude, we study the implication of the second alternative further. By using \cref{thm:HO-transfer}, this can be transformed:
        \begin{align}
            & X \affects Y \given \{ \doo(Z), W \} \\
            \implies &
            X \affects Y \given W \ \lor \ Z \affects Y \given W \ \lor \ Z \affects Y \given \{ \doo(X), W \} \\
            \implies &
            \exists \tilde{s}_Z \subseteq Z: \ \tilde{s}_Z \affects Y \given W \ \lor \ \tilde{s}_Z \affects Y \given \{ \doo(X), W \} \ \text{irreducible}
        \end{align}
        Here we have used that as $X \affects Y \given \{ \doo(Z), W \}$ is strongly indecreasable, $X \naffects Y \given W$,\footnote{
            If we have only general indecreasability, we can transform to
            $e_Z \affects Y \given \{ \doo(Z \setminus e_Z, W \} \ \lor \ e_Z \affects Y \given \{ \doo(X Z \setminus e_Z), W \}$.
            However, applying \hyperref[def:compat-strong-indecreasable]{\textbf{compat1-irreducible}} to this relation gives a trivial constraint.}
        as well as \cref{def:reducible} to arrive at an irreducible affects relation.
        Since we know that in this case $\RC_{\tilde{s}_Z} = \RC_Z$, we can derive using \hyperref[def:compat]{\textbf{compat-irreducible}}:
        \begin{equation}
            \RC_{YW} \subseteq \RC_Z \ \lor \ \RC_{YW} \cap \RC_X \subseteq \RC_Z
            \implies \ \RC_{YW} \cap \RC_X \subseteq \RC_Z \, .
        \end{equation}
        Therefore, for both cases we get \hyperref[def:compat-strong-indecreasable]{\textbf{compat-strong-indecreasable}}.
    \end{proof}
\end{theorem}

\begin{corollary}
    Let $\TT$ show location symmetry, and $\mathcal{E}$ be an embedding of a set of affects relations which satisfies \hyperref[def:compat]{\textbf{compat-irreducible}}, but not \hyperref[def:compat-indecreasable]{\textbf{compat-indecreasable}}. Then, for any affects relation $(X \affects Y \given \{ \doo(Z), W \}) \in \mathscr{A}$ individually not satisfying \hyperref[def:compat-indecreasable]{\textbf{compat-indecreasable}}, all elements of $Z$ share the same location in $\TT$.
\end{corollary}

Therefore, in any poset $\TT$ with location symmetry, all prior interventions which are not done \enquote{before} studying an affects relation need to be performed at a single point in spacetime.
We will consider an example to showcase the different implications for strong and weak indecreasability.

\begin{example}
    For a set of RVs $S = \{ X, Y, Z_1, Z_2 \}$, consider an embedding $\mathcal{E}$ into a poset $\TT$, satisfying an affects relation $X \affects Y \given \doo(Z_1 Z_2)$ that is irreducible.
    If this relation is weakly indecreasable and $O(Z_1) = O(Z_2)$, \hyperref[def:compat]{\textbf{compat-irreducible}} poses no stronger restriction than $\RC_Y \cap \RC_{Z_1 Z_2} \subseteq \RC_X$.
    Hence, it is possible for the nodes $Z_1 Z_2$ that are intervened on to be in the future of both $X$ and $Y$: $O(X) \prec O(Z_1) = O(Z_2) \succ O(Y)$.
    If the affects relation is strongly irreducible, we additionally get the restriction $\RC_Y \cap \RC_X \subseteq \RC_{Z_1 Z_2}$, ruling out that possibility.
\end{example}

We continue by showing that location symmetry indeed poses a necessary condition for \cref{thm:compat-weak-to-strong} to hold.

\begin{lemma}
    For an embedding into a generic poset $\TT$, \hyperref[def:compat]{\textbf{compat-irreducible}} does not imply \hyperref[def:compat-strong-indecreasable]{\textbf{compat-(strong-)indecreasable}}.
    \begin{proof}
        Consider the example of a causal model over the binary RVs $X, Y, Z, B$ uniformly distributed with $Y \cong X \oplus Z \oplus B$.
        Then it is clear that $X \affects Y \given \doo(Z B)$ is irreducible and (strongly) indecreasable.
        Now consider an embedding $\mathcal{E}$ in 1+1-Minkowski spacetime where $X$ and $Z$ are spacelike separated, $Z$ is in the lightlike future of $B$ and $Y$ is in the future of $X$ but spacelike to $Z$ and $B$. (An example for an embedding satisfying this is depicted in \cref{fig:ACL7-compat}.)
        Then \hyperref[def:compat]{\textbf{compat-irreducible}} is satisfied since $\Fut (\YY\ZZ) \subseteq \Fut (\YY) \subseteq \Fut( \XX)$.
        However, $\Fut (\YY) = \Fut_s (\XX \YY) \not\subseteq \Fut_s ( \ZZ \BB)$, so \hyperref[def:compat]{\textbf{compat-(strong-)indecreasable}} fails.
    \end{proof}
\end{lemma}

We continue by using \cref{thm:compat-weak-to-strong} to derive a corollary about restrictions on the accessible regions imposed for affects relations which are both irreducible and (strongly) indecreasable, if \hyperref[def:compat-strong-indecreasable]{\textbf{compat-(strong-)indecreasable}} holds for a given embedding.

\begin{corollary}
    \label{thm:compat-strong-sets}
    \hyperref[def:compat-strong-indecreasable]{\textbf{compat-(strong-)indecreasable}} implies the following properties for an affects relation $(X \affects Y \given \{ \doo(Z), W \}) \in \mathscr{A}$ which is both irreducible and (strongly) indecreasable:
    \begin{enumerate}
        \item $\RC_{YW} \cap \RC_X = \RC_{YW} \cap \RC_Z$
        \item $\RC_{YW} \cap \RC_X \subseteq \RC_X \cap \RC_Z \supseteq \RC_{YW} \cap \RC_Z$
    \end{enumerate}
    \begin{proof}
        This follows from \hyperref[def:compat-strong-indecreasable]{\textbf{compat1-irreducible}} and \hyperref[def:compat-indecreasable]{\textbf{compat1-indecreasable}} by $\RC_{YW} \cap \RC_Z \subseteq \RC_X$ and $\RC_{YW} \cap \RC_X \subseteq \RC_Z$, respectively.
        By elementary set theory, this implies
        \begin{align}
            \RC_{YW} \cap \RC_Z \subseteq \RC_X \cap \RC_{YW} \ &\land \ \RC_{YW} \cap \RC_X \subseteq \RC_Z \cap \RC_{YW} \quad and \\
            \RC_{YW} \cap \RC_Z \subseteq \RC_X \cap \RC_Z \ &\land \ \RC_{YW} \cap \RC_X \subseteq \RC_Z \cap \RC_X \, .
        \end{align}
        These are precisely the first and second claim.
    \end{proof}
\end{corollary}

With this, we get the remarkable result that the information in $X$ and $Z$ for $X \affects Y \given \doo(Z)$ irreducible and indecreasable does not only bear a full symmetry with regard to the implied causal relations as shown in \cref{thm:affects-to-cause-HO}, but also appear symmetric when considering non-degenerate embeddings in a spacetime with location symmetry.
That this symmetry seems to be a property which is fulfilled in our current model of physical spacetime, but not in general posets, is remarkable in itself.

We will top this off by deriving a compatibility condition of the form $\RC_{YW} \subset \RC_X \cap \RC_Z$ for a certain class of spacetimes in the non-degenerate case, allowing to drop the distinction between $X$ and $Z$ here. Thereby, we will conclude the analogy to \cref{thm:irr-indec}.
Otherwise, by including the case of degenerate embeddings, we retain an actual duality between $X$ and $Z$, which however only holds for strongly indecreasable affects relations.

\begin{lemma}
    \label{thm:irr-indec-compat}
    Let $\TT$ show conicality (cf. \cref{def:conical}) and location symmetry (cf. \cref{def:location-symmetry}) and $\mathcal{E}$ be a non-degenerate embedding in $\TT$ satisfying \hyperref[def:compat-indecreasable]{\textbf{compat-indecreasable}}. Then
    \begin{equation}
        \label{eq:irr-indec-compat}
        (X \affects Y \given \{ \doo(Z), W \}) \in \mathscr{A} \ \text{irreducible and indecreasable}
        \quad \implies \quad
        \RC_Y \cap \RC_W \subseteq \RC_X \cap \RC_Z \, .
    \end{equation}
    \begin{proof}
        For the given affects relation, \cref{thm:compat-strong-sets}.1 implies that $\RC_{YW} \cap \RC_X = \RC_{YW} \cap \RC_Z$.
        By \hyperref[def:compat-strong-indecreasable]{\textbf{compat2}}, the claim can be rewritten to $\Fut_s (\YY\WW\XX) = \Fut_s (\YY\WW\ZZ)$.
        Using location symmetry, we arrive at
        \begin{equation}
            \Fut_s (\YY\WW) \subseteq \Fut_s (\XX\ZZ) \ \lor \ \exists s_{\XX} \subseteq \XX, \, s_{\ZZ} \subseteq \ZZ \ \text{non-empty} \ \colon \ \Fut_s (s_{\XX}) = \Fut_s (s_{\ZZ}) \, .
        \end{equation}
        By conicality, the second option implies the embedding to be degenerate, in contrast to the assumption. Therefore, the claim, corresponding to the first option, follows.
    \end{proof}
\end{lemma}

Effectively, this gives that here, the interventional data given by $X$ and $Z$, which can be associated with the preparation of an experiment, must be entirely given before outcome which is signaled due to this preparation.
Also, in contrast to \cref{sec:compat} and reusing the notation defined by \cref{not:affects-tuple}, the transformation
\begin{equation}
    (X, Y, Z, W) \mapsto (XZ, YW, \emptyset, \emptyset)
\end{equation}
actually allows to transform general irreducible and indecreasable conditional higher-order affects relations to simple affects relations, while keeping both the implied causal structure and compatibility unchanged.

Therefore, we have achieved our goal to construct a parallel between the form of causal relations inferable from a set of affects relations, and the compatibility conditions necessary on an embedding into a spacetime poset $\mathcal{T}$, for the special case of a non-degenerate embedding and Minkowski-like spacetimes.
We conclude by generalizing any proof ruling out non-degenerate embeddings for causal loops built from 0th-order affects relations to the higher-order case.

\begin{corollary}
    Assume there exists a proof ruling out the presence of causal loops implied by a set of irreducible 0th-order affects relations $\mathscr{A}'$ for non-degenerate compatible embeddings into a certain poset $\TT$.
    Then, for any $\TT$ that shows conicality and location symmetry, the same proof generalizes to a set of higher-order affects relations $\mathscr{A}$ which are both irreducible and indecreasable.
    \begin{proof}
        Consider a transformation of $\mathscr{A}$ to a set of 0th-order unconditional affects relations $\mathscr{A}'$, replacing each affects relation
        $X \affects Y \given \{ \doo(Z), W \}$ irreducible and indecreasable with $XZ \affects YW$ irreducible.
        Through \cref{thm:irr-indec} this implies that $\mathscr{A}'$ implies the presence of a causal loop if and only if $\mathscr{A}$ does.
        Further, by \cref{thm:irr-indec-compat}, this poses no additional restriction for compatibility, thereby excluding the possibility that there might be compatible non-degenerate embeddings (potentially, with a compatible embedding for causal loops) which can not be reached using this transformation.
        With both causal information and compatibility invariant, this concludes the proof.
    \end{proof}
\end{corollary}

\newpage
\part{Conclusion}
\section{Discussion} \label{sec:discussion}
Within this work, we have plunged into a wide variety of directions, yielding various new results. As these are abstract, yet highly operationally relevant, they may have various applications for causal inference in and beyond physics.
While this variety of results is not that unexpected when working within a formalism still at its infancy, it is still remarkable to see how many directions are fruitful regarding new insights.
Therefore, many of our results are not set and done, but rather at an intermediate state from which to continue further research.

\subsection{Causal Structures, Interventions, Affects Relations and Causal Inference}

While the results of the first part of this thesis may appear rather self-contained, there is actually a large amount of questions that warrant future research.

First, there is research left to be done on the framework of cyclic causal models itself, as we have argued in \cref{sec:cyclic-structures}. While for the classical case, it is actually known that d-separation is not the most general suitable compatibility condition for cyclic causal models, it appears that no such results exist for the non-classical case. Using a more general property here could provide valuable insight either in widening the applicability of our research regarding causal inference or in giving examples of causal models which violate properties we have technically only shown to hold using d-separation -- as is the case for our causal inference statements.

Second, it might be valuable to try to generalize the notion of (classical) interventions to non-classical nodes, as we have suggested in \cref{sec:completeness-general}. Allowing to intervene on hidden nodes, even if we are unable to associate them with a probability distribution, can have a clear operational meaning, as for example by replacing a quantum state. This might provide a valuable tool to provide additional causal inference. Thinking of local operations, one might imagine a concept of partial interventions.
If the theory in question shows reversible dynamics, we might even consider a dual concept to interventions, acting backwards on the causal structure. To do so, we could reverse the arrow directions for a given causal structure, and perform interventions on it afterwards.
This would allow to (counterfactually) set an output for a causal process and \enquote{observe} how the inputs change.
Further ideas regarding the study of such causal reversals have been studied in \cite{arxiv.1902.00129}\cite{arxiv.2104.00071}.

Third, we may wonder how various restrictions of generality in our causal models influence the power of affects relations to infer information regarding the causal structure. As we have already seen in \cref{sec:completeness-general}, in absence of unobserved nodes, information on the presence of affects relations suffices to uncover the full causal structure even in the presence of fine-tuning.
Similarly, it is interesting to consider possible stronger statements for causal inference for general acyclic causal models with unobserved nodes, or for faithful cyclic causal models.

Fourth, there is a wide variety of ideas to further develop the toolbox of causal inference within this framework. As we have already seen in \cref{sec:affects-new}, there exist powerful ways to do causal inference using the absence of affects relations, and each of the lemmata to transform one affects relation to a disjunction of affects relations effectively forms a blueprint to lead to further techniques for inference.
Moreover, it seems plausible that further transformative properties exist, leading to even more inferential statements. In particular, scenarios like $X \affects Y \given \doo(Z) \ \land \ X \naffects Y \given \doo(ZA)$ come to mind here.
Therefore, it might be worthwhile to systematically review these possibilities.

Fifth, conjoining information about affects relations with a study of the correlations in the pre- and post-intervention graph may lead to further such tools. In particular, for theories with unobserved nodes, it is still unknown whether correlations might allow to detect causal loops which are not detectable using affects relations (alone).
While we have mostly abstained from incorporating correlations in this work, more generally it poses an interesting question whether a pre-intervention correlation between two random variables in a causal model does actually imply the presence of certain affects relations.
Up until now, the only thing known here is that there is no simple correspondence: $X$ and $Y$ being correlated does not imply the presence of an affects relation between $X$ and $Y$ \cite{VVC}.
Nonetheless, we deem it probable that research of the connections between affects relations and correlations would lead to more interesting insights when fixing a particular theory (or a family thereof) than when retaining full generality.

This leads us to our sixth and final point: Of course, it is highly desirable to apply this framework and its results to a wide class of physical models, to find connections and attain a better understanding of signalling in the respective theories. Here, appealing examples would be multiple recent quantum causal models \cite{arxiv.1906.10726}\cite{Henson2014}\cite{Barrett2021}\cite{VVR}.

\subsection{Compatibility, Stability and Minkowski Spacetime}

While working on this thesis, the tenacity of questions of embeddability for causal loops has proven to be surprising: While over the course of this project, proofs ruling out all causal loops for higher-dimensional Minkowski spacetime seemed to be within our grasp multiple times, even the introduction of the causal loop graph in \cref{sec:graph} has not lead us to such results up until now.
However, there still has been considerable progress on some questions:
In \cref{sec:general-loops}, we have established that no stable embeddings exist for causal loops built from complete affects chains, and have constricted the space of possible embeddings further.
Additionally, we have shown in \cref{sec:compat-indec} that any proof for 0th-order unconditional affects relations should generalize to any set of affects relations.

Accordingly, we give multiple suggestions of directions how to continue the quest for proofs in this direction. First, it might be possible to generalize the conditions of non-degeneracy, to apply not only to individual ORVs, but, similarly to support stability, also to sets of ORVs, enforcing $\Fut_s (\AA) \neq \Fut_s (\BB)$ for a larger set in $\AA, \BB \in \SC$. Here, considerable attention is needed to find a suitable condition for $\AA$ and $\BB$. In particular, location symmetry (cf. \cref{def:location-symmetry}) gives an example where instead, equality is expected to hold for non-disjoint sets.
Further, as an intermediate step, it might be worthwhile to find proofs for known classes of incomplete affects chains, which have not been studied explicitly in this work.

However, some further research directions are apparent here as well: First and foremost, a thorough and mathematically precise study of spacetime at a purely order-theoretic level comes to mind here.
While there is a significant amount of research focusing on order-theoretic properties of timelike curves and larger compact subsets, we were unable to find any analysis on the order-theoretic properties of the individual points in spacetime with regard to the causal structure. While we have suggested a bunch of such properties in \cref{sec:minkowski-prop}, we do not presume this list to be comprehensive.
Further, the respective proof sketches require significantly more work to reach a level where they are formally precise.

Beyond that, we have seen in \cref{sec:compat-indec} that even the absence of affects relations may impact compatibility, and it would be interesting to see how this generalizes.
In particular, for the case of indecreasable affects relations, we have shown that at least for a wide class of physically relevant spacetimes, the absence of affects relations yields symmetric implications for both the causal information and compatibility, substantiating the choice of compatibility condition.
Hence, it poses a particularly intriguing question if for all such cases we find a similar symmetry, or if this symmetry can be broken for some physically relevant scenarios.

\newpage
\part{Appendix}
\begin{appendices}
\section{References} \label{sec:ref}
\printbibliography[heading=none]

\newpage
\section{List of Figures}
\listoffigures

\section{Abbreviations}
\begin{tabular}{l|l}
    ACL    & Affects Causal Loop \\
    aQFT   & algebraic Quantum Field Theory \\
    Causet & Causal set \\
    CTC    & Closed Timelike Curve \\
    DAG    & Directed Acyclic Graph \\
    DG     & Directed Graph \\
    dim.   & dimensional \\
    GPT    & Generalized Probabilistic Theory \\
    HCL    & Hidden Causal Loop \\
    HO     & higher-order \\
    ORV    & Ordered Random Variable  \\
    Poset  & Partially ordered set \\
    RV     & Random Variable \\
\end{tabular}



\end{appendices}

\end{document}